\documentclass[
a4paper,%
12pt%
]{article}

\usepackage[dvips]{graphicx}
\title{Search of a general form of superpotential in hierarchy with
discrete energy spectrum}

\author{Sergei~P.~Maydanyuk
\thanks{E-mail: maidan@kinr.kiev.ua} \\
\small\emph{Institute for Nuclear Research,
National Academy of Sciences of Ukraine}\\%
\small\emph{prosp. Nauki, 47, Kiev-28, 03680, Ukraine}}

\date{\small\today}

\begin{document}
\begin{sloppypar}

\maketitle

\begin{abstract}
In paper a generalized definition of superpotential has proposed,
which connects two one-dimensional potentials $V_{1}$ and $V_{2}$
with discrete energy spectra completely and where:
{\small
\begin{itemize}

\vspace{-1mm}
\item
for definition of an energy of factorization an arbitrary level of
the energy spectrum of the given $V_{1}$ is used and a function of
factorization is defined concerning a bound (ground or excited) state
at this energy level,

\vspace{-1mm}
\item
for the definition of the energy of factorization an arbitrary
energy (which can be not coincident with levels of the spectrum of
$V_{1}$) is used and the function of factorization is defined
concerning an unbound (or non-normalizable) state at this energy.
\end{itemize}}

\vspace{-2mm}
\noindent
It has shown, that for the unknown superpotential such its definition
follows directly from a solution of Riccati equation at the given
potential $V_{1}$.

Using the arbitrary bound state in the construction of the
superpotential, the SUSY QM methods at their effectiveness in
construction of new exactly solvable potentials on the basis of the
given potential $V_{1}$, in detail calculations of spectral
characteristics under the deformation of $V_{1}$ have been coming
to a level of methods of inverse problem.
So, if as the given potential $V_{1}$ to choose a rectangular well
with finite width and infinitely high walls, then we reconstruct by
SUSY QM approach all pictures of deformation
(without displacement of the levels in the spectrum) of this
potential and its wave functions of the lowest bound states, which
were obtained early 
by the methods of the inverse problem.
Here, interdependence between parameters of the deformation for the
methods of SUSY QM and the inverse problem has found, an analysis of
a behavior of wave functions and the potential under the deformation
has fulfilled, a classification has proposed for zero-points of the
potential, nodes of the deformed wave functions, points, where wave
functions are not deformed, an analysis of angles of wave functions
leaving from such points has fulfilled.

Using the unbound and non-normalizable states at the arbitrary energy
of factorization in the definition of the superpotential, we obtain
new types of deformations.
So, using only one superpotential, one can join two potentials, which
have the real energy spectra with own bound states and without
coincident levels.
\end{abstract}



\newpage
\tableofcontents

\newpage
\section*{Designations
\label{sec.0}}

In paper we shall use the following designations.
\begin{itemize}

\item
The first potential $V_{1}$, on the basis of which we construct
new potentials using SUSY-transformations, we shall name as
\emph{the starting potential with number ``1''}.

\item
$\varphi_{i}^{(k)}$ is wave function (WF or WFs --- in the plural case,
partial or general solution) of the potential $V_{k}$ with the number
$k$, where instead of index $i$ at the bottom we shall use one index
from the following:
{\small
\begin{itemize}

\item
$n$ --- as the number of the level of discrete energy spectrum of the
potential $V_{k}$ with number $k$, if this level does not coincide
with \emph{energy of factorization} $\cal E$
(i.~e. the energy, at which the superpotential is defined);

\item
$m$ --- as the number of the level of the discrete energy spectrum of
the potential $V_{k}$ with the number $k$, if this level coincides
with the energy of factorization $\cal E$;

\item
$w$ --- as the index at energy, which coincides with the energy of
factorization $\cal E$ and does not coincide with any level of the
energy spectrum of the starting potential $V_{1}$.
\end{itemize}}

\item
$E_{i}^{(k)}$ is the level with the number $i$ (corresponding to WF
$\varphi_{i}^{(k)}$) of the energy spectrum for the potential $V_{k}$.
We shall number the levels of the discrete spectrum so:
{\small
\begin{itemize}

\item
the lowest level, concerned with the ground bound state, by number
\emph{``1''};

\item
the next levels, located higher and described excited bound states,
by numbers \emph{``2''} and so on.
\end{itemize}}

\item
$W_{i}^{(k)}$ is the superpotential, for which as the energy of
factorization we shall choose the energy with index $i$,
and as a \emph{function of factorization} we shall choose WF
$\varphi_{i}^{(k)}$ (of bound or unbound state) of the potential
$V_{k}$.
If the function $W_{i}^{(k)}$ will be considered with only one number
$k$ in the text, then we omit this number sometimes.

\item
We shall choose such definition of operators $A_{i}$ and $A_{i}^{+}$:
\begin{equation}
\begin{array}{cc}
  A_{i} = \displaystyle\frac{d}{dx} + W_{i}(x), &
  A_{i}^{+} = -\displaystyle\frac{d}{dx} + W_{i}(x).
\end{array}
\label{eq.0.1}
\end{equation}

\item
$\bar{\varphi_{i}}^{(k)}$ is
WF of unbound or non-normalizable state, which we shall denote by
stroke on top.

\end{itemize}

\newpage
\section{Introduction 
\label{sec.introduction}}

Supersymmetric theories are ones from the most modern and vigorously
developed topics of physics of particles and interactions.
Usually, under a word \emph{``supersymmetry''} we have in mind a
relation established between bosons and fermions, which allows to
consider the boson and the fermion as two different manifestations
of an unified particle \cite{Wess.1974.NPB}.
It is interesting to note, that according to one of main principles 
of quantum mechanics (without applying a formalism of field theories),
a conception of ``wave'' (which we relate to interactions in the
simplest understanding) and a conception of ``particle'' (under
which we keep in mind the matter in the simplest understanding) can
be considered as two different manifestations of only one object ---
``particle-wave''.
Mathematically, for description of the particles and interactions
the supersymmetric models (for example, such as \emph{Minimal
Supersymmetric Standard Model}
\cite{Kazakov.hep-ph.0001257,Kazakov.hep-ph.0012288}), based on a
modern apparatus of quantum field theory mainly with application of
algebras of supersymmetry \cite{Salam.1975.PRD}, are used.
It makes study of properties of interdependence between fields of
the matter (as the fermion fields) and fields of the interactions
(as the boson fields) enough complicated.

In 1981 E.~Witten in his famous paper \cite{Witten.1981.NuclPhys}
constructed a simple example of one-dimensional supersymmetric system,
which has no any reference to the theory of fields at all!
It had opened a possibility to study absolutely independently
properties of supersymmetric transformations between partners on the
basis of the simplest examples, without using formalism of the field
theories (and without formalism for description of \emph{spin}).
Since then, after publishing of some fundamental papers, a new topic
of physics had been created and has been developed vigorously, which
is named as \emph{supersymmetric quantum mechanics (SUSY QM)}
\cite{Cooper.1995.PRPLC}.

An essential interest in SUSY QM can be explained by that it allows to
study in details the properties of the supersymmetric transformations,
using the simplest quantum systems. Of cause, then such results can be
applied in the field theories up to cosmology \cite{Yurov.2004.TMP}
and quantum gravity.
So, in the paper \cite{Sukumar.1985.JPAGB.p2917} (see p.~2917--2918)
C.~V.~Sukumar clearly explained such connection so 
(also see~\cite{Freedman.1983.APNYA,Takeda.1984} of other authors):
\emph{``It is well known that $d=1$ quantum mechanics is formally
equivalent to the $d=1$ quantum field theory with the identification
$x \to \varphi$, $p \to \partial_{t}\varphi$ and canonical
quantisation of the field $\varphi$ leads to the usual commutation
relation between $x$ and $p$ \ldots''}

From another side, the second important perspective of SUSY QM has
been opened:
it improves essentially methods of quantum mechanics directly.
Keeping in the frameworks of quantum mechanics, SUSY QM proposes to
researcher its own simple mathematically and clear enough methods for
calculation of spectral characteristics of quantum systems,
it gives answers to questions and vagueness, putted early by 
\emph{methods of direct} and \emph{inverse problems} of quantum
mechanics. First of all, one note a \emph{method of factorization}
(principally, introduced at the first time by
A.~A.~Andrianov, N.~V.~Borisov and M.~V.~Ioffe
in~\cite{Andrianov.1984} and independently by C.~V.~Sukumar
at first in~\cite{Sukumar.1985.JPAGB.L57}
and then in~\cite{Sukumar.1985.JPAGB.p2917,%
Sukumar.1985.JPAGB.p2937,%
Sukumar.1986.JPAGB.p2229,%
Sukumar.1986.JPAGB.p2297,%
Sukumar.1987.JPAGB.2461,%
Sukumar.1988.JPAGB.L455} where a main formalism of such a method was
constructed,
however, the algorithm of factorization, started else by P.~A.~M.~Dirac
in \cite{Dirac.1935} and E.~Schr\"{o}dinger in \cite{Schrodinger.1940},
can be found earlier for solution of the spectral problem with some
simplest one-dimensional potentials in quantum mechanics,
for example, see \cite{Infeld.1951.RMP,Ui.1984.PTP}),
formulation of which has been causing a large number of papers in
SUSY QM,
search of new methods and opening of new types of exactly solvable
potentials on the basis of intensive study of \emph{Darboux
transformations} (first of all, one can point out papers of
B.~F.~Samsonov, V.~G.~Bagrov with their colleagues at development
of a general formalism of these transformations with their
application in a number of different problems
\cite{Bagrov.1997.PPN,Bagrov.quant-ph/9804032,Samsonov.quant-ph.9904009,%
Samsonov.2002.PRC,Samsonov.2002.PLA},
it is necessary to note papers of other researchers at their
investigations of these transformations with application to theory
of \emph{solitons}
\cite{Matveev.1979.LMP,Matveev.1979.LMP.2,Matveev.1991},
in approaches of \emph{nonlinear supersymmetry}
\cite{Andrianov.1995.PLA,Andrianov.2003.NuclPhys,Andrianov.2004.JPAGB},
in solution of \emph{matrix Schr\"{o}dinger equations}
\cite{Samsonov.2004.JPA},
in solution of \emph{system of coupled discrete Schr\"{o}dinger equations}
\cite{Suzko.2002.PAN},
in generalization of the formalism of these transformations for
\emph{three-dimensional and many-dimensional spaces}
\cite{Andrianov.1984,Andrianov.1984.TMP,Humi.1988.JPA,Gonzalez-Lopez.1998.JGP},
in \emph{scattering theory}
\cite{Cannata.1993.JPA,Bagrov.quant-ph/9804032,Samsonov.2002.PRC},
in improvement of the \emph{methods of the inverse problem} on the
basis of these transformations \cite{Rybin.1991.JPA},
in study of interdependence between these transformations, the method
of factorization and the methods of the inverse problem 
\cite{Andrianov.1984.TMP,Luban.1986.PRD,Pursey.1986.PRD},
in intriguing and having large perspectives \emph{non-stationary
approaches} \cite{Nikitin.1992.JPAGB,Schulze-Halberg.2006.IJMP},
one can note \emph{approach of the inverse Darboux transformations}
in description of some types of exactly solvable potentials
\cite{Gomez-Ullate.2004.JPA},
and also interesting review about such Darboux transformations 
\cite{Rosu.1998} and historical paper \cite{Darboux.1882}).

An essential progress was achieved early in development of the
methods of the inverse problem in construction of the new exactly
solvable potentials on the basis of known ones
(see \cite{Zakhariev.1990.PEPAN,Zakhariev.1994.PEPAN,Zakhariev.1999.PEPAN}).
A vigorous accumulation of information about new types of the
potentials with spectral characteristics, which have explicit
analytical form, little by little forms an \emph{unified theory of
exactly solvable models (or potentials)}
(see~\cite{Zakhariev.1990.PEPAN}, p.~915).
The methods of SUSY QM, after their opening of original solutions for
new types of such potentials (for example, \emph{shape invariant
potentials} or \emph{SIP-potentials}
\cite{Gendenshtein.1983.JETPL,Cooper.1995.PRPLC}
with \emph{translations of parameters}
\cite{Cooper.1987.PHRVA,Dutt.1988.AJPIA,Barclay.1991.PHLTA},
with \emph{scaling of the parameters}
\cite{Barclay.1993.PHRVA,Khare.1993.JPAGB},
with \emph{rotations of parameters} \cite{Maydanyuk.2005.QTS-5},
\emph{class of the potentials of Shabat and Spiridonov}
\cite{Shabat.1992.INPEE,Spiridonov.1992.PRLTA,Cooper.1987.PHRVA},
which includes \emph{soliton-like reflectionless solutions} without
tunneling,
search of the reflectionless potentials with barriers 
\cite{Maydanyuk.2005.Surveys_in_HEP,Maydanyuk.2005.APNYA,Maydanyuk.2004.PAST.refl}
(where an \emph{effect of reflectionless tunneling} can appear,
existence of which is discussed up to now),
the potentials in approach of \emph{nonlinear supersymmetry}
\cite{Andrianov.2003.NuclPhys,Andrianov.2004.JPAGB}),
have proved their right to originality, and their effectiveness in
detail calculations of spectral characteristics of the potentials
comes little by little to a level of the methods of the inverse
problem.
During last two decades, one can see a continuous competition between
the methods of the inverse problem and SUSY QM methods, which
supplement each other
\cite{Sukumar.1985.JPAGB.p2937,Sukumar.1986.JPAGB.p2297,Sukumar.1988.JPAGB.L455}.
However, perhaps, increase of number of the new SUSY QM methods looks
more intensive, that one can explain by their comparative simplicity
and availability.

These arguments emphasize the interest in the further development
of the new SUSY QM methods, which give to researcher the enlarger
possibilities in construction of the new exactly solvable potentials
(on the basis of known potentials) in comparison with early known
methods.

In this paper we shall restrict ourselves by consideration of
one-dimensional potentials with discrete spectra of energy
completely.
Here, we shall analyze more generalized definition of a
superpotential, which we construct in the region of real values only
and which connects two potentials SUSY-partners in frameworks of the
most widely known approach of SUSY QM (for example, as in formalism 
\cite{Sukumar.1985.JPAGB.p2917,Cooper.1995.PRPLC}),
adding an arbitrariness into a choice of a bound state at an
\emph{energy of factorization}, which coincides with the ground or
excited level of the energy spectrum of the given potential $V_{1}$,
and introducing an arbitrary unbound and non-normalizable states at
the arbitrary energy of factorization, which can be not coincident
with the levels of the energy spectrum of $V_{1}$.
We note, that similar generalizations on the construction of the
superpotential were considered early
(see \cite{Sukumar.1986.JPAGB.p2297} --- more for the potentials
with continuous energy spectra;
\cite{Zakhariev.2002.PEPAN} p.~378--386 --- for the superpotential
and the energy of factorization in the region of complex values).
However, we see, that this question for the potentials with discrete
energy spectra has not been studied yet practically, and our analysis
has shown an extreme availability of taking into account of different
types of states at the arbitrary energy of factorization in the
definition of the superpotential, even if to find it in the region
of real values only.

It turns out, that such generalization follows directly from a form
of the superpotential, if to find it as an \emph{exact solution of
Riccati equation} at a given potential $V_{1}$, own independent
variant of which we present in Sec.~\ref{sec.2} (see also original
papers \cite{Sukumar.1985.JPAGB.p2917,Sidharth.physics.0305083,%
Rosu.quant-ph.9809021,Junker.1998.AP,Ferreira.math-ph.9807017,%
Common.1990,Chau.1987,Kumar.1986,Rogers.1985,Harnad.1983,Snyder.1983}
and monographs~\cite{Weigelhofer.1999,Hartman.1982}).
It is necessary to note, that the Riccati equation admits an
arbitrariness in a choice of boundary conditions in the definition
of the superpotential, while the standard approach to the construction
of the superpotential uses wave function of the bound (mainly, ground)
state only and, therefore, it is partial.

The generalization for definition of the superpotential requires
introduction of the \emph{unbound} or \emph{non-normalizable states}
at the arbitrary energy of factorization, and a formalism of such
states we present in Sec.~\ref{sec.4}
(also see
\cite{Sukumar.1985.JPAGB.p2917} p.~2925--2934,
\cite{Cooper.1995.PRPLC} p.~324--325,
\cite{Zakhariev.1999.PEPAN} p.~286--290, 293--297, 309--318,
\cite{Zakhariev.2002.PEPAN} p.~378--382, 382--389).
In Sec.~\ref{sec.5} we improve a method of construction of
\emph{$n$-parametric family of isospectral potentials},
developed early in~\cite{Cooper.1995.PRPLC} (see p.~326--328).

Generalizing the definition of the superpotential on the basis of
wave function of the arbitrary excited bound state of the given
potential $V_{1}$, we have achieved practically full analogous
between SUSY QM methods and the methods of the inverse problem. 
So, in Sec.~\ref{sec.7} we reconstruct all pictures of the deformation
(without of displacement of the levels) of a rectangular well with
finite width and infinitely high walls and its wave functions of the
lowest bound states, which were obtained early in the review
\cite{Zakhariev.1990.PEPAN} (see p.~916--919) on the basis of the
methods of the inverse problem.
Here, we find interdependence between main parameters of deformation
for the SUSY QM methods and the methods of the inverse problem, find
answers (confirmed by analytical expressions) on some questions,
putted early in this review.
These arguments make the SUSY QM methods practically such effective
as the methods of the inverse problem.

We see, that use of the unbound states in the definition of the
superpotential (in the energy region of the real values)
opens new original perspectives for the SUSY QM methods.
It turns out, that if to do not restrict oneself by the bound states
only in the definition of the superpotential, but to construct it on
the basis of a function, which is a general (or partial) solution
(without implying the boundary conditions for bound states on it)
of the Schr\"{o}dinger equation with the starting potential $V_{1}$
at the arbitrary energy (which can be not coincident with the energy
levels of $V_{1}$), then we obtain new exactly solvable potentials
which have own energy spectra with the bound states.
In particular, using only one superpotential, one can join two
potentials, energy spectra of which (with own bound states) have no
one coincident level (it has found at the first time in the
supersymmetric approach)!
As a demonstration, in Sec.~\ref{sec.8} we present two new
approaches of construction of the new potentials, where as the
starting potential $V_{1}$ we use the rectangular well with infinitely
high external walls.
This opens a possibility to deform at once whole energy spectrum for
the given potential (perhaps, by a given rule, for example,
as for SIP-potentials with scaling of the parameters).
Perhaps, such a way will open a possibility to connect together
boson and fermion components with absolutely different energy
spectra in the supersymmetric quantum theory of fields, using only
one superpotential.

\section{An exact analytical solution of a superpotential at the
given potential
\label{sec.2}}

To clarify, which the most general form an unknown superpotential
can have, when we know only a form of the potential, concerning
with this superpotential in the most prevailing formalism (for example,
as in~\cite{Cooper.1995.PRPLC} p.~275--277,
\cite{Sukumar.1985.JPAGB.p2917} p.~2922-2923, 2925--2927,
\cite{Andrianov.1984} p.~19--20,
at an energy of factorization, located not higher then the lowest
level of energy spectrum of the given potential), we inevitably come
to a \emph{problem of a solution of Riccati equation}, where as
unknown function the superpotential is used.

\subsection{The superpotential as the solution of the Riccati equation 
\label{sec.2.1}}

Let's consider a quantum system with a potential $V_{1}(x)$ and a
discrete energy spectrum completely.
For such a system one can write down the Schr\"{o}dinger equation:
\begin{equation}
\begin{array}{ccl}
  \hat{h}_{1} \varphi^{(1)}_{m}(x) & = &
    \biggl(-\displaystyle\frac{d^{2}}{dx^{2}} + V_{1}(x) \biggr)
      \varphi^{(1)}_{m}(x) =
      E^{(1)}_{m} \varphi^{(1)}_{m}(x),
\end{array}
\label{eq.2.1.1}
\end{equation}
where
$\hat{h}_{1}$ is Hamiltonian of the system,
$\varphi^{(1)}_{m}(x)$ is wave function (WF) of a state with a number
$m$ of this system,
$E_{m}^{(1)}$ is energy level, corresponding to the state with 
the number $m$,
$E_{1}^{(1)}$ is the lowest level of the energy spectrum with the
number $1$.
For the given potential one can introduce a superpotential $W(x)$,
defining it by such a condition 
(for example, according to \cite{Cooper.1995.PRPLC}, p.~287--289):
\begin{equation}
  V_{1}(x) = 
    W^{2}(x) - \displaystyle\frac{d W(x)}{dx} + E^{(1)}_{1}.
\label{eq.2.1.2}
\end{equation}

Let the potential $V_{1}(x)$ be \emph{exactly solvable}, i.~e. one
can write WFs for all states and the energy spectrum of the system
with such a potential in the explicit analytical form. Let's assume,
that we know a form of the potential $V_{1}(x)$, its WFs and the
energy spectrum. Also we assume, that we do not know the
superpotential $W(x)$ and we shall find it.
The condition (\ref{eq.2.1.2}), defining the unknown superpotential
$W(x)$, is \emph{the Riccati equation} (see in the Introduction
citations on some original papers and monographs). Let's find the
form of the superpotential, solving this equation.

We fulfill a substitution of variable $x \to u = u(x)$, defining it so:
\begin{equation}
  f(x) = \displaystyle\frac{d u(x)}{dx},
\label{eq.2.1.3}
\end{equation}
where on the function $f(x)$ we impose arbitrariness in its choice.
We obtain:
\begin{equation}
\begin{array}{ll}
  d u(x) = f dx, &
  \displaystyle\frac{d}{dx} = f \displaystyle\frac{d}{du}.
\end{array}
\label{eq.2.1.4}
\end{equation}
Then one can rewrite the equation (\ref{eq.2.1.2}) through new
variables:
\begin{equation}
  V_{1}(u) - E^{(1)}_{1} =
    W^{2}(u) - f(u) \displaystyle\frac{d W(u)}{du}.
\label{eq.2.1.5}
\end{equation}

Now let's consider a function of a from:
\begin{equation}
\begin{array}{l}
  W_{R}(u) =
  \left\{
    \begin{array}{cl}
      -\displaystyle\frac{\alpha}{u-u_{0}}, & \mbox{at } u \le 0, \\
      -\displaystyle\frac{\alpha}{u+u_{0}}, & \mbox{at } u \ge 0,
    \end{array}
  \right.
\end{array}
\label{eq.2.1.6}
\end{equation}
where $\alpha>0$ and $u_{0}>0$ are constant real parameters.
By direct substitution one can make sure, that at $\alpha=1$ the potential
\begin{equation}
  V_{R}(u) = W_{R}^{2}(u) - \displaystyle\frac{d W_{R}(u)}{du}
\label{eq.2.1.7}
\end{equation}
becomes zero
(according to~\cite{Maydanyuk.2005.Surveys_in_HEP,Maydanyuk.2005.APNYA},
the function of the form (\ref{eq.2.1.6}) is only one from five possible
solutions for the superpotential $W_{R}$, on the basis of which one
can construct zero potential (\ref{eq.2.1.7});
therefore, the choice (\ref{eq.2.1.6}) further will give us only
\underline{a partial solution} in the definition of the superpotential
$W(x)$;
therefore, for example, substitutions (A3) and $y = 1/\varphi$
in~\cite{Sukumar.1985.JPAGB.p2917}, perhaps, lead to the partial
definition of the superpotential $U$ in the form (A12) at the energy
of factorization $E$ located not higher then the lowest level of the
energy spectrum of the potential $V$).
In the definition of the function $W_{R}(u)$ in the form
(\ref{eq.2.1.6}) its divergence at point $u=0$ is excluded by
introduction of the non-zero parameter $u_{0}$;
however, the function $W_{R}(u)$ has \underline{discontinuity at zero},
nevertheless constructed on its basis the function $V_{R}(u)$ is
\underline{continuous at this point}, and, therefore, on all axis $u$
(for example, in contrast to the function $y(x)$ in limit
$\varphi(x) \to 0$ at substitution $y=1/\varphi$
in~\cite{Sukumar.1985.JPAGB.p2917}, p.~2934;
exclusion of such a divergence for the function $y(x)$ and/or
$\varphi(x)$ at passing of the variable $x$ through the extreme
value $x=0$ can be interesting, because it allows to consider
oscillations of the functions $y(x)$ and $\varphi(x)$; 
it is important to study a question of passing of these functions
through their zero values or ``nodes'', nevertheless it turns out,
that it is possible to define the superpotential at the energy of
factorization, which coincides not only with the lowest level of the
energy spectrum for the given potential or below it, but and higher
then such level).
Here, in future investigations, it is interesting to use a formalism
for obtaining solutions for
\emph{``the regulized one-dimensional form of the centrifugal potential
$V(x) = \alpha/(x+i\epsilon)^{2}$''
where $\alpha$ is a real strength and $\epsilon$ a real constant that
removes the singularity at the origin},
proposed in~\cite{Cannata.quant-ph.0606129} (see p.~32--34),
as a possible further complex extensions of a definition of the
function $W_{R}(u)$ in search of new types of exact solutions of
Riccati equation by the approach proposed in this section.

For simplicity of analysis further we shall consider the function
$V_{R}(u)$ on the positive semi-axis $u > 0$.
Subtracting this function from the expression (\ref{eq.2.1.5}), we
obtain:
\begin{equation}
  V_{1}(u) - E^{(1)}_{1} =
    W^{2}(u) - f(u) \displaystyle\frac{d W(u)}{du} -
    f^{2}(u) \biggl(W_{R}^{2}(u) -
      \displaystyle\frac{d W_{R}(u)}{du} \biggr),
\label{eq.2.1.8}
\end{equation}
here multiplication of $V_{R}(u)$ on square of the function $f(u)$
leaves this function as zero.

Now fix $f(u)$ by such a condition:
\begin{equation}
  f(u) W_{R}(u) = W(x).
\label{eq.2.1.9}
\end{equation}
Then from (\ref{eq.2.1.8}) we find:
\begin{equation}
\begin{array}{c}
  V_{1}(x) - E^{(1)}_{1} =
    f^{2}(u) W_{R}^{2}(u) - f(u) \displaystyle\frac{d W(u)}{du} -
    f^{2}(u) W_{R}^{2}(u) + f^{2}(u) \displaystyle\frac{d W_{R}(u)}{du} = \\

    = - f(u) \displaystyle\frac{d}{du} \biggl(f(u) W_{R}(u) \biggr) +
    f^{2}(u) \displaystyle\frac{d W_{R}(u)}{du} = 

    - f(u) \displaystyle\frac{d f(u)}{du} W_{R}(u) - \\
    - f^{2}(u) \displaystyle\frac{d W_{R}(u)}{du} +
    f^{2}(u) \displaystyle\frac{d W_{R}(u)}{du} = 

    -\displaystyle\frac{1}{2} \displaystyle\frac{d f^{2}(u)}{du} W_{R}(u).
\end{array}
\label{eq.2.1.10}
\end{equation}
From here we find $f(u)$:
\begin{equation}
  \displaystyle\frac{d f^{2}(u)}{du} =
    \displaystyle\frac{2 \Bigl( E^{(1)}_{1} - V_{1}(x) \Bigr)}
    {W_{R}(u)}.
\label{eq.2.1.11}
\end{equation}
Thus, we have found the function $f(u)$ in dependence on the known
potential $V_{1}(x)$ and the superpotential $W_{R}(u)$. However, it
is failed to integrate explicitly this expression else, because
implicit dependence of the function $W_{R}(u)$ on the variable $f$ is
included into the right part of (\ref{eq.2.1.11}), and $V_{1}(x)$
depends on $x$.

Taking into account (\ref{eq.2.1.4}), we rewrite (\ref{eq.2.1.11})
by such a way:
\[
\begin{array}{ll}
  W_{R}(u) \displaystyle\frac{d f^{2}(u)}{du} =
    2 W_{R}(u) f(u) \displaystyle\frac{d}{du} f(u) =
    2 W_{R}(u) \displaystyle\frac{d f(u)}{dx} =
    2 \Bigl(E^{(1)}_{1} - V_{1}(x) \Bigr).
\end{array}
\]
\begin{equation}
  W_{R}(u) \displaystyle\frac{d f(u)}{dx} = E^{(1)}_{1} - V_{1}(x).
\label{eq.2.1.12}
\end{equation}
We have obtained the equation, in which the left and right parts
depend only on one variable $u$ or $x$. Further, using the solution
(\ref{eq.2.1.6}) for non-zero superpotential $W_{R}$ (at $u>0$), we
obtain:
\[
\begin{array}{ll}
  -\displaystyle\frac{\alpha}{u+u_{0}} \displaystyle\frac{d f(u)}{dx} =
  -\displaystyle\frac{\alpha}{u+u_{0}}
   \displaystyle\frac{d}{dx} \displaystyle\frac{d u(x)}{dx} =
  -\displaystyle\frac{\alpha}{u+u_{0}}
    \displaystyle\frac{d^{2} u(x)}{dx^{2}} =
    E^{(1)}_{1} - V_{1}(x)
\end{array}
\]
or (at $\alpha=1$)
\begin{equation}
  \biggl(-\displaystyle\frac{d^{2}}{dx^{2}} + V_{1}(x)\biggr)
    \Bigl(u(x)+u_{0}\Bigr)=
    E^{(1)}_{1} \Bigl(u(x)+u_{0}\Bigr).
\label{eq.2.1.13}
\end{equation}
Thus, we have obtained the Schr\"{o}dinger equation for the potential 
$V_{1}(x)$!

Solving the Schr\"{o}dinger equation at selected energy, one can find
a general form of the function $u(x)+u_{0}$. Usually, for potentials
with infinitely high external walls the bound states cause a physical
interest, WFs of which tend to zero in asymptotic space regions. Such
boundary condition, imposed on WFs, introduces restrictions on possible 
energy values, defining the discrete spectrum.

The level $E^{(1)}_{1}$ is the lowest level of the energy spectrum for
the potential $V_{1}(x)$. Therefore, this level is eigenvalue of the
Hamiltonian $\hat{H}^{(1)}$, for which there is non-zero solution for
the function $u(x)+u_{0}$, being the WF. According to set of the
problem, the value $E^{(1)}_{1}$ and WF concerned with it are known
to us. One can write:
\begin{equation}
\begin{array}{ccl}
  u(x) & = & C_{0} \varphi^{(1)}_{1}(x) - u_{0}, \\
  f(x) & = &
    \displaystyle\frac{du(x)}{dx} =
    C_{0} \displaystyle\frac{d \varphi^{(1)}_{1}(x)}{dx}, \\
  W_{R}(x) & = & 
    -\displaystyle\frac{1}{u(x)+u_{0}} =
    -\displaystyle\frac{1}{C_{0} \varphi^{(1)}_{1}(x)}, \\
  W(x) & = & f(x) W_{R}(u(x)) =
    -\displaystyle\frac{1}{\varphi^{(1)}_{1}(x)}
    \displaystyle\frac{d \varphi^{(1)}_{1}(x)}{dx} =
    -\displaystyle\frac{d}{dx}
     \log \Bigl(\varphi^{(1)}_{1}(x) \Bigr), \\
  C_{0} & = & const.
\end{array}
\label{eq.2.1.14}
\end{equation}
Note, that the found solution for the superpotential $W_{n}(x)$ has
obtained only in such interval of values $x$, where $u(x)>0$ ($u(x)$
is the continuous function in this interval), and therefore such a
determination of the superpotential does not contain divergences.
It agrees with a behavior of WF of the bound state at the lowest
energy level, which has no nodes.

One can define a form of a potential-partner $V_{2}(x)$ so:
\begin{equation}
  V_{2}(x) =
    W^{2}(x) + \displaystyle\frac{d W^{2}(x)}{dx} +
    E^{(1)}_{1}.
\label{eq.2.1.15}
\end{equation}

Usually, if the superpotential $W(w)$ is defined in the form of fourth 
expression from (\ref{eq.2.1.14}), then the energy level $E^{(1)}_{1}$
is named as an \emph{energy of factorization} (let's denote it as 
$\cal E$), and the function $\varphi^{(1)}_{1}(x)$, defining the
superpotential by such a way, is named as a \emph{function of
factorization}.

\vspace{5mm}
\noindent
\underline{\bf Conclusion:}

\noindent
\emph{%
The solution for the superpotential was obtained by resolving the
Riccati equation.
In obtaining of the equation (\ref{eq.2.1.13}) for determination of
the superpotential we \underline{do not use the boundary conditions},
which can be imposed on the function $u(x)$.
Further, applying the boundary conditions, we define a state,
described by the function $u(x)$, as bound one at the energy of
factorization, coincident with the lowest energy level $E^{(1)}_{1}$.
Therefore, the solution (\ref{eq.2.1.14}), constructed on the basis
of WF of the bound state at the energy of factorization, coincident
with the lowest level $E^{(1)}_{1}$, can be a partial solution in
definition of the superpotential $W(x)$.}

\subsection{Arbitrariness in a choice of the energy of factorization
\label{sec.2.2}}

Now let's assume, that for considered above quantum system with the
potential $V_{1}(x)$ one can define the superpotential $W(x)$
instead of the condition (\ref{eq.2.1.2}) by the following:
\begin{equation}
  V_{1}(x) = 
    W^{2}(x) - \displaystyle\frac{d W(x)}{dx} +
    E^{(1)}_{1} + \Delta E^{(1)},
\label{eq.2.2.1}
\end{equation}
where $\Delta E^{(1)}$ is arbitrary real number.

Assuming WF and energy spectrum for the potential $V_{1}$ as known
to us, we find the superpotential in this case. Then in the approach,
presented in the previous section, we obtain instead of
(\ref{eq.2.1.13}) the following equation:
\begin{equation}
  \biggl(-\displaystyle\frac{d^{2}}{dx^{2}} + V_{1}(x)\biggr)
    \Bigl(u(x)+u_{0}\Bigr)=
    \Bigl( E^{(1)}_{1} + \Delta E^{(1)} \Bigr)
    \Bigl(u(x)+u_{0}\Bigr).
\label{eq.2.2.2}
\end{equation}
We again obtain the Schr\"{o}dinger equation with the potential
$V_{1}(x)$. From (\ref{eq.2.2.2}) we find:
\begin{itemize}

\item
If the energy $E^{(1)}_{1} + \Delta E^{(1)}$ coincides with arbitrary 
level $E^{(1)}_{m}$ with a number $m$ of the energy spectrum for the
potential $V_{1}$, then it is eigenvalue of the Hamiltonian
$\hat{H}^{(1)}$, for which there is non-zero solution $u(x)+u_{0}$ of
the equation (\ref{eq.2.2.2}), described a bound state for $V_{1}$ at
the selected value $\Delta E^{(1)}$. Then the solution for the
superpotential $W(x)$ exists, where $E^{(1)}_{1} + \Delta E^{(1)}$ is
the energy of factorization:
\begin{equation}
\begin{array}{ccl}
  u(x) & = & C_{m} \varphi^{(1)}_{m}(x) - u_{0}, \\
  f(x) & = &
    \displaystyle\frac{du(x)}{dx} =
    C_{m} \displaystyle\frac{d \varphi^{(1)}_{m}(x)}{dx}, \\
  W_{R}(x) & = & 
    -\displaystyle\frac{1}{u(x)+u_{0}} =
    -\displaystyle\frac{1}{C_{m} \varphi^{(1)}_{m}(x)}, \\
  W(x) & = & f(x) W_{R}(u(x)) =
    -\displaystyle\frac{1}{\varphi^{(1)}_{m}(x)}
    \displaystyle\frac{d \varphi^{(1)}_{m}(x)}{dx} =
    -\displaystyle\frac{d}{dx}
     \log {\bigl|\varphi^{(1)}_{m}(x) \bigr|}, \\
  C_{m} & = & const.
\end{array}
\label{eq.2.2.3}
\end{equation}

\item
If the value $E^{(1)}_{1} + \Delta E^{(1)}$ does not coincide with
any level $E^{(1)}_{m}$ from the energy spectrum of the potential
$V_{1}$, then it cannot be eigenvalue of the Hamiltonian
$\hat{h}_{1}$ and, therefore, there is no any non-zero solution
$u(x)+u_{0}$ of the equation (\ref{eq.2.2.2}), described the bound
state for $V_{1}$ at the selected value $\Delta E^{(1)}$ (with
interdependence (\ref{eq.2.2.1}) between the potential $V_{1}(x)$ and
the superpotential $W(x)$).

\item
Each separate known solution for wave function with its level of the
energy spectrum of the potential $V_{1}$ gives a new independent
solution for the superpotential $W(x)$, and concerned with it the
energy of factorization.

\end{itemize}

However, so defined superpotential can have divergent values at
points-nodes of WFs of exited bound states for $V_{1}$, it turns out,
that by use of it one can construct new exactly solvable potentials
with bound states and without divergences.
Therefore, further in this paper we shall suppose a possibility of
definition of the superpotential with divergences at isolated points.

\subsection{An exact solution for the superpotential with next number
in the hierarchy 
\label{sec.2.3}}

Let's consider three Hamiltonians $\hat{h}_{n}$, $\hat{h}_{n+1}$ and
$\hat{h}_{n+2}$ for three potentials $V_{n}(x)$, $V_{n+1}(x)$ and
$V_{n+2}(x)$ with discrete energy spectra completely:
\begin{equation}
\begin{array}{lclcl}
  \hat{h}_{n} \varphi^{(n)}_{l}(x) & = &
    \biggl(-\displaystyle\frac{d^{2}}{dx^{2}} + V_{n}(x) \biggr)
      \varphi^{(n)}_{l}(x) & = &
      E^{(n)}_{l} \varphi^{(n)}_{l}(x), \\

  \hat{h}_{n+1} \varphi^{(n+1)}_{m}(x) & = &
    \biggl(-\displaystyle\frac{d^{2}}{dx^{2}} + V_{n+1}(x) \biggr)
      \varphi^{(n+1)}_{m}(x) & = &
      E^{(n+1)}_{m} \varphi^{(n+1)}_{m}(x), \\

  \hat{h}_{n+2} \varphi^{(n+2)}_{k}(x) & = &
    \biggl(-\displaystyle\frac{d^{2}}{dx^{2}} + V_{n+2}(x) \biggr)
      \varphi^{(n+2)}_{k}(x) & = &
      E^{(n+2)}_{k} \varphi^{(n+2)}_{k}(x),
\end{array}
\label{eq.2.3.1}
\end{equation}
where
$\varphi^{(n)}_{l}(x)$, $\varphi^{(n+1)}_{m}(x)$ and
$\varphi^{(n+2)}_{k}(x)$ are WFs for states with 
numbers $l$, $m$ and $k$ for three Hamiltonians;
$E_{l}^{(n)}$, $E_{m}^{(n+1)}$ and $E_{k}^{(n+2)}$ are levels of the
energy spectra of these Hamiltonians.
Let's assume, that the Hamiltonians $\hat{h}_{n}$ and $\hat{h}_{n+1}$
connected between themselves by a superpotential $W^{(n)}(x)$, and
the Hamiltonians $\hat{h}_{n+1}$ and $\hat{h}_{n+2}$ --- by a
superpotential $W^{(n+1)}(x)$, where the superpotentials are defined
so:
\begin{equation}
\begin{array}{lcl}
  V_{n}(x) & = &
    \Bigl(W^{(n)} (x)\Bigr)^{2} -
    \displaystyle\frac{d W^{(n)}(x)}{dx} + E^{(n)}_{1}, \\
  V_{n+1}(x) & = &
    \Bigl(W^{(n+1)} (x)\Bigr)^{2} -
    \displaystyle\frac{d W^{(n+1)}(x)}{dx} + E^{(n+1)}_{1},
\end{array}
\label{eq.2.3.2}
\end{equation}
where
$E_{1}^{(n)}$ and $E_{1}^{(n+1)}$ are the lowest levels of the energy
spectra. From such definition we obtain:

\begin{equation}
\begin{array}{lcl}
  V_{n+1}(x) & = &
    \Bigl(W^{(n)} (x)\Bigr)^{2} +
    \displaystyle\frac{d W^{(n)}(x)}{dx} + E^{(n)}_{1}, \\
  V_{n+2}(x) & = &
    \Bigl(W^{(n+1)} (x)\Bigr)^{2} +
    \displaystyle\frac{d W^{(n+1)}(x)}{dx} + E^{(n+1)}_{1}.
\end{array}
\label{eq.2.3.3}
\end{equation}
From (\ref{eq.2.3.2}) and (\ref{eq.2.3.3}) one can find equation for
connection the superpotentials $W^{(n)}$ and $W^{(n+1)}$ with
neighboring numbers $n$ and $n+1$ in the hierarchy:
\begin{equation}
\begin{array}{lcl}
  \Bigl(W^{(n)} (x)\Bigr)^{2} +
    \displaystyle\frac{d W^{(n)}(x)}{dx} =
    \Bigl(W^{(n+1)} (x)\Bigr)^{2} -
    \displaystyle\frac{d W^{(n+1)}(x)}{dx} + k_{n+1},
\end{array}
\label{eq.2.3.4}
\end{equation}
where $k_{n+1} = E^{(n+1)}_{1} - E^{(n)}_{1}$.
This equation is widely used in different topics of SUSY QM,
and, in particular, in \emph{parasupersymmetric quantum mechanics}
or \emph{PSUSY}
(one can see a brief review of PSUSY with a literature list
in~\cite{Cooper.1995.PRPLC} p.~370--377,
at fist time PSUSY of order 2 was introduced in~\cite{Rubakov.1993.MPLAE},
PSUSY of arbitrary order $p$ was proposed
in~\cite{Khare.1992.JPA,Khare.1993.JMP}).

Let's we know the superpotential $W^{(n)}(x)$ and WF of the potential
$V_{n}(x)$ at the level, coincident with the energy of factorization
of the second superpotential $W^{(n)}(x)$. Then as it was shown in
the previous section, the equation (\ref{eq.2.3.4}) can be resolved
exactly analytically concerning the unknown superpotential
$W^{(n+1)}$ with the next number $n+1$.
\emph{We can find exactly analytically as many new solutions for 
the superpotential $W^{(n+1)}$, as many WFs with levels corresponding
to them (and not coincident with the energy of factorization of the
first superpotential $W^{(n)}$) for the potential $V_{n}(x)$ we know 
(it improves the early known approach for determination of exact
solutions for the superpotential with the next number in the
hierarchy on the basis of known superpotential with the previous
number).
So, if the potential $V_{n}(x)$ is exactly solvable and we know all
its spectral characteristics, then the method in sec.~\ref{sec.2}
gives a whole set of the new exact solutions for the superpotential
$W^{(n+1)}$}.
Partially, such approach can be used for construction of new exactly
solvable models in PSUSY QM.

\section{A variety of potentials-partners $V_{2}$ to one potential $V_{1}$
\label{sec.3.1}}

Arbitrariness in the choice of the energy of factorization gives a set 
of different potentials $V_{2}$ partners to one potential $V_{1}$.
Let's consider this question in more details, when superpotential is
constructed on the basis of wave function of an arbitrary bound state
of $V_{1}$.
For the given $V_{1}$ we have a Hamiltonian $\hat{h}_{1}$ of a form:
\begin{equation}
  \hat{h}_{1} \varphi^{(1)}_{n}(x) =
    \biggl(-\displaystyle\frac{d^{2}}{dx^{2}} + V_{1}(x) \biggr)
      \varphi^{(1)}_{n}(x) =
      E^{(1)}_{n} \varphi^{(1)}_{n}(x).
\label{eq.3.1.1}
\end{equation}
According to analysis in the previous section, for the same $V_{1}$
one can construct the superpotential $W$ by different ways:
\begin{equation}
  V_{1}(x) = 
    W^{2}(x) - \displaystyle\frac{d W(x)}{dx} +
      E_{1}^{(1)} =
    \bar{W}^{2}(x) - \displaystyle\frac{d \bar{W}(x)}{dx} +
      E_{1}^{(1)} + \Delta E^{(1)}.
\label{eq.3.1.2}
\end{equation}
Including operators:
\begin{equation}
\begin{array}{ccl}
  A = \displaystyle\frac{d}{dx} + W(x), &
  \bar{A} = \displaystyle\frac{d}{dx} + \bar{W}(x), \\
  A^{+} = -\displaystyle\frac{d}{dx} + W(x), &
  \bar{A}^{+} = -\displaystyle\frac{d}{dx} + \bar{W}(x),
\end{array}
\label{eq.3.1.3}
\end{equation}
one can write:
\begin{equation}
\begin{array}{ccl}
  \hat{h}_{1} \varphi^{(1)}_{n}(x) & = &
    \biggl(A^{+} A + E_{1}^{(1)} \biggr)
    \varphi^{(1)}_{n}(x) = \\
  & = & \biggl(\bar{A}^{+} \bar{A} +
    E_{1}^{(1)} + \Delta E^{(1)} \biggr) \varphi^{(1)}_{n}(x) =
      E^{(1)}_{n} \varphi^{(1)}_{n}(x).
\end{array}
\label{eq.3.1.4}
\end{equation}
Using the found $W$ and $\bar{W}$, one can define new $V_{2}$ and
$\bar{V}_{2}$, which are partners to the same $V_{1}$:
\begin{equation}
\begin{array}{lcl}
  V_{2}(x) & = &
    W^{2}(x) + \displaystyle\frac{d W(x)}{dx} +
      E_{1}^{(1)}, \\
  \bar{V}_{2}(x) & = &
    \bar{W}^{2}(x) + \displaystyle\frac{d \bar{W}(x)}{dx} +
      E_{1}^{(1)} + \Delta E^{(1)}.
\end{array}
\label{eq.3.1.5}
\end{equation}

\vspace{3mm}
\noindent
\underline{\bf Conclusion:}

\noindent
\emph{%
For one given potential $V_{1}$ one can define a whole set of
superpotentials $W$ (equals to number of known energy levels in the
spectrum for $V_{1}$), each of which defines the new potential $V_{2}$,
which is the partner to the same $V_{1}$.}

\subsection{Whether are the potentials $V_{2}$ and $\bar{V}_{2}$
exactly solvable and what are their spectral characteristics?
\label{sec.3.2}}

Let's find out, whether the potentials $V_{2}$ and $\bar{V}_{2}$,
constructed as the partners to the same $V_{1}$ by use of two
different superpotentials $W$ and $\bar{W}$, are exactly solvable.

For analysis of the potential $V_{2}$, constructed at the energy of
factorization ${\cal E} = E_{1}^{(1)}$, we use expression for the
Hamiltonian $h_{1}$ through operators $A$ and $A^{+}$:
\begin{equation}
  \hat{h}_{1} \varphi^{(1)}_{n}(x) =
    \biggl(A^{+} A + E_{1}^{(1)} \biggr)
      \varphi^{(1)}_{n}(x) =
    E_{n}^{(1)} \varphi^{(1)}_{n}(x).
\label{eq.3.2.1}
\end{equation}
Let's act on this equation by the operator $A$ from the left:
\begin{equation}
\begin{array}{c}
  A \Bigl(\hat{h}_{1} \varphi^{(1)}_{n}(x) \Bigr) =
  A \biggl(A^{+} A + E_{1}^{(1)} \biggr)
    \varphi^{(1)}_{n}(x) = \\
  = \biggl(A A^{+} \biggr)
    \Bigl( A \varphi^{(1)}_{n}(x) \Bigr) +
    E_{1}^{(1)} \Bigl(A \varphi^{(1)}_{n}(x) \Bigr) = \\
  = \biggl(A A^{+} + E_{1}^{(1)} \biggr)
    \Bigl(A \varphi^{(1)}_{n}(x) \Bigr) =
    \hat{h}_{2} \Bigl(A \varphi^{(1)}_{n}(x) \Bigr) =
    E_{m}^{(1)} \Bigl( A \varphi^{(1)}_{n}(x) \Bigr).
\end{array}
\label{eq.3.2.2}
\end{equation}
From (\ref{eq.3.2.2}) one can see, that values $E_{n}^{(1)}$ represent
eigenvalues of the Hamiltonian $\hat{h}_{2}$, and functions
$A\varphi^{(1)}_{n}$ represent eigenfunctions of this operator to
normalizing constant. It proves, that new potential $V_{2}$ is
exactly solvable.
We write down wave functions and the energy spectrum of this potential:
\begin{equation}
\begin{array}{ll}
  V_{2} (x) =
    W^{2}(x) + \displaystyle\frac{d W(x)}{dx} + E_{1}^{(1)}, &
  E_{m}^{(2)} = E_{n}^{(1)}, \\
  \varphi^{(2)}_{m}(x) = C_{m} A \varphi^{(1)}_{n}(x), &
  C_{m} = const.
\end{array}
\label{eq.3.2.3}
\end{equation}

For the analysis of the potential $\bar{V}_{2}$ at the energy of
factorization ${\cal E} = E_{1}^{(1)} + \Delta E^{(1)}$, we use
expression for the Hamiltonian $\hat {h}_{1}$ through operators
$\bar{A}$ and $\bar{A}^{+}$:
\begin{equation}
  \hat{h}_{1} \varphi^{(1)}_{n}(x) =
    \biggl(\bar{A}^{+} \bar{A} +
    E_{1}^{(1)} + \Delta E^{(1)} \biggr)
      \varphi^{(1)}_{n}(x) =
    E_{n}^{(1)} \varphi^{(1)}_{n}(x).
\label{eq.3.2.4}
\end{equation}
Acting on this equation from the left by the operator $\bar{A}$, one
can obtain the Schr\"{o}dinger equation again. From here we find, that
new potential $\bar{V}_{2}$ is exactly solvable also, and its wave
functions $\bar{\varphi}^{(2)}_{m}$ and energy spectrum
$\bar{E}_{m}^{(2)}$ have forms:
\begin{equation}
\begin{array}{ll}
  \bar{V}_{2} (x) =
    \bar{W}^{2}(x) + \displaystyle\frac{d \bar{W}(x)}{dx} +
    E_{1}^{(1)} + \Delta E^{(1)}, &
  \bar{E}_{m}^{(2)} = E_{n}^{(1)}, \\
  \bar{\varphi}^{(2)}_{m}(x) =
    \bar{C}_{m} \bar{A} \varphi^{(1)}_{n}(x), &
  \bar{C}_{m} = const.
\end{array}
\label{eq.3.2.5}
\end{equation}

\vspace{3mm}
\noindent
\underline{\bf Conclusion:}

\noindent
\emph{%
The set of potentials $V_{2}$ is exactly solvable \underline{completely}.
Irrespective of the arbitrariness in the choice of the energy of
factorization, all potentials $V_{2}$ have coincided energy spectra
(with possible exception of selected levels), and their wave functions
are determined explicitly from (\ref{eq.3.2.3}) or (\ref{eq.3.2.5}).}

\subsection{Absence of additional levels in the spectra of the
potentials $V_{2}$ and $\bar{V}_{2}$, and 
exception of a level coincided with the energy of factorization
\label{sec.3.3}}

In the beginning we shall analyze wave functions for levels of the
energy spectra of the potentials $V_{2}$ and $\bar{V}_{2}$ in the case
when these levels do not coincide with the energy of factorization
$\cal E$, i.~e. at $E_{m}^{(2)} \ne {\cal E} = E_{1}^{(1)}$ for
$V_{2}$ and at
$\bar{E}_{k}^{(2)} \ne {\cal E} = E_{1}^{(1)} + \Delta E^{(1)}$ for
$\bar{V}_{2}$.
According to (\ref{eq.3.2.3}) and (\ref{eq.3.2.5}), for each level
$E_{n}^{(1)}$ of potential $V_{1}$ there is as a minimum one non-zero
solution for wave function at the level $E_{m}^{(2)}$, which coincides
with $E_{n}^{(1)}$, for the potential $V_{2}$ and there is as a minimum
one non-zero solution for wave function at the level $\bar{E}_{k}^{(2)}$,
which coincides with $E_{n}^{(1)}$, for the potential $\bar{V}_{2}$.
Acting by the operators $A^{+}$ and $\bar{A}^{+}$ on the equations:
\begin{equation}
\begin{array}{ll}
  h_{2} \varphi^{(2)}_{m}(x) =
    \biggl(A A^{+} + E_{1}^{(2)} \biggr)
      \varphi^{(2)}_{m}(x) =
    E_{m}^{(2)} \varphi^{(2)}_{m}(x), \\
  \bar{h}_{2} \bar{\varphi}^{(2)}_{k}(x) =
    \biggl(\bar{A} \bar{A}^{+} + \bar{E}_{1}^{(2)} \biggr)
      \bar{\varphi}^{(2)}_{k}(x) =
    \bar{E}_{k}^{(2)} \bar{\varphi}^{(2)}_{k}(x),
\end{array}
\label{eq.3.3.1}
\end{equation}
we find, that for each level $E_{m}^{(2)}$ for $V_{2}$ and for each
level $\bar{E}_{k}^{(2)}$ for $\bar{V}_{2}$ there is as a minimum one
non-zero solution for wave function for $V_{1}$ at the level
$E_{n}^{(1)}$, which coincides with them.
Thus, we come to \emph{exact one-to-one correspondence between all
levels (with non-zero wave functions) of the spectra of energy for
the potentials $V_{1}$, $V_{2}$ and $\bar{V}_{2}$ (i.~e. there are no
any additional level for $V_{2}$ and $\bar{V}_{2}$, which do not exist
in the spectrum of $V_{1}$, at the chosen definition of the
superpotential):}
\begin{equation}
  E_{n}^{(1)} = E_{m}^{(2)} = \bar{E}_{k}^{(2)} \ne {\cal E}.
\label{eq.3.3.2}
\end{equation}

Now let's consider the levels of the potentials $V_{2}$ and
$\bar{V}_{2}$, which coincide with the energy of factorization $\cal E$,
i.~e. the level $E_{m}^{(2)} = {\cal E}$ for $V_{2}$ at
${\cal E} = E_{1}^{(1)}$ and the level $\bar{E}_{k}^{(2)} = {\cal E}$
for $\bar{V}_{2}$ at ${\cal E} = E_{1}^{(1)} + \Delta E^{(1)}$.
Analyzing all possible states at these levels, we conclude:
\begin{itemize}

\item
\emph{
The Hamiltonian $\hat{h}_{2}$ cannot have a state with non-zero wave
function at the energy level $E_{m}^{(2)}$, which coincides with the
energy of factorization ${\cal E} = E_{1}^{(1)}$.
Therefore, the energy spectrum of the potential $V_{2}$ coincides
completely with the energy spectrum of the potential $V_{1}$ with
possible exception of the single level $E_{1}^{(1)}$.}

\item
\emph{
The Hamiltonian $\bar{h}_{2}$ cannot have a state with non-zero wave
function at the energy level $E_{k}^{(2)}$, which coincides with the
energy of factorization ${\cal E} = E_{1}^{(1)} + \Delta E^{(1)}$.
Therefore, the energy spectrum of the potential $\bar{V}_{2}$
coincides completely with the energy spectrum of the potential $V_{1}$
with possible exception of the singe level $E_{1}^{(1)} + \Delta E^{(1)}$.}
\end{itemize}

\subsection{How do shapes of the potentials $V_{2}$ and $\bar{V}_{2}$
differ?
\label{sec.3.4}}

Let's find out, how do the shapes of the potentials $V_{2}$ and
$\bar{V}_{2}$ differ.
Taking into account (\ref{eq.3.1.2}) and (\ref{eq.3.1.5}), we obtain:
\begin{equation}
  V_{1} = 
    V_{2} - 2 \displaystyle\frac{d W}{dx} =
    \bar{V}_{2} - 2 \displaystyle\frac{d \bar{W}}{dx}.
\label{eq.3.4.1}
\end{equation}
From here find:
\begin{equation}
  \bar{V}_{2} =
    V_{2} +
      2\displaystyle\frac{d}{dx} \Bigl(\bar{W} - W\Bigr).
\label{eq.3.4.2}
\end{equation}
According to (\ref{eq.2.1.14}) and (\ref{eq.2.2.3}), we see, that the
superpotentials $W$ and $\bar{W}$ are expressed through different
wave functions and, therefore, the difference $\bar{W} - W$ is not
reduced to constant. One can write down:
\begin{equation}
  \bar{V}_{2} =
    V_{2} +
      2\displaystyle\frac{d^{2}}{dx^{2}}
      \log \Biggl|\displaystyle\frac{\varphi^{(1)}_{1}}
        {\bar{\varphi}^{(1)}_{m}} \Biggr|.
\label{eq.3.4.3}
\end{equation}

\vspace{3mm}
\noindent
\underline{\bf Conclusion:}

\noindent
\emph{%
The potential $ \bar {V}_{2}$ at its shape does not coincide with
the potential $V_{2}$, and, therefore, it is the new exactly solvable
potential.}

\section{New unbound and non-normalizable states in discrete spectra
\label{sec.4}}

\subsection{A partial solution of wave function of the unbound state
\label{sec.4.1}}

According to the approach from sec.~\ref{sec.2.1}, the superpotential
$W$ at the given potential $V_{1}$ can be constructed at arbitrary
energy of factorization $\cal E$.
If the energy of factorization coincides with arbitrary selected level
$E_{m}^{(1)}$ of the bound state of $V_{1}$ (as in sec.~\ref{sec.3.1}),
then we write down:
\begin{equation}
  W_{m} = W_{m}^{(1)} =
    -\displaystyle\frac{d}{dx} \log{|\varphi_{m}^{(1)}|},
\label{eq.4.1.1}
\end{equation}
where $\varphi_{m}^{(1)}$ is WF of the bound state with number $m$ for
$V_{1}$ and the index $m$ is added to the superpotential. The
superpotential is connected with the potential $V_{1}$ so:
\begin{equation}
  V_{1} =
    W_{m}^{2} - \displaystyle\frac{d W_{m}}{dx} + E_{m}^{(1)}
\label{eq.4.1.2}
\end{equation}
and a new potential $V_{2}$ can be defined so:
\begin{equation}
  V_{2} =
    W_{m}^{2} + \displaystyle\frac{d W_{m}}{dx} + E_{m}^{(1)}.
\label{eq.4.1.3}
\end{equation}

According to sec.~\ref{sec.3.3}, the potential $V_{2}$ has no bound
states at the level $E_{m}^{(1)}$. However, such a function:
\begin{equation}
  \bar{\varphi_{m}}^{(2,1)} =
    \displaystyle\frac{1}{\varphi_{m}^{(1)}}
\label{eq.4.1.4}
\end{equation}
is a solution of the Schr\"{o}dinger equation for this potential at
this level. The wave function $\varphi_{m}^{(1)}$ describes the bound
state for $V_{1}$, tending to zero in both directions in asymptotic
regions. Therefore, the function $\bar{\varphi_{m}}^{(2)}$ tends to its
infinite values in asymptotic limits and can not describe a bound
state for $V_{2}$.
Usually, such states are not used in practice, because they do not
contain obvious physical sense for potentials with discrete energy
spectra.
However, it turns out, that on the basis of such states one can
improve essentially earlier known supersymmetric methods of
construction of new exactly solvable potentials at the given potential
$V_{1}$, building new potentials with own bound states. Therefore,
further we shall name such states as \emph{unbound (or non-normalizable)
states in the discrete energy spectrum}, and the functions, describing
them, as wave functions of such states.
If the function $\varphi_{m}^{(1)}$ has nodes, then the function
$\bar{\varphi_{m}}^{(2)}$ has divergences at coordinates of these
nodes. In such a case one can name the state, described by the
function $\bar{\varphi_{m}}^{(2)}$, as \emph{discontinuous (broken)
state}.

From (\ref{eq.4.1.1}) and (\ref{eq.4.1.4}) one can write down:
\begin{equation}
  W_{m}^{(1)} =
    \displaystyle\frac{d}{dx} \log{|\bar{\varphi}_{m}^{(2,1)}|}.
\label{eq.4.1.5}
\end{equation}

\vspace{3mm}
\noindent
\underline{\bf Conclusion:}

\noindent
\emph{%
SUSY-algorithm of construction of new potential $V_{2}$ with the next
number in hierarchy with the given potential $V_{1}$ destroys one
bound state at the level, which coincides with the energy of
factorization, and introduces a new unbound or discontinuous state
(both partial, and general solutions for its WF) at the same level}.

Note that the unbound states in such \emph{partial} definition in the
form (\ref{eq.4.1.4}) were introduced early (perhaps, for the first
time by C.~V.~Sukumar
in~\cite{Sukumar.1985.JPAGB.p2917}, see p.~2925--2934)
but, mainly, when the energy of factorization (when it has real value)
locates not higher then lowest energy level $E_{1}^{(1)}$ of $V_{1}$
(for example,
see \cite{Sukumar.1987.JPAGB.2461},
\cite{Cooper.1995.PRPLC} p.~323--325 ---
when $V_{1}$ has discrite energy spectrum completely,
see \cite{Sukumar.1986.JPAGB.p2297} ---
when $V_{1}$ has energy spectrum with discrete and continuous parts)
or has complex value
(for example, see \cite{Baye.1996.NPA,Andrianov.1999.IJMP}).
Of cause, an important progress has been achieved by
B.~N.~Zakhariev and V.~M.~Chabanov 
in development of the main formalism of the unbound states:
in \cite{Zakhariev.1999.PEPAN} (see p.~286--290) ---
for inversion of $\delta$-potentials (${\cal E} < 0$),
in \cite{Zakhariev.1999.PEPAN} (see p.~293--297) ---
for analysis of deformations and for finding out their rules for
rectangular well and oscillator after inclusion of one new unbound
state (at ${\cal E} < E_{1}^{(1)}$),
in \cite{Zakhariev.1999.PEPAN} (see p.~309--318) ---
for generalization of the main formalism of such states into a
\emph{multicannel case},
in \cite{Zakhariev.2002.PEPAN} (see p.~378--382, 382--389) and
in~\cite{Chabanov.2001.IP} ---
for generalization of the formalism of \emph{double Darboux
transformations} using unbound states with complex energies of
factorization
(for \emph{non-hermitian Schr\"{o}dinger operators with complex
potentials and real energy spectra}), allowing to shift selected
levels in spectrum into complex region of values.
In complex consideration of solutions for superpotential,
potentials (when energy spectra have real or complex values),
it needs to point out such intriguing direction in SUSY QM as
\emph{PT-symmetric quantum mechanics}
(for example, see
\cite{Bender.1998.PRL,%
Znojil.1999.PLA,%
Bender.1999.JP,%
Andrianov.1999.IJMP,%
Znojil.2000.PLB,%
Znojil.2000.MG,%
Znojil.2001.PLA,%
Levai.2001.JPA,%
Levai.2002.JPA,%
Deb.2003.PLA,%
Cannata.quant-ph.0606129}).

\subsection{A general solution of wave function of unbound state
\label{sec.4.2}}

The second partial solution of wave function of the unbound state for
the potential $V_{2}$ at the level $E_{m}^{(1)}$ can be written down
as (it is taken from \cite{Cooper.1995.PRPLC} p.~323--324, as a
generalization of a solution of WF of the ground state at level
$E_{1}^{(1)}$;
\emph{in contrast to \cite{Cooper.1995.PRPLC} we define the unbound
state at arbitrary level $E_{m}^{(1)}$}):
\begin{equation}
  \bar{\varphi}_{m}^{(2,2)} =
    \displaystyle\frac{I_{m}} {\varphi_{m}^{(1)}},
\label{eq.4.2.1}
\end{equation}

\begin{equation}
  I_{m}(x) =
    \int\limits_{-\infty}^{x} \Bigl( \varphi_{m}^{(1)} (x') \Bigr)^{2} dx'.
\label{eq.4.2.2}
\end{equation}
Indeed, substituting such function into the Schr\"{o}dinger equation
with the potential (\ref{eq.4.1.3}), we obtain equality at the level
$E_{m}^{(1)}$.
Let's analyze, what boundary conditions this function satisfies to.
As $\varphi_{m}^{(1)}(x)$ describes the bound state, the function
$I_{m}(x)$ is finite at any $x$ (at $x \to +\infty$ one can connect
the function $I_{m}(x)$ with a finite normalizing constant for
$\varphi_{m}^{(1)}$).
Therefore, at $x \to 0$ or $x \to a$ the second partial solution
$\bar{\varphi}_{m}^{(2,2)}$ tends to infinity, i.~e. it describes
the unbound state at $E_{m}^{(1)}$.

Let's compose a \emph{general solution of WF of the unbound state}
from two partial solutions (\ref{eq.4.1.4}) and (\ref{eq.4.2.1}):
\begin{equation}
  \bar{\varphi}_{m}^{(2)} =
    \displaystyle\frac{I_{m} + \lambda_{m}} {\varphi_{m}^{(1)}}.
\label{eq.4.2.3}
\end{equation}
Taking into account (\ref{eq.4.2.3}) and (\ref{eq.4.1.5}), we write
the superpotential as:
\begin{equation}
\begin{array}{c}
  W_{m}(\lambda_{m}) =
    \displaystyle\frac{d}{dx} \log{|\bar{\varphi}_{m}^{(2)}|} =
  \displaystyle\frac{d}{dx} \log{\biggl|
    \displaystyle\frac{I_{m} + \lambda_{m}} {\varphi_{m}^{(1)}} \biggr|} =
  W_{m}^{(1)} +
    \Delta W_{m}(\lambda_{m}), \\

  \Delta W_{m}(\lambda_{m}) =
    \displaystyle\frac{d}{dx} \log{\bigl|I_{m} + \lambda_{m} \bigr|}.
\end{array}
\label{eq.4.2.4}
\end{equation}

\vspace{3mm}
\noindent
\underline{\bf Conclusion:}

\noindent
Changing the parameter $\lambda_{m}$, one can deform the
superpotential $W_{m}$, but it does not change the form of the
potential $V_{2}$. From here we obtain:
\begin{itemize}

\item
\emph{%
WFs of all bound states and arrangement of levels in the spectrum
for the potential $V_{2}$ are not changed in variations of the
parameter $\lambda_{m}$!}

\item
\emph{%
We obtain the whole set of new potentials $V_{1}(\lambda_{m})$
(which can be considered as deformations of the given $V_{1}$ by the
parameter $\lambda_{m}$), which all are partners to one $V_{2}$.}
\end{itemize}

Note that the \emph{general solutions} for WF for unbound states in
the formalism similar to (\ref{eq.4.2.1})--(\ref{eq.4.2.3})
(perhaps, which was introduced for the first time by C.~V.~Sukumar
in~\cite{Sukumar.1985.JPAGB.p2917} p.~2926) can be found in
literature also
(one can meet with it essentially more rarely then with the partial
solutions in the form (\ref{eq.4.1.4}), but which \underline{improves
essentially} the SUSY QM methods of deformations),
but again mainly, when the energy of factorization locates not higher
then lowest energy level $E_{1}^{(1)}$ of $V_{1}$
(for example, in~\cite{Cooper.1995.PRPLC} p.~323--326 ---
the most universal and convenient variant,
in \cite{Sukumar.1986.JPAGB.p2297,Sukumar.1987.JPAGB.2461} ---
a very improved formalism with analysis of different peculiarities
described in details,
in \cite{Zakhariev.1999.PEPAN} p.~286--288 ---
a simple original variant with use of needed boundary conditions
and its application for inversion of the given potential $V_{1}$
and in \cite{Zakhariev.1999.PEPAN} p.~309--313 ---
a generalization of such formalism for the multichannel case for
some special types of potentials,
in \cite{Zakhariev.1994.PEPAN} p.~1571--1572 ---
use of unbound states for insertion of new levels into spectrum
in the formalism of the inverse problem).

\section{Isospectral potentials
\label{sec.5}}

\subsection{1-parametric potentials
\label{sec.5.1}}

Under changing the parameter $\lambda_{m}$, the solution
(\ref{eq.4.2.3}) keeps interdependence between WF of the unbound
state for $V_{2}$ and WF, dependent on $\lambda_{m}$, of the
bound state for $V_{1}$ at the level $E_{m}^{(1)}$, which coincides
with the energy of factorization $\cal E$. Thus, we obtain a
possibility to deform the potential $V_{1}$ and all its WFs
(without a displacement
of levels in the spectrum) with the help of one parameter $\lambda_{m}$
at the selected energy of factorization. The deformed WF at
$E_{m}^{(1)}$ for $V_{1}$ has a form:
\begin{equation}
  \varphi_{m}^{(1)} (\lambda_{m}) =
    \displaystyle\frac{1}{\bar{\varphi}_{m}^{(2)}} =
    \displaystyle\frac{\varphi_{m}^{(1)}} {I_{m} + \lambda_{m}}.
\label{eq.5.1.1}
\end{equation}
Let's find the deformed potential $V_{1}(\lambda_{m})$, using its
connection with $W_{m}(\lambda_{m})$ and taking into account the
deformation:
\[
\begin{array}{lcl}
  \left.
  \begin{array}{lcl}
    V_{1}(\lambda_{m}) & = &
      V_{2} - 2 \displaystyle\frac{d W_{m}(\lambda_{m})}{dx} \\
    V_{2} & = &
      V_{1} + 2 \displaystyle\frac{d W_{m}}{dx} \\
  \end{array}
  \right \}
  & \to &
  V_{1}(\lambda_{m}) =
    V_{1} - 2 \displaystyle\frac{d}{dx}
    \Bigl(W_{m}(\lambda_{m}) - W_{m}\Bigr) =
    V_{1} - 2 \displaystyle\frac{d}{dx}
    \Delta W_{m} (\lambda_{m}),
\end{array}
\]
and, therefore:
\begin{equation}
\begin{array}{lcl}
  V_{1}(\lambda_{m}) & = &
    V_{1} + \Delta V_{1}(\lambda_{m}), \\

  \Delta V_{1}(\lambda_{m}) & = &
    - 2 \displaystyle\frac{d}{dx} \Delta W_{m}(\lambda_{m}) =

    - 2\displaystyle\frac{d^{2}}{dx^{2}}
      \log{ \Bigl|I_{m} + \lambda_{m} \Bigr|}.
\end{array}
\label{eq.5.1.2}
\end{equation}
From (\ref{eq.5.1.1}) and (\ref{eq.5.1.2}) one can see, how using
the parameter $\lambda_{m}$ to deform WF of the bound state at the
level $E_{m}^{(1)}$, which coincides with the energy of factorization,
and the shape of the potential $V_{1}$.

Note, that we have obtained 1-parametric set of new exactly solvable
isospectral potentials $V_{1}(\lambda_{m})$ on the basis of
deformation of wave function of the arbitrary bound state, in contrast
to \cite{Cooper.1995.PRPLC} (see p.~324--326), where for the
deformation the wave function of the ground bound state is used only.
Such a generalization allows us further in frameworks of the
SUSY-approach to construct all pictures of deformation of a rectangular
well and its WFs (without a displacement of levels), which were
obtained early by the approach of inverse problem
(see~\cite{Zakhariev.1990.PEPAN}, p.~916--927).

\subsection{Deformation of wave functions of other bound states
\label{sec.5.2}}

Now let's find, how the WFs of other bound states for $V_{1}$ at
levels, which do not coincide with the energy of factorization
$\cal E$, are deformed by the change of the parameter $\lambda_{m}$.
WFs of the potentials $V_{1}$ and $V_{2}$ for coinciding levels 
$E_{n_{1}}^{(1)} = E_{n_{2}}^{(2)} \ne {\cal E} = E_{m}^{(1)}$ are
connected so (see also \cite{Cooper.1995.PRPLC}, p.~287--289):
\begin{equation}
\begin{array}{lcl}
  \varphi_{n_{2}}^{(2)} & = &
    |E_{n_{1}}^{(1)} - E_{m}^{(1)}|^{-1/2}
    A_{m} \varphi_{n_{1}}^{(1)}, \\
  \varphi_{n_{1}}^{(1)} & = &
    |E_{n_{1}}^{(1)} - E_{m}^{(1)}|^{-1/2}
    A_{m}^{+} \varphi_{n_{2}}^{(2)},
\end{array}
\label{eq.5.2.1}
\end{equation}
where
\begin{equation}
\begin{array}{cc}
  A_{m} = \displaystyle\frac{d}{dx} + W_{m}, &
  A_{m}^{+} = -\displaystyle\frac{d}{dx} + W_{m}.
\end{array}
\label{eq.5.2.2}
\end{equation}
Here, in calculations of the normalizing constants for these WFs we
use a condition of their normalization (as for bound states). Taking
into account, that WFs of the potential $V_{2}$ are not deformed under
variations of the parameter $\lambda_{m}$, we obtain:
\[
\begin{array}{lcl}
  \varphi_{n_{1}}^{(1)}(\lambda_{m}) & = &
    |E_{n_{1}}^{(1)} - E_{m}^{(1)}|^{-1}
    A_{m}^{+}(\lambda_{m}) A_{m}  \;
    \varphi_{n_{1}}^{(1)} = \\

  & = &
    |E_{n_{1}}^{(1)} - E_{m}^{(1)}|^{-1}
    \biggl(-\displaystyle\frac{d}{dx} + W_{m}(\lambda_{m}) \biggr)
    A_{m} \; \varphi_{n_{1}}^{(1)} = \\

  & = &
    |E_{n_{1}}^{(1)} - E_{m}^{(1)}|^{-1}
    \biggl(-\displaystyle\frac{d}{dx} + W_{m} +
          \Delta W(\lambda_{m}) \biggr)
    A_{m} \; \varphi_{n_{1}}^{(1)} = \\

  & = &
    |E_{n_{1}}^{(1)} - E_{m}^{(1)}|^{-1}
    A_{m}^{+} A_{m} \varphi_{n_{2}}^{(2)} +
    |E_{n_{1}}^{(1)} - E_{m}^{(1)}|^{-1}
      \Delta W(\lambda_{m}) A_{m} \; \varphi_{n_{1}}^{(1)} = \\

  & = &
    \varphi_{n_{1}}^{(1)} +
    |E_{n_{1}}^{(1)} - E_{m}^{(1)}|^{-1}
    \Delta W(\lambda_{m}) A_{m} \; \varphi_{n_{1}}^{(1)},
\end{array}
\]
or
\begin{equation}
\begin{array}{lcl}
  \varphi_{n_{1}}^{(1)}(\lambda_{m}) & = &

  \biggl(1 + \displaystyle\frac{\Delta W(\lambda_{m})}
    {|E_{n_{1}}^{(1)} - E_{m}^{(1)}|} A_{m} \biggr)
    \varphi_{n_{1}}^{(1)} =

  \varphi_{n_{1}}^{(1)} + \Delta \varphi_{n_{1}}^{(1)}(\lambda_{m}),
\end{array}
\label{eq.5.2.3}
\end{equation}
where
\begin{equation}
\begin{array}{lcl}
  \Delta\varphi_{n_{1}}^{(1)}(\lambda_{m}) & = &
    \displaystyle\frac{\Delta W(\lambda_{m})}
    {|E_{n_{1}}^{(1)} - E_{m}^{(1)}|} A_{m} 
    \varphi_{n_{1}}^{(1)}, \\

  \Delta W(\lambda_{m}) & = &
    \displaystyle\frac{d}{dx} \log{\Bigl| I_{m} + \lambda_{m} \Bigr|}.
\end{array}
\label{eq.5.2.4}
\end{equation}
Expressions (\ref{eq.5.2.3})--(\ref{eq.5.2.4}) describe, how under
change of the parameter $\lambda_{m}$ WF of the bound state for
$V_{1}$ is deformed at the arbitrary level $E_{n_{1}}^{(1)}$, which
does not coincide with the energy of factorization $\cal E$.

\vspace{3mm}
\noindent
\underline{\bf Analysis:}

\vspace{-3mm}
\begin{itemize}
\item
Analyzing a possibility of the potential $V_{1}$ to change its shape
with keeping the shape of the potential $V_{2}$, we come to the
following assumption:
\emph{Each separate level of the energy spectrum contains one degree
of freedom for deformation of the potential and its spectral
characteristics
(expressed mathematically through the parameter $\lambda_{m}$).}

\vspace{-1mm}
\item
From (\ref{eq.5.2.4}) we find:
\emph{The distance between levels $E_{n_{1}}^{(1)}$ and 
${\cal E} = E_{m}^{(1)}$ is larger, the influence of the parameter 
$\lambda_{m}$ on the deformation of WF at the level $E_{n_{1}}^{(1)}$
is smaller.}
\end{itemize}

%
\subsection{$n$-parametric potentials
\label{sec.5.3}}

One can generalize the method of deformation of the potential $V_{1}$
without displacement of the levels in the spectrum by use of the
parameter of deformation $\lambda_{m}$, described in sec.~\ref{sec.5.1}:
in the beginning we fulfill a transition from the given potential
$V_{1}$ into a new potential $V_{n}$ by applying SUSY-transformation
many times, which transforms consistently the bound states for selected
$n$ levels for the potential $V_{1}$ into unbound states for the new
potential $V_{n}$, and then we fulfill an inverse transition from
the obtained potential $V_{n}$ into a new deformed potential $V_{1}$
(which will depend on $n$ parameters $\lambda_{i}$, $i = 1\ldots n$),
transforming consistently the found unbound states into the bound
states (in the assumption, that they exist).
By such transformation we obtain a whole set of new exactly solvable
potentials with identical energy spectra, which depend on $n$
parameters of deformation and one can name them as
\emph{$n$-parametrical family of isospectral potentials}.
Note, that \emph{here we use arbitrary numbers of the levels of the
potential $V_{1}$ and their sequence for the $n$-multiple application
of SUSY-transformations (in contrast to \cite{Cooper.1995.PRPLC},
p.~326--328, where the lowest $n$ levels are used
\underline{consistently} only)}.

Let's consider a construction of the family of 2-parametrical
isospectral potentials (as it is done in~\cite{Cooper.1995.PRPLC},
p.~326--328 for two lowest levels).
Let's we have a potential $V_{1}$ with known WFs and energy
spectrum. In such spectrum we choose two levels $E_{m_{1}}^{(1)}$ and
$E_{m_{2}}^{(1)}$ with numbers $m_{1}$ and $m_{2}$.
In the beginning we fulfill a transition from the potential $V_{1}$
into a new potential $V_{2}$ applying SUSY-transformation, for which
we define a superpotential $W_{m_{1}}^{(1)}$ with energy of
factorization ${\cal E}_{m_{1}}$, which coincides with
$E_{m_{1}}^{(1)}$ (as it is done in sec.~\ref{sec.5.1}).
As we know, at this level there is WF of the unbound state for the
new potential $V_{2}$:
\begin{equation}
\begin{array}{cc}
  \bar{\varphi}_{m_{1}}^{(2)} (\lambda_{1}) =
    \displaystyle\frac{I_{m_{1}}^{(1)} + \lambda_{1}}
      {\varphi_{m_{1}}^{(1)}}, &
  I_{m_{1}}^{(1)} (x) = 
    \displaystyle\int\limits_{-\infty}^{x}
    \Bigl(\varphi_{m_{1}}^{(1)}(x') \Bigr)^{2} dx',
\end{array}
\label{eq.5.3.1}
\end{equation}
$\varphi_{m_{1}}^{(1)}$ is WF for $V_{1}$ at the level
$E_{m_{1}}^{(1)}$ and $\lambda_{1}$ is an arbitrary constant.
One can find WFs of the bound states for other levels of the
potential $V_{2}$ from the first expression in (\ref{eq.5.2.1}) and
from (\ref{eq.5.2.2}) at $m=m_{1}$, where:
\begin{equation}
  W^{(1)}_{m_{1}} =
    - \displaystyle\frac{d}{dx} \log | \varphi_{m_{1}}^{(1)} |,
\label{eq.5.3.2}
\end{equation}
and one can obtain a shape of the potential $V_{2}$ from:
\begin{equation}
  V_{2} =
    V_{1} + 2 \displaystyle\frac{d}{dx} W^{(1)}_{m_{1}} =
    V_{1} - 2 \displaystyle\frac{d^{2}}{dx^{2}} \log |\varphi_{m_{1}}^{(1)}|.
\label{eq.5.3.3}
\end{equation}

Further, we fulfill the transition from the potential $V_{2}$ into a
new potential $V_{3}$ applying next SUSY-transformation, for which we
 define the second superpotential $W_{m_{2}}^{(2)}$ with the energy of
factorization ${\cal E}_{m_{2}}$, which coincides with another level
$E_{m_{2}}^{(1)}$. For the new potential $V_{3}$, we obtain WF of
unbound state for this level $E_{m_{2}}^{(1)}$:
\begin{equation}
  \bar{\varphi}_{m_{2}}^{(3)} (\lambda_{2}) =
    \displaystyle\frac{I_{m_{2}}^{(2)} + \lambda_{2}}
      {\varphi_{m_{2}}^{(2)}},
\label{eq.5.3.4}
\end{equation}
and WFs of bound states for other levels:
\begin{equation}
\begin{array}{cc}
  \varphi_{n_{3}}^{(3)} =
    |E_{n_{2}}^{(2)} - E_{m_{2}}^{(1)}|^{-1/2}
    A_{m_{2}}^{(2)} \varphi_{n_{2}}^{(2)}, &
  \mbox{at } E_{n_{2}}^{(2)} = E_{n_{3}}^{(3)},
\end{array}
\label{eq.5.3.5}
\end{equation}
where
\begin{equation}
\begin{array}{ccc}
  I_{m_{2}}^{(2)} =
    \displaystyle\int\limits_{-\infty}^{x}
    \Bigl(\varphi_{m_{2}}^{(2)}(x') \Bigr)^{2} dx', &
  A_{m_{2}}^{(2)} =
    \displaystyle\frac{d}{dx} + W_{m_{2}}^{(2)}, &
  W_{m_{2}}^{(2)} =
    - \displaystyle\frac{d}{dx} \log |\varphi_{m_{2}}^{(2)}|
\end{array}
\label{eq.5.3.6}
\end{equation}
and a new constant $\lambda_{2}$ is introduced.
We calculate the shape of the potential $V_{3}$ as:
\begin{equation}
  V_{3} =
    V_{2} + 2 \displaystyle\frac{d}{dx} W^{(2)}_{m_{2}} =
    V_{2} - 2 \displaystyle\frac{d^{2}}{dx^{2}} \log |\varphi_{m_{2}}^{(2)}|.
\label{eq.5.3.7}
\end{equation}
One can see, that a function
\begin{equation}
\begin{array}{ccc}
  \bar\varphi_{m_{1}}^{(3)}(\lambda_{1}) =
    |E_{m_{1}}^{(1)} - E_{m_{2}}^{(1)}|^{-1/2}
    A_{m_{2}}^{(2)} \bar{\varphi}_{m_{1}}^{(2)}(\lambda_{1})
\end{array}
\label{eq.5.3.8}
\end{equation}
describes unbound state for the potential $V_{3}$ at the level
$E_{m_{1}}^{(1)}$ also (here we have introduced a ``normalizing''
factor). We see, that \emph{consecutive $n$-multiple application of
SUSY-transformation to the given potential $V_{1}$ at the selected
$n$ levels reduces number of the bound states by $n$ and increases
number of unbound states by $n$ also.}

Now we fulfill an inverse transition, coming back from the potential
$V_{3}$ to a new potential $V_{1}$ with consecutive transformation of
the unbound states into bound states at the levels $E_{m_{1}}^{(1)}$
and $E_{m_{2}}^{(1)}$.
However, in definition of superpotentials for two inverse
SUSY-transformation we can change sequence of a choice of the
energies of factorization ${\cal E}_{m_{1}}$ and ${\cal E}_{m_{2}}$
(in contrast to \cite{Cooper.1995.PRPLC}, p.~326--328).
Let's introduce new designations $\bar{m}_{i}$ and $\bar{\lambda}_{i}$
for numbers of the levels and the parameters by such condition:
\begin{equation}
\begin{array}{ccc}
  (\bar{m}_{1} = m_{1}, \bar{m}_{2} = m_{2}, 
  \bar{\lambda}_{1} = \lambda_{1}, \bar{\lambda}_{2} = \lambda_{2}) &
  \mbox{ or } &
  (\bar{m}_{1} = m_{2}, \bar{m}_{2} = m_{1}, 
  \bar{\lambda}_{1} = \lambda_{2}, \bar{\lambda}_{2} = \lambda_{1}).
\end{array}
\label{eq.5.3.9}
\end{equation}

For definition of the energy of factorization (let's denote it as 
${\cal E}_{\bar{m}_{1}}$) in the construction of the superpotential
for the transition from $V_{3}$ into $V_{2}$ we choose the level
$E_{\bar{m}_{1}}^{(1)}$ with the number $\bar{m}_{1}$. The unbound
state ${\bar{\varphi}_{\bar{m}_{1}}^{(3)}(\bar{\lambda}_{1})}$ of
the potential $V_{3}$ at this level transforms into bound state
$\varphi_{\bar{m}_{1}}^{(2)}(\bar{\lambda}_{1})$ of the potential
$V_{2}$, dependent on one parameter $\bar{\lambda}_{1}$ (a question
of existence of divergences is important in obtaining of the solution
for WF of such ``bound'' state; but, perhaps, one can exclude them
as it will be done in sec.~\ref{sec.7} in deformation of a rectangular
well with infinitely high external walls with the energy of
factorization, located not below than the first exited level of the
spectrum of such potential):
\begin{equation}
  \varphi_{\bar{m}_{1}}^{(2)} (\bar{\lambda}_{1}) =
    \displaystyle\frac{1}
    {\bar{\varphi}_{\bar{m}_{1}}^{(3)} (\bar{\lambda}_{1})} =
    \left\{
    \begin{array}{ll}
      \displaystyle\frac{1}{\bar{\varphi}_{m_{2}}^{(3)} (\lambda_{2})} =
      \displaystyle\frac{\varphi_{m_{2}}^{(2)}}
        {I_{m_{2}}^{(2)} + \lambda_{2}},
        & \mbox{at } \bar{m}_{1} = m_{2}, \\
      \displaystyle\frac{1}{\bar{\varphi}_{m_{1}}^{(3)} (\lambda_{1})} =
        \displaystyle\frac
          {\sqrt{|E_{m_{1}}^{(1)} - E_{m_{2}}^{(1)}|}}
          {A_{m_{2}}^{(2)} \bar{\varphi}_{m_{1}}^{(2)} (\lambda_{1})},
        & \mbox{at } \bar{m}_{1} = m_{1}.
    \end{array}
  \right.
\label{eq.5.3.10}
\end{equation}
The other unbound state
$\bar{\varphi}_{\bar{m}_{2}}^{(3)}(\bar{\lambda}_{2})$ of the
potential $V_{3}$ at the level $E_{\bar{m}_{2}}^{(2)}$ with the
number $\bar{m}_{2}$ in the transition forms a new unbound state
$\bar{\varphi}_{\bar{m}_{2}}^{(2)}(\bar{\lambda}_{1}, \bar{\lambda}_{2})$
for the potential $V_{2}$:
\begin{equation}
  \bar{\varphi}_{\bar{m}_{2}}^{(2)}
    (\bar{\lambda}_{1}, \bar{\lambda}_{2}) =
    |E_{m_{1}}^{(1)} - E_{m_{2}}^{(1)}|^{-1/2}
    A_{\bar{m}_{1}}^{(2),+} (\bar{\lambda}_{1})
    \bar{\varphi}_{\bar{m}_{2}}^{(3)} (\bar{\lambda}_{2}).
\label{eq.5.3.11}
\end{equation}
For WFs of bound states for the other levels we obtain:
\begin{equation}
\begin{array}{cc}
  \varphi_{n_{2}}^{(2)}(\bar{\lambda}_{1}) =
    |E_{n_{2}}^{(2)} - E_{\bar{m}_{1}}^{(1)}|^{-1/2}
    A_{\bar{m}_{1}}^{(2),+} (\bar{\lambda}_{1})
    \varphi_{n_{3}}^{(3)}, &
  \mbox{at } E_{n_{2}}^{(2)} = E_{n_{3}}^{(3)},
\end{array}
\label{eq.5.3.12}
\end{equation}
where
\begin{equation}
\begin{array}{cc}
  A_{\bar{m}_{1}}^{(2),+} (\bar{\lambda}_{1}) =
    -\displaystyle\frac{d}{dx} +
    W_{\bar{m}_{1}}^{(2)}(\bar{\lambda}_{1}), &
  W_{\bar{m}_{1}}^{(2)}(\bar{\lambda}_{1}) =
    -\displaystyle\frac{d}{dx} \log
      | \varphi_{\bar{m}_{1}}^{(2)}(\bar{\lambda}_{1}) | =
    \displaystyle\frac{d}{dx} \log
      | \bar{\varphi}_{\bar{m}_{1}}^{(3)}(\bar{\lambda}_{1}) |.
\end{array}
\label{eq.5.3.13}
\end{equation}
The new potential $V_{2}(\bar{\lambda}_{1})$ already depends on one
parameter $\bar{\lambda}_{1}$ and has a form:
\begin{equation}
  V_{2}(\bar{\lambda}_{1}) =
    V_{3} - 2 \displaystyle\frac{d}{dx}
      W_{\bar{m}_{1}}^{(2)}(\bar{\lambda}_{1}) =
    V_{3} + 2 \displaystyle\frac{d^{2}}{dx^{2}} \log
      | \varphi_{\bar{m}_{1}}^{(2)}(\bar{\lambda}_{1}) |.
\label{eq.5.3.14}
\end{equation}

For the last transition from $V_{2}$ to $V_{1}$, for the definition
of the energy of factorization ${\cal E}_{\bar{m}_{2}}$ for the
second superpotential we choose the level with the number
$\bar{m}_{2}$. In result, we obtain:
\begin{itemize}

\item
the unbound state of the potential $V_{2}$ at the level
$E_{\bar{m}_{2}}^{(1)}$ transforms into bound state of the potential
$V_{1}$:
\begin{equation}
  \varphi_{\bar{m}_{2}}^{(1)} (\bar{\lambda}_{1}, \bar{\lambda}_{2}) =
    \displaystyle\frac{1}{\bar{\varphi}_{\bar{m}_{2}}^{(2)}
    (\bar{\lambda}_{1}, \bar{\lambda}_{2})};
\label{eq.5.3.15}
\end{equation}

\item
the bound states of the potential $V_{2}$ at the other levels
transform into the bound states of the potential $V_{1}$:
\begin{equation}
\begin{array}{cc}
  \varphi_{n_{1}}^{(1)}(\bar{\lambda}_{1},\bar{\lambda}_{2}) =
    |E_{n_{1}}^{(1)} - E_{\bar{m}_{2}}^{(1)}|^{-1/2}
    A_{\bar{m}_{2}}^{(1),+} (\bar{\lambda}_{1},\bar{\lambda}_{2})
    \varphi_{n_{2}}^{(2)}(\bar{\lambda}_{1}), &
  \mbox{at } E_{n_{1}}^{(1)} = E_{n_{2}}^{(2)},
\end{array}
\label{eq.5.3.16}
\end{equation}
where
\begin{equation}
\begin{array}{cc}
  A_{\bar{m}_{2}}^{(1),+} (\bar{\lambda}_{1},\bar{\lambda}_{2}) =
    -\displaystyle\frac{d}{dx} +
    W_{\bar{m}_{2}}^{(1)}(\bar{\lambda}_{1},\bar{\lambda}_{2}), &
  W_{\bar{m}_{2}}^{(1)}(\bar{\lambda}_{1},\bar{\lambda}_{2}) =
    -\displaystyle\frac{d}{dx} \log
    | \varphi_{\bar{m}_{2}}^{(1)} (\bar{\lambda}_{1},\bar{\lambda}_{2}) |;
\end{array}
\label{eq.5.3.17}
\end{equation}

\item
the new potential $V_{1}$ already dependes on two parameters
$\bar{\lambda}_{1}$ and $\bar{\lambda}_{2}$, and has a form:
\begin{equation}
  V_{1}(\bar{\lambda}_{1},\bar{\lambda}_{2}) =
    V_{2}(\bar{\lambda}_{2}) - 2 \displaystyle\frac{d}{dx}
      W_{\bar{m}_{2}}^{(1)}(\bar{\lambda}_{1},\bar{\lambda}_{2}) =
    V_{2}(\bar{\lambda}_{2}) + 2\displaystyle\frac{d^{2}}{dx^{2}} \log
    | \varphi_{\bar{m}_{2}}^{(1)} (\bar{\lambda}_{1},\bar{\lambda}_{2}) |.
\label{eq.5.3.18}
\end{equation}
Taking into account (\ref{eq.5.3.3}), (\ref{eq.5.3.7}),
(\ref{eq.5.3.14}) and (\ref{eq.5.3.18}), we obtain:
\begin{equation}
  V_{1}(\bar{\lambda}_{1},\bar{\lambda}_{2}) =
    V_{1} - 2 \displaystyle\frac{d^{2}}{dx^{2}} \log \Biggl|
      \displaystyle\frac{
      \varphi_{m_{1}}^{(1)}
      \varphi_{m_{2}}^{(2)}}
      {\varphi_{\bar{m}_{1}}^{(2)}(\bar{\lambda}_{1})
      \varphi_{\bar{m}_{2}}^{(1)} (\bar{\lambda}_{1},\bar{\lambda}_{2})}
      \Biggr|.
\label{eq.5.3.19}
\end{equation}
\end{itemize}

Let's generalize the found result into a case of deformation of the
potential $V_{1}$ by use of $n$ parameters $\lambda_{i}$. For this
purpose, in the beginning we choose $n$ levels in the energy
spectrum of this potential. Let's these levels are numbered by two
independent sequences of numbers $m_{1}$, $m_{2}$ \ldots $m_{n}$
and $\bar{m}_{1}$, $\bar{m}_{2}$ \ldots $\bar{m}_{n}$, accordingly.
Then for the deformed potential we find:
\begin{equation}
  V_{1}(\bar{\lambda}_{1} \ldots \bar{\lambda}_{n}) =
    V_{1} - 2 \displaystyle\frac{d^{2}}{dx^{2}} \log \Biggl|
    \displaystyle\frac
      {\varphi_{m_{1}}^{(1)} \ldots \varphi_{m_{n}}^{(1)}}
      {\varphi_{\bar{m}_{1}}^{(n)} (\bar{\lambda}_{1})
       \varphi_{\bar{m}_{2}}^{(n-1)}
       (\bar{\lambda}_{1}, \bar{\lambda}_{2})
      \ldots
      \varphi_{\bar{m}_{n}}^{(1)}
      (\bar{\lambda}_{1} \ldots \bar{\lambda}_{n})}
      \Biggr|
\label{eq.5.3.20}
\end{equation}
or
\begin{equation}
  V_{1}(\bar{\lambda}_{1} \ldots \bar{\lambda}_{n}) =
    V_{1} - 2 \displaystyle\frac{d^{2}}{dx^{2}} \log \biggl|
      \varphi_{m_{1}}^{(1)} \ldots \varphi_{m_{n}}^{(1)}
        \cdot
      \bar{\varphi}_{\bar{m}_{1}}^{(n+1)} (\bar{\lambda}_{1})
      \bar{\varphi}_{\bar{m}_{2}}^{(n)}
        (\bar{\lambda}_{1}, \bar{\lambda}_{2})
      \ldots
      \bar{\varphi}_{\bar{m}_{n}}^{(2)}
      (\bar{\lambda}_{1} \ldots \bar{\lambda}_{n})
      \biggr|.
\label{eq.5.3.21}
\end{equation}
WFs $\varphi_{m_{2}}^{(2)}$ \ldots $\varphi_{m_{n}}^{(n)}$ of the
bound states for the potentials $V_{1}$ \ldots $V_{n}$ can be found
on the basis of $\varphi_{m_{2}}^{(1)}$ \ldots
$\varphi_{m_{n}}^{(1)}$ of the bound states of the started potential
$V_{1}$
(we write new WFs up to normalizing constants, denoting them as $C$):
\begin{equation}
\begin{array}{lcllcl}
  \varphi_{m_{2}}^{(2)} & = &
    C A_{m_{1}}^{(1)} \varphi_{m_{2}}^{(1)}, &
  \varphi_{m_{i}}^{(i)} & = &
    C A_{m_{i-1}}^{(i-1)} \ldots A_{m_{1}}^{(1)}
    \varphi_{m_{i}}^{(1)}, \\
  \varphi_{m_{3}}^{(3)} & = &
    C A_{m_{2}}^{(2)} A_{m_{1}}^{(1)}
    \varphi_{m_{3}}^{(1)}, &
  \varphi_{m_{n}}^{(n)} & = &
    C A_{m_{n-1}}^{(n-1)} \ldots A_{m_{1}}^{(1)}
    \varphi_{m_{n}}^{(1)}.
\end{array}
\label{eq.5.3.22}
\end{equation}
The deformed WFs $\varphi_{\bar{m}_{i}}^{(n-i+1)}
(\bar{\lambda}_{1} \ldots \bar{\lambda}_{i})$
of the bound states with the parameters $\lambda_{i}$ can be
calculated on the basis of WFs $\bar{\varphi}_{m_{1}}^{(n+1)}$ \ldots
$\bar{\varphi}_{m_{n}}^{(n+1)}$ of the unbound states of the
potential $V_{n+1}$ (at $i=2 \ldots n$):
\begin{equation}
\begin{array}{l}
  \varphi_{\bar{m}_{1}}^{(n)} (\bar{\lambda}_{1}) =
    \displaystyle\frac{1}
    {\bar{\varphi}_{\bar{m}_{1}}^{(n+1)} (\bar{\lambda}_{1})}, \\
  \varphi_{\bar{m}_{2}}^{(n-1)}
      (\bar{\lambda}_{1}, \bar{\lambda}_{2}) =
    \displaystyle\frac{1}
    {\bar{\varphi}_{\bar{m}_{2}}^{(n)}
    (\bar{\lambda}_{1}, \bar{\lambda}_{2})}, \\
  \varphi_{\bar{m}_{i}}^{(n-i+1)}
      (\bar{\lambda}_{1}, \bar{\lambda}_{2} \ldots \bar{\lambda}_{i}) =
    \displaystyle\frac{1}
    {\bar{\varphi}_{\bar{m}_{i}}^{(n-i+2)}
    (\bar{\lambda}_{1}, \bar{\lambda}_{2} \ldots \bar{\lambda}_{i})}, \\
  \varphi_{\bar{m}_{n}}^{(1)}
      (\bar{\lambda}_{1}, \bar{\lambda}_{2} \ldots \bar{\lambda}_{n}) =
    \displaystyle\frac{1}
    {\bar{\varphi}_{\bar{m}_{n}}^{(2)}
    (\bar{\lambda}_{1}, \bar{\lambda}_{2} \ldots \bar{\lambda}_{n})}
\end{array}
\label{eq.5.3.23}
\end{equation}
or

\begin{equation}
\begin{array}{lcl}
  \bar{\varphi}_{\bar{m}_{2}}^{(n)}
    (\bar{\lambda}_{1}, \bar{\lambda}_{2}) & = &
    C A_{\bar{m}_{1}}^{(n),+} (\bar{\lambda}_{1})   \;
    \bar{\varphi}_{\bar{m}_{2}}^{(n+1)} (\bar{\lambda}_{2}), \\

  \bar{\varphi}_{\bar{m}_{3}}^{(n-1)}
    (\bar{\lambda}_{1}, \bar{\lambda}_{2}, \bar{\lambda}_{3}) & = &
    C A_{\bar{m}_{2}}^{(n-1),+}
      (\bar{\lambda}_{1}, \bar{\lambda}_{2})
      A_{\bar{m}_{1}}^{(n),+} (\bar{\lambda}_{1})   \;
    \bar{\varphi}_{\bar{m}_{3}}^{(n+1)} (\bar{\lambda}_{3}), \\

  \bar{\varphi}_{\bar{m}_{i}}^{(n-i+2)}
    (\bar{\lambda}_{1} \ldots \bar{\lambda}_{i}) & = &
    C A_{\bar{m}_{i-1}}^{(n-i+2),+}
      (\bar{\lambda}_{1}, \bar{\lambda}_{2} \ldots \bar{\lambda}_{i-1})
      \ldots
      A_{\bar{m}_{1}}^{(n),+} (\bar{\lambda}_{1})   \;
    \bar{\varphi}_{\bar{m}_{i}}^{(n+1)} (\bar{\lambda}_{i}), \\

  \bar{\varphi}_{\bar{m}_{n}}^{(2)}
    (\bar{\lambda}_{1} \ldots \bar{\lambda}_{n}) & = &
    C A_{\bar{m}_{n-1}}^{(2),+}
      (\bar{\lambda}_{1}, \bar{\lambda}_{2} \ldots \bar{\lambda}_{n-1})
      \ldots
      A_{\bar{m}_{1}}^{(n),+} (\bar{\lambda}_{1})   \;
    \bar{\varphi}_{\bar{m}_{n}}^{(n+1)} (\bar{\lambda}_{n}).
\end{array}
\label{eq.5.3.24}
\end{equation}
WFs of the unbound states of the potential $V_{n+1}$ have forms:
\begin{equation}
\begin{array}{lcl}
  \bar{\varphi}_{m_{1}}^{(n+1)} (\lambda_{1}) & = &
    C A_{m_{n}}^{(n)} \ldots  A_{m_{2}}^{(2)}   \;
    \bar{\varphi}_{m_{1}}^{(2)} (\lambda_{1}), \\
  \bar{\varphi}_{m_{2}}^{(n+1)} (\lambda_{2}) & = &
    C A_{m_{n}}^{(n)} \ldots  A_{m_{3}}^{(3)}   \;
    \bar{\varphi}_{m_{2}}^{(3)} (\lambda_{2}), \\
  \bar{\varphi}_{m_{i-1}}^{(n+1)} (\lambda_{i-1}) & = &
    C A_{m_{n}}^{(n)} \ldots  A_{m_{i}}^{(i)} \;
    \bar{\varphi}_{m_{i-1}}^{(i)} (\lambda_{i-1}), \\
  \bar{\varphi}_{m_{n-1}}^{(n+1)} (\lambda_{n-1}) & = &
    C A_{m_{n}}^{(n)} \;
    \bar{\varphi}_{m_{n-1}}^{(n)} (\lambda_{n-1}).
\end{array}
\label{eq.5.3.25}
\end{equation}
WFs $\bar{\varphi}_{m_{i}}^{(i+1)} (\lambda_{i})$ must be
calculated by such formulas (at $i=1 \ldots n$):
\begin{equation}
\begin{array}{cc}
  \bar{\varphi}_{m_{i}}^{(i+1)} (\lambda_{i}) =
    \displaystyle\frac{I_{m_{i}}^{(i)} + \lambda_{i}}
    {\varphi_{m_{i}}^{(i)}}, &
  I_{m_{i}}^{(i)} (x) = 
    \displaystyle\int\limits_{-\infty}^{x}
    \Bigl(\varphi_{m_{i}}^{(i)}(x') \Bigr)^{2} dx'.
\end{array}
\label{eq.5.3.26}
\end{equation}
The non-deformed and deformed operators $A$ and $A^{+}$ have forms
(at $i=1 \ldots n$):
\begin{equation}
\begin{array}{cc}
  A_{m_{i}}^{(i)} = \displaystyle\frac{d}{dx} + W_{m_{i}}^{(i)}, &

  A_{\bar{m}_{i}}^{(n-i+1),+}
      (\bar{\lambda}_{1} \ldots \bar{\lambda}_{i}) =
    -\displaystyle\frac{d}{dx} +
    W_{\bar{m}_{i}}^{(n-i+1)}
      (\bar{\lambda}_{1} \ldots \bar{\lambda}_{i}),
\end{array}
\label{eq.5.3.27}
\end{equation}
where
\begin{equation}
\begin{array}{cc}
  W_{m_{i}}^{(i)} =
    -\displaystyle\frac{d}{dx} \log {|\varphi_{m_{i}}^{(i)} |}, &

  W_{\bar{m}_{i}}^{(n-i+1)}
      (\bar{\lambda}_{1} \ldots \bar{\lambda}_{i}) =
    -\displaystyle\frac{d}{dx} \log
    |\varphi_{\bar{m}_{i}}^{(n-i+1)}
      (\bar{\lambda}_{1} \ldots \bar{\lambda}_{i})|.
\end{array}
\label{eq.5.3.28}
\end{equation}
The deformed WF of arbitrary bound state for the potential
$V_{1} (\bar{\lambda}_{1} \ldots \bar{\lambda}_{n})$ at the level
$E_{k}$ with the number $k$,
where $E_{k} \ne E_{m_{1}}$, \ldots $E_{k} \ne E_{m_{n}}$, has a form:
\begin{equation}
  \varphi_{k}^{(1)} (\bar{\lambda}_{1} \ldots \bar{\lambda}_{n}) =
    C A_{\bar{m}_{n}}^{(1),+}
      (\bar{\lambda}_{1} \ldots \bar{\lambda}_{n}) \ldots
    A_{\bar{m}_{1}}^{(n),+} (\bar{\lambda}_{1})
    \cdot
    A_{m_{n}}^{(n)} \ldots A_{m_{1}}^{(1)}
    \varphi_{k}^{(1)}.
\label{eq.5.3.29}
\end{equation}

\subsection{New approaches of construction of isospectral potentials
\label{sec.5.4}}

It turns out, that if in definition of the function of factorization
for construction of the superpotential to do not restrict oneself
only to the bound states of the given potential $V_{1}$, but to use
an arbitrary unbound state with a selected level for this potential,
then such a way allows to construct new types of potentials, which can
be isospectral to $V_{1}$.
Here, use of the unbound states is caused by rejection of imposing
boundary conditions (for bound states) on WF of such state, that does
not destroy interdependence (but, rather, expands it) between two
potentials-partners.
\emph{Such consideration inevitably leads to an absolute arbitrariness
in a choice of a value of the energy of factorization (which can be
not coincident with the levels of the energy spectrum of the given
potential $V_{1}$)}.
Let's consider this question in more details.

Let's we know the wave function $\bar{\varphi}^{(1)}_{w}$ of the
unbound state at the level $E_{w}$ for the given potential $V_{1}$.
Then a function of a form:
\begin{equation}
\begin{array}{cc}
  f_{w}^{(0)} (\bar{\lambda}_{w}) =
    \displaystyle\frac
    {\bar{I}_{w}^{(1)} + \bar{\lambda}_{w}}
    {\bar{\varphi}_{w}^{(1)}}, &
  \bar{I}_{w}^{(1)} (x) = 
    \displaystyle\int\limits_{-\infty}^{x}
    \Bigl(\bar{\varphi}_{w}^{(1)}(y) \Bigr)^{2} dy
\end{array}
\label{eq.5.4.1}
\end{equation}
is a \emph{general} solution of the Schr\"{o}dinger equation at the
same level $E_{w}$ for a potential $V_{0}$ with a previous number $0$
(which can be named as the number ``down'';
see ~\cite{Maydanyuk.2005.APNYA}, p.~448--451), which has a form:
\begin{equation}
  V_{0} =
    V_{1} - 2 {W^{(1)}}' =
    \Bigl( W^{(1)} \Bigr)^{2} -
    {\Bigl( W^{(1)} \Bigl)}' + E_{w},
\label{eq.5.4.2}
\end{equation}
where
\begin{equation}
  W^{(1)} = \displaystyle\frac{d}{dx} \log |\bar{\varphi}_{w}^{(1)}|
\label{eq.5.4.3}
\end{equation}
and $\bar{\lambda}_{w}$ is arbitrary constant.

\vspace{5mm}
{\footnotesize
\noindent
\underline{\bf Proof:}

For the function $\bar{I}_{w}^{(1)} (x)$ we have:
\[
\begin{array}{cc}
  \Bigl( \bar{I}_{w}^{(1)} \Bigr)^{\prime} = 
  \displaystyle\frac{d \bar{I}_{w}^{(1)}}{dx} = 
    \Bigl( \bar{\varphi}_{w}^{(1)} \Bigr)^{2}, &

  \Bigl( \bar{I}_{w}^{(1)} \Bigr)^{\prime\prime} = 
  \displaystyle\frac{d^{2} \bar{I}_{w}^{(1)}}{dx^{2}} = 
    2 \bar{\varphi}_{w}^{(1)}
      \Bigl(\bar{\varphi}^{(1)}_{w} \Bigr)^{\prime}.
\end{array}
\]
Then we obtain:
\[
\begin{array}{c}
  \Biggl(
    \displaystyle\frac{\bar{I}_{w}^{(1)}}
      {\bar{\varphi}^{(1)}_{w}} \Biggr)^{\prime\prime} =

   \displaystyle\frac{\bar{I}_{w}^{(1)}}
        {\bar{\varphi}^{(1)}_{w}}
     \Biggl\{ 2 \Biggl(
       \displaystyle\frac
         { \Bigl( \bar{\varphi}_{w}^{(1)} \Bigr)' }
         {\bar{\varphi}^{(1)}_{w}}
       \Biggr)^{2} -
       \displaystyle\frac
         {\Bigl(\bar{\varphi}^{(1)}_{w} \Bigr)^{\prime\prime}}
         {\bar{\varphi}^{(1)}_{w}}
     \Biggr \},
\end{array}
\]
\[
\begin{array}{c}
  \Biggl( \displaystyle\frac{\bar{\lambda}_{w}}
    {\bar{\varphi}^{(1)}_{w}} \Biggr)^{\prime\prime} =

  \bar{\lambda}_{w}
    \Biggl( 
      -\displaystyle\frac
      {\Bigl( \bar{\varphi}^{(1)}_{w} \Bigr)^{\prime}}
      {\bigl(\bar{\varphi}^{(1)}_{w} \bigr)^{2}}
    \Biggr)^{\prime} =

  \bar{\lambda}_{w} \Biggl \{
    2 \displaystyle\frac
      {\Bigl(\bar{\varphi}^{(1)}_{w} \Bigr)^{\prime, 2}}
      {\Bigl(\bar{\varphi}^{(1)}_{w} \Bigr)^{3}}
    -\displaystyle\frac
      {\Bigl(\bar{\varphi}^{(1)}_{w} \Bigr)^{\prime\prime}}
      {\Bigl( \bar{\varphi}^{(1)}_{w} \Bigr)^{2}}
      \Biggr \} =

  \displaystyle\frac
    {\bar{\lambda}_{w}} {\bar{\varphi}^{(1)}_{w}}
    \Biggl \{
    2 \Biggl ( \displaystyle\frac
        {\Bigl(\bar{\varphi}^{(1)}_{w} \Bigr)^{\prime}}
        {\bar{\varphi}^{(1)}_{w}}
      \Biggr)^{2}
    -\displaystyle\frac
      {\Bigl(\bar{\varphi}^{(1)}_{w} \Bigr)^{\prime\prime}}
      {\bar{\varphi}^{(1)}_{w}}
    \Biggr \},
\end{array}
\]
\begin{equation}
\begin{array}{c}
  \Bigl( f_{w}^{(0)} (\bar{\lambda}_{w})
    \Bigr)^{\prime\prime} =

  \Biggl( \displaystyle\frac
    {\bar{I}_{w}^{(1)} + \bar{\lambda}_{w}}
    {\bar{\varphi}^{(1)}_{w}} \Biggr)^{\prime\prime} =

  \displaystyle\frac
    {\bar{I}_{w}^{(1)} + \bar{\lambda}_{w}}
    {\bar{\varphi}^{(1)}_{w}}

    \cdot

    \Biggl \{
    2 \Biggl ( \displaystyle\frac
        {\Bigl(\bar{\varphi}^{(1)}_{w} \Bigr)^{\prime}}
        {\bar{\varphi}^{(1)}_{w}}
      \Biggr)^{2}
    -\displaystyle\frac
      {\Bigl(\bar{\varphi}^{(1)}_{w} \Bigr)^{\prime\prime}}
      {\bar{\varphi}^{(1)}_{w}}
    \Biggr \} =

  f_{w}^{(0)} (\bar{\lambda}_{w})
    \Biggl \{
    2 \Biggl ( \displaystyle\frac
        {\Bigl(\bar{\varphi}^{(1)}_{w} \Bigr)^{\prime}}
        {\bar{\varphi}^{(1)}_{w}}
      \Biggr)^{2}
    -\displaystyle\frac
      {\Bigl(\bar{\varphi}^{(1)}_{w} \Bigr)^{\prime\prime}}
      {\bar{\varphi}^{(1)}_{w}}
    \Biggr \}.
\end{array}
\label{eq.5.4.4}
\end{equation}
Let's analyze, whether the function $f_{w}^{(0)} (\bar{\lambda}_{w})$
satisfies to the Schr\"{o}dinger equation with the potential $V_{0}$
at the level $E_{w}$. We write:
\[
  - \Bigl( f_{w}^{(0)} (\bar{\lambda}_{w})
    \Big)^{\prime\prime} +
    V_{0} f_{w}^{(0)} (\bar{\lambda}_{w}) =
    E_{w} f_{w}^{(0)} (\bar{\lambda}_{w}).
\]
Taking into account (\ref{eq.5.4.4}), we obtain:
\[
\begin{array}{c}
  f_{w}^{(0)} (\bar{\lambda}_{w})  
    \Biggl \{ \displaystyle\frac
      {\Bigl(\bar{\varphi}^{(1)}_{w} \Bigr)^{\prime\prime}}
      {\bar{\varphi}^{(1)}_{w}} -
    2 \Biggl ( \displaystyle\frac
        {\Bigl(\bar{\varphi}^{(1)}_{w} \Bigr)^{\prime}}
        {\bar{\varphi}^{(1)}_{w}}
      \Biggr)^{2}
    \Biggr \} +

    V_{0} f_{w}^{(0)} (\bar{\lambda}_{w}) =
    E_{w} f_{w}^{(0)} (\bar{\lambda}_{w}).
\end{array}
\]
One can see, that this expression becomes identity, if to define the
potential $V_{0}$ as (\ref{eq.5.4.2}) with taking into account
(\ref{eq.5.4.3}). Therefore, we have proved, that the function
$f_{w}^{(0)}$ in the form (\ref{eq.5.4.1}) is the solution of the
Schr\"{o}dinger equation for the potential $V_{0}$ at the level $E_{w}$.
One can see also, that the solution (\ref{eq.5.4.1}) consists of two
linearly independent partial solutions and, therefore, it is a
\emph{general solution}.
}

\vspace{5mm}
\noindent
\underline{\bf Consequences:}
\begin{itemize}

\item
The potentials $V_{0}$ and $V_{1}$ are SUSY-partners with the
superpotential of the form (\ref{eq.5.4.3}).

\item
Variations of the parameter $\bar{\lambda}_{w}$ with the given form
of the wave function $\bar{\varphi}^{(1)}_{w}$ do not deform the
superpotential $\bar{W}^{(1)}_{w}$, and, therefore, the potential
$V_{0}$.

\item
From (\ref{eq.5.4.1}) one can see, that by use of the parameter
$\bar{\lambda}_{w}$ one can deform the function $f^{(0)}_{w}$
(with keeping of the shape of the potential $V_{0}$).

\item
The found solution for $f^{(0)}_{w}$ describes the bound state of
the potential $V_{0}$ at the level $E_{w}$ not always, but only in
the case, when it satisfies to the boundary conditions for the bound
states and a condition of a continuity of this function inside whole
region of its definition (which can be found from an analysis of
possible divergences of this function).
\end{itemize}

\vspace{3mm}
In dependence on whether such a value of the parameter
$\bar{\lambda}_{w}$ exists, at which the function $f^{(0)}_{w}$
satisfies to the bounding conditions, we have two cases (if the
energy of factorization $E_{w}$ does not coincide with any level
$E_{n}^{(1)}$ of the spectrum for $V_{1}$):
\begin{itemize}

\item
If the function $f^{(0)}_{w}$ satisfies to the bounding conditions,
then it defines a new bound state for the potential $V_{0}$ at the
level $E_{w}$.
Then one can consider this function as WF of the bound state for
$V_{0}$ (let's denote it as $\varphi^{(0)}_{w}$), and the spectrum of
the new potential $V_{0}$ differs on the spectrum of the given
potential $V_{1}$ by the additional new level $E_{w}$.

\item
If the function $f^{(0)}_{w}$ does not satisfy to the bounding
conditions, then it defines a new unbound state for the potential
$V_{0}$ at the level $E_{w}$ (let's consider this function as WF of
the unbound state and denote it as $\bar{\varphi}^{(0)}_{w}$).
Then if the superpotential for SUSY-transformations keeps all other
states for $V_{0}$ as bound ones, then the spectra of the potentials
$V_{0}$ and $V_{1}$ are equal absolutely, and these potentials are
isospectral.

\end{itemize}
The second case gives us a new way of construction of the
isospectral potentials on the basis of the known WF of the unbound
state of the potential $V_{1}$ (note, that this way is some easier
in comparison with the approach from sec.~\ref{sec.5.1}, however here
it needs to pay more attention to an analysis of divergences).
In the new approach one can deform the found isospectral potentials
$V_{0}$ not only by variation of the parameter $\bar{\lambda}_{w}$
(as in the approach from sec.~\ref{sec.5.1}), but also by change of
the value of the energy of factorization $E_{w}$ in a possible range
and by deformation of WF shape of the unbound state
$\bar{\varphi}_{w}^{(1)}$.

\vspace{5mm}
Now let's redefine the superpotential on the basis of the function
$f^{(0)}_{w} (\bar{\lambda}_{w})$ so:
\begin{equation}
  W^{(1)} (\bar{\lambda}_{w}) =
    -\displaystyle\frac{d}{dx} \log |f_{w}^{(0)} (\bar{\lambda}_{w})|.
\label{eq.5.4.5}
\end{equation}
By use of this superpotential one can construct the new potentials 
$V_{0}(\bar{\lambda}_{w})$ and $V_{1}(\bar{\lambda}_{w})$. These
potentials will be SUSY-partners and, therefore, they should have
the energy spectra, which coincide completely with the energy spectra
of the old potentials $V_{0}$ and $V_{1}$ (with a possible exception
of the level $E_{w}$).

The definition (\ref{eq.5.4.5}) introduces the dependence on the
parameter $\bar{\lambda}_{w}$ into the superpotential (and, as a
result, into the potentials $V_{0} (\bar{\lambda}_{w})$ and
$V_{1} (\bar{\lambda}_{w})$).
In result, we obtain the \emph{second} new approach of the construction
of new isospectral potentials, which can be some like to the method
from sec.~\ref{sec.5.1}. A difference between these two methods
consists in the following:
\begin{itemize}

\item
If the method in sec.~\ref{sec.5.1} is constructed with use of the
superpotential, which is defined on the basis of WF of the bound
state of the starting potential $V_{1}$, here we define the
superpotential on the basis of WF of the unbound state of the
starting potential $V_{1}$. Such a way is richer essentially and
it gives more large class of new isospectral potentials.

\item
In contrast to the method in sec.~\ref{sec.5.1} for the construction
of the new isospectral potentials with the transition of their numbers
\emph{``up-down} ($V_{1} \to V_{2} \to V_{1} (\lambda)$), here we
fulfill the inverse transition of the numbers \emph{``down-up}
($V_{1} \to V_{0} \to V_{1} (\bar{\lambda}_{w})$).

\end{itemize}

\vspace{5mm}
\noindent
\underline{\bf Conclusions:}
\emph{
We have obtained two new methods of the construction of the set of
the isospectral potentials on the basis of the given WF of the
unbound state for the given potential $V_{1}$ at the given level
$E_{w}$. In contrast to the known approach in sec.~\ref{sec.5.1},
where as the parameter of deformation we use $\lambda$, in the new
approaches as two additional parameters of deformation the function
of factorization (defined on the basis of WF of the unbound state)
and the energy of factorization $E_{w}$ are used, that makes a class
of the new isospectral potentials larger, and methods richer.}

\vspace{3mm}
\noindent
\underline{\bf The first approach:}
\begin{equation}
\begin{array}{lcl}
  V_{0} =
    V_{1} - 2 \Bigl(W^{(1)} \Bigr)' =
    \Bigl( W^{(1)} \Bigr)^{2} - \Bigl( W^{(1)} \Bigr)' + E_{w}, \\

  W^{(1)} = 
    \displaystyle\frac{d}{dx} \log |\bar{\varphi}_{w}^{(1)}|, \\
  E_{w}, \bar{\varphi}^{(1)}_{w}
    \mbox{ --- the parameters of deformation}.
\end{array}
\label{eq.5.4.6}
\end{equation}

\vspace{3mm}
\noindent
\underline{\bf The second approach:}
\begin{equation}
\begin{array}{l}
  V_{1}
    (E_{w}, \bar{\varphi}^{(1)}_{w}, \bar{\lambda}_{w}) =
    V_{1} +
    \Delta V_{1}
      (E_{w}, \bar{\varphi}^{(1)}_{w}, \bar{\lambda}_{w}), \\

  \Delta V_{1}
    (E_{w}, \bar{\varphi}^{(1)}_{w}, \bar{\lambda}_{w}) =
    - 2\displaystyle\frac{d^{2}}{dx^{2}}
    \log{ \Bigl|\bar{I}_{w}^{(1)} +
    \bar{\lambda}_{w} \Bigr|}, \\

  \bar{I}_{w}^{(1)} (x) =
    \displaystyle\int\limits_{-\infty}^{x}
    \Bigl(\bar{\varphi}_{w}^{(1)}(y) \Bigr)^{2} dy, \\

  E_{w}, \bar{\varphi}^{(1)}_{w}, \bar{\lambda}_{w}
    \mbox{ --- the parameters of deformation}.
\end{array}
\label{eq.5.4.7}
\end{equation}

\vspace{5mm}
\noindent
\underline{\bf Consequence:}
On the basis of the second approach one can generalize the method
from sec.~\ref{sec.5.3} for the construction of the $n$-parametrical
family of the isospectral potentials with taking into account of a
possibility to define the function of factorization as WF of the
unbound state.

\section{A rectangular well with infinitely high walls
\label{sec.7}}

An important progress has been achieved in development of methods
in the approach of inverse problem for construction of new exactly
solvable potentials on the basis of known ones. One can say, that
today the methods of the inverse problem come in the vanguard in
forming an \emph{unified theory of exactly solvable models}
(see~\cite{Zakhariev.1990.PEPAN}, p.~916).
Thus, in~\cite{Zakhariev.1990.PEPAN,Zakhariev.1999.PEPAN}
a nice analysis of a deformation of a rectangular well (and also
some other potentials) under variations of spectral characteristics,
allowing to observe an influence of these characteristics on the
potential shape, was fulfilled on the basis of such methods.
One can explain an interest to such papers by words of authors 
(see~\cite{Zakhariev.1990.PEPAN}, p.~916):
\emph{``Approach of inverse problem is remarkable by that it allows
from another side to see on connection of forces with physical
characteristics of systems, where they act. A possibility appear
to construct the systems with needed properties...''}
(in the original text:
\emph{``Подход обратной задачи замечателен тем, что позволяет
посмотреть с другой стороны на связь сил с физическими характеристиками
систем, в которых они действуют. Появляется возможность конструировать
системы с требуемыми свойствами...''}).

It turns out, that supersymmetric methods, developed at full their
power, allow to achieve this also, not yielding to the methods of
the inverse problem and having own attractive aspects.
Let's consider, how one can deform the rectangular well with
infinitely high walls (which we shall consider as a starting
potential $V_{1}$) by use of the method from sec.~\ref{sec.5.1},
where in construction of superpotential an energy of factorization
coincides with arbitrary selected level of discrete spectrum of the
given $V_{1}$, and a function of factorization is FW of a bound
state at such level.

Let's define this potential so:
\begin{equation}
\begin{array}{lcl}
  V_{1}(x) =
  \left \{
  \begin{array}{cl}
    0,       & \mbox{at } 0<x<a; \\
    +\infty, & \mbox{at } x<0 \mbox{ or } x>a.
  \end{array}
  \right.
\end{array}
\label{eq.7.1.1}
\end{equation}

The problem of determination of the energy spectrum and WFs for this
potential is, perhaps, the task, which is included into any textbook
at quantum mechanics. Solving Schr\"{o}dinger equation with the
potential (\ref{eq.7.1.1}) at the energy $E$ higher then a bottom of
the well ($E \ge 0$), we find a general solution of this equation:
\begin{equation}
  \varphi^{(1)}(x) = A \sin(kx) + B \cos(kx),
\label{eq.7.1.2}
\end{equation}
where
$A$, $B$ are arbitrary (complex) constants and
$k = \sqrt{E}$.

For extracting from the solutions (\ref{eq.7.1.2}) WF, which describes
the bound states, we use boundary conditions (providing continuity of
WF in the whole region of its definition):
\begin{equation}
\begin{array}{ccc}
  \varphi^{(1)}(0) = 0 & \mbox{ and } & \varphi^{(1)}(a) = 0.
\end{array}
\label{eq.7.1.3}
\end{equation}
Application of the first condition gives $B=0$. From the second
condition we obtain:
\begin{equation}
\begin{array}{ccc}
  (ka = \pi n) \to
  \biggl(k = k_{n} = \displaystyle\frac{\pi n}{a} = k_{1} n \biggr) \to
  \biggl(E_{n} = k_{n}^{2} = \displaystyle\frac{\pi^{2}}{a^{2}} n^{2} =
         E_{1} n^{2} \biggr).
\end{array}
\label{eq.7.1.4}
\end{equation}
at $n = 0, 1, 2 \ldots$.
So, we obtain the discrete levels of the energy spectrum $E$ and then
to each level with its WF we add corresponding index $n$. The constant
$A$ for WF with the number $n$ can be found from the condition of its
normalization:
\begin{equation}
  \displaystyle\int\limits_{0}^{a} |\varphi^{(1)}_{n} (x)|^{2} dx = 1,
\label{eq.7.1.5}
\end{equation}
From here we obtain:
\[
\begin{array}{c}
  \displaystyle\int\limits_{0}^{a} |\varphi_{n}^{(1)}(x)|^{2} dx =
  \displaystyle\int\limits_{0}^{a} |A|^{2} |\sin(k_{n}x)|^{2} dx =
  |A|^{2} \displaystyle\int\limits_{0}^{a}
    \displaystyle\frac{1}{2} \bigl(1 - \cos(2k_{n}x)\bigr)^{2} dx = \\

  = \displaystyle\frac{|A|^{2}}{2}
    \biggl(x - \displaystyle\frac{\sin(2k_{n}x)}{2k_{n}} \biggr)
    \Biggr|_{x=0}^{x=a} =
  \displaystyle\frac{|A|^{2}}{2} a = 1
\end{array}
\]
or
\begin{equation}
  A = |A| = \sqrt{\displaystyle\frac{2}{a}}.
\label{eq.7.1.6}
\end{equation}
We see, that the coefficient $A$ is determined \emph{up to phase
factor} and we choose: $A = |A|$.
This gives us only real values for WFs for all states. Note, that
the coefficient $A$ is the same for different states with the numbers
$n$. Finally, write the solution for the \emph{normalizable WF} of
the bound state with the number $n$:
\begin{equation}
  \varphi_{n}^{(1)}(x) =
    \sqrt{\displaystyle\frac{2}{a}} \sin(k_{n} x) =
    \sqrt{\displaystyle\frac{2}{a}} \sin(k_{1} n x).
\label{eq.7.1.7}
\end{equation}

From (\ref{eq.7.1.4}) and (\ref{eq.7.1.7}) we find coordinates of
\emph{nodes for WF with the number $n$}:
\begin{equation}
\begin{array}{cc}
  x_{l} = \displaystyle\frac{\pi l}{k_{n}} =
    \displaystyle\frac{al}{n},
  & l = 0, 1, 2 \ldots n.
\end{array}
\label{eq.7.1.8}
\end{equation}
One can see, that \emph{the nodes for WF with arbitrary number $n$
are located on equal distances one from another, which decrease in
increasing of the number $n$}.

\subsection{The deformation of the rectangular well
\label{sec.7.2}}

Let's find a function $I_{m}^{(1)}(x)$:
\[
\begin{array}{c}
  I_{m}^{(1)} (x) =
  \displaystyle\int\limits_{-\infty}^{x}
    (\varphi_{m}^{(1)}(x'))^{2} dx' =
  \displaystyle\frac{2}{a}
    \displaystyle\int\limits_{0}^{x} \sin^{2}(k_{m}x) dx' =
  \displaystyle\frac{2}{a}
    \displaystyle\int\limits_{0}^{x}
    \displaystyle\frac{1}{2} \Bigl(1 - \cos(2k_{m}x') \Bigr) dx' = \\

  = \displaystyle\frac{1}{a}
    \biggl(x' - \displaystyle\frac{\sin(2k_{m}x')}{2k_{m}} \biggr)
    \Biggr|_{x'=0}^{x'=x} =
    \displaystyle\frac{1}{a}
    \biggl(x - \displaystyle\frac{\sin(2k_{m}x)}{2k_{m}} \biggr).
\end{array}
\]
or
\begin{equation}
  I_{m}^{(1)} (x) =
  \displaystyle\frac{1}{a}
    \biggl(x - \displaystyle\frac{\sin(2k_{m}x)}{2k_{m}} \biggr).
\label{eq.7.2.1}
\end{equation}

For the deformed potential $V_{1}(\lambda_{m})$ we obtain:
\[
\begin{array}{c}
  V_{1}(x, \lambda_{m}) = 
  -2 \displaystyle\frac{d^{2}}{dx^{2}} \log |I_{m} + \lambda_{m}\bigr| =
  -2 \displaystyle\frac{d^{2}}{dx^{2}} \log
    \Biggl|\displaystyle\frac{1}{a}
    \biggl(x - \displaystyle\frac{\sin{2k_{m}x}}{2k_{m}} \biggr) +
    \lambda_{m} \Biggr| = \\

  = -2 \displaystyle\frac{d^{2}}{dx^{2}} \log
    \Bigl|2k_{m}x - \sin{2k_{m}x} + 2k_{m}a\lambda_{m} \Bigr|
\end{array}
\]
or
\begin{equation}
  V_{1}(x, \lambda_{m}) = 
  -2 \displaystyle\frac{d^{2}}{dx^{2}} \log
    \Bigl|2k_{m}x - \sin{2k_{m}x} + 2k_{m}a\lambda_{m} \Bigr|.
\label{eq.7.2.2}
\end{equation}
From here one can see, that at 
$\lambda_{m} \to +\infty$ (or $\lambda_{m} >> 1$)
the potential $V_{1}(\lambda_{m})$ tends to the starting undeformed
potential $V_{1}$.

We find an explicit form of the deformed potential $V_{1}(\lambda_{m})$.
From (\ref{eq.7.2.2}) we write:
\begin{equation}
  V_{1}(x, \lambda_{m}) = 
    -32 k_{m}^{2} \displaystyle\frac
      {\sin{k_{m}x}
        \Bigl(k_{m} (x + a\lambda_{m}) \cos{k_{m}x} - \sin{k_{m}x} \Bigr)}
      {\Bigl(2k_{m}x - \sin{2k_{m}x} + 2k_{m}a\lambda_{m} \Bigr)^{2}}.
\label{eq.7.2.3}
\end{equation}
\emph{Let's name points, where the potential $V_{1}$ equals to zero,
as zero-points or ``nodes'' of this potential}. Then according to
(\ref{eq.7.2.3}), all zero-points of the deformed potential
$V_{1}(\lambda_{m})$ can be separated into two types (for the
rectangular well such a property has found for the first time):

\begin{itemize}
\item
Zero-points, which coincide with the nodes of WF of the bound state
at the level with the number $m$, coincident with the energy of
factorization. These zero-points do not displaced under deformation
of the potential by use of the parameter $\lambda_{m}$ and their
coordinates equal (they determined from (\ref{eq.7.1.7}) with change
of the index $n$ into $m$):
\begin{equation}
\begin{array}{cc}
  x_{l} = \displaystyle\frac{\pi l}{k_{m}} =
    \displaystyle\frac{al}{m},
  & l = 0, 1, 2 \ldots m.
\end{array}
\label{eq.7.2.4}
\end{equation}

\item
Zero-points, which do not coincide with the nodes of WF of the bound
state at the level with the number $m$, coincident with the energy of
factorization. These zero-points are displaced under the deformation
of the potential by use of the parameter $\lambda_{m}$ and their
coordinates are determined from:
\begin{equation}
  \tan{k_{m}x} = k_{m} (x + a\lambda_{m}).
\label{eq.7.2.5}
\end{equation}

\end{itemize}

\subsection{The superpotential under the deformation
\label{sec.7.3}}

Let's find the superpotential for the given potential $V_{1}$:
\[
  W_{m}(x) = 
  -\displaystyle\frac{d}{dx} \log {|\varphi_{m}^{(1)}|} =
  -\displaystyle\frac{d \sin{k_{m}x}}{dx} \Big/ \sin{(k_{m}x)} =
  -k_{m} \cot{k_{m}x}
\]
or
\begin{equation}
  W_{m}(x) = -k_{m} \cot{k_{m}x}.
\label{eq.7.3.1}
\end{equation}
Note, that the superpotential has divergences at the nodes of WF with
the number $m$ (see~(\ref{eq.7.1.7})).

\subsection{The deformation of wave function of the bound state at
the level, coincident with the energy of factorization
\label{sec.7.4}}

For the deformed potential $V_{1}(\lambda_{m})$ we find WF of the
bound state at the level with the number $m$, coincident with the
energy of factorization. The expression (\ref{eq.5.1.1}) determines
such function up to a normalizing constant. One can write:
\[
\begin{array}{c}
  \varphi_{m}^{(1)} (x,\lambda_{m}) =

  C_{m} \sqrt{\displaystyle\frac{2}{a}} \sin(k_{m}x)
    \Biggl(\displaystyle\frac{1}{a}
    \Bigl(x - \displaystyle\frac{\sin{k_{m}x}\cos{k_{m}x}} {k_{m}}
    \Bigr) + \lambda_{m} \Biggr)^{-1} = \\

  = C_{m} \sqrt{\displaystyle\frac{2}{a}} \sin(k_{m}x)
    \displaystyle\frac{k_{m}a}
    {k_{m}x - \sin{k_{m}x} \cos{k_{m}x} + k_{m}a\lambda_{m}} =

  C_{m} \displaystyle\frac{\sqrt{2a} k_{m} \sin(k_{m}x)}
    {k_{m}x - \sin{k_{m}x} \cos{k_{m}x} + k_{m}a\lambda_{m}}
\end{array}
\]
or
\begin{equation}
  \varphi_{m}^{(1)} (x,\lambda_{m}) =
  C_{m} \displaystyle\frac{\sqrt{2a} k_{m} \sin(k_{m}x)}
    {k_{m}x - \sin{k_{m}x} \cos{k_{m}x} + k_{m}a\lambda_{m}}.
\label{eq.7.4.1}
\end{equation}

The constant $C_{m}$ can be found from the normalizing condition of
the found WF:
\begin{equation}
  \displaystyle\int\limits_{0}^{a} 
    |\varphi_{m}^{(1)} (x,\lambda_{m})|^{2} dx = 1
\label{eq.7.4.2}
\end{equation}
or
\[
\begin{array}{lcl}
  1 & = &
  \displaystyle\int\limits_{0}^{a} |C_{m}|^{2}
    \displaystyle\frac{2a k_{m}^{2} \sin^{2}(k_{m}x)}
    {\Bigl(k_{m}x - \sin{k_{m}x} \cos{k_{m}x} +
     k_{m}a\lambda_{m} \Bigr)^{2}} dx = \\

  & = &
    |C_{m}|^{2} \cdot 2a k_{m}^{2}
    \displaystyle\int\limits_{0}^{a}
    \displaystyle\frac{\sin^{2}(k_{m}x)}
    {\Bigl(k_{m}x - \frac{1}{2} \sin{2k_{m}x} +
      k_{m}a\lambda_{m} \Bigr)^{2}} dx = \\

  & = &
    |C_{m}|^{2} \cdot a k_{m}^{2}
    \displaystyle\int\limits_{0}^{a}
    \displaystyle\frac{1 - \cos{(2k_{m}x)}}
    {\Bigl(k_{m}x - \frac{1}{2} \sin{2k_{m}x} +
      k_{m}a\lambda_{m} \Bigr)^{2}} dx.
\end{array}
\]
Substitution:
\[
\begin{array}{llcl}
  y  = k_{m}x - \frac{1}{2} \sin{2k_{m}x} + k_{m}a\lambda_{m}, &
    (x=0) & \to & y = k_{m}a\lambda_{m}, \\
  dy = k_{m} \bigl(1 - \cos{2k_{m}x} \bigr) dx, &
    (x=a) & \to & (y = k_{m}a(1+\lambda_{m})).
\end{array}
\]
Then we obtain:
\[
\begin{array}{lcl}
  1 & = &
    |C_{m}|^{2} \cdot a k_{m}
    \displaystyle\int\limits_{x=0}^{x=a}
      \displaystyle\frac{dy}{y^{2}} =

    -|C_{m}|^{2} \cdot a k_{m} \displaystyle\frac{1}{y} \bigg|_{x=0}^{x=a} =

    |C_{m}|^{2} \cdot a k_{m}
    \biggl( \displaystyle\frac{1}{k_{m}a\lambda_{m}} -
      \displaystyle\frac{1}{k_{m}a (\lambda_{m}+1)} \biggr) = \\

  & = &
    |C_{m}|^{2}
    \biggl( \displaystyle\frac{1}{\lambda_{m}} -
      \displaystyle\frac{1}{\lambda_{m}+1} \biggr) =

    |C_{m}|^{2}
    \displaystyle\frac{1}{\lambda_{m} (\lambda_{m}+1)}.
\end{array}
\]
From here we find:
\begin{equation}
  |C_{m}| = \sqrt{\lambda_{m} (\lambda_{m}+1)}.
\label{eq.7.4.3}
\end{equation}
We see, that the normalizing constants $C_{m}$ are different for the
different $m$ and $\lambda_{m}$. From (\ref{eq.7.4.1}) and taking into
account (\ref{eq.7.4.3}), we obtain a final expression for the
normalizable WF (we select $C_{m}=|C_{m}|$):
\begin{equation}
  \varphi_{m}^{(1)} (x,\lambda_{m}) =
    \sqrt{\lambda_{m} (\lambda_{m}+1)}
    \displaystyle\frac{\sqrt{2a} k_{m} \sin(k_{m}x)}
    {k_{m}x - \sin{k_{m}x} \cos{k_{m}x} + k_{m}a\lambda_{m}}.
\label{eq.7.4.4}
\end{equation}

\vspace{3mm}
\noindent
\underline{\bf Property 1:}
\emph{nodes of this WF do not displaced at its deformation by the
parameter $\lambda_{m}$ for arbitrary $m$ and they coincide with the
nodes (\ref{eq.7.1.8}) of the undeformed WF.} Let's find a derivative
of this WF:
\[
\begin{array}{c}
  \displaystyle\frac{d \varphi_{m}^{(1)} (x,\lambda_{m})}{dx} =

  \sqrt{\lambda_{m} (\lambda_{m}+1)}
    \displaystyle\frac{d}{dx}
    \displaystyle\frac{\sqrt{2a} k_{m} \sin(k_{m}x)}
    {k_{m}x - \sin{k_{m}x} \cos{k_{m}x} + k_{m}a\lambda_{m}} = \\

  = \sqrt{\lambda_{m} (\lambda_{m}+1)} \sqrt{2a} k_{m}^{2}
    \displaystyle\frac{\cos{k_{m}x}
        \Bigl(k_{m}x - \sin{k_{m}x}\cos{k_{m}x} + k_{m}a\lambda_{m}\Bigr) -
      \sin{k_{m}x}
        \Bigl(1 - \cos^{2}{k_{m}x} + \sin^{2}{k_{m}x} \Bigr)}
    {\Bigl(k_{m}x - \sin{k_{m}x}\cos{k_{m}x} + k_{m}a\lambda_{m}\Bigr)^{2}} = \\

  = \sqrt{2a\lambda_{m} (\lambda_{m}+1)} k_{m}^{2}
    \displaystyle\frac{\cos{k_{m}x}
        \bigl(k_{m}x - \sin{k_{m}x}\cos{k_{m}x} + k_{m}a\lambda_{m}\bigr) -
      2\sin^{3}{k_{m}x}}
    {\Bigl(k_{m}x - \sin{k_{m}x}\cos{k_{m}x} + k_{m}a\lambda_{m}\Bigr)^{2}}
\end{array}
\]
or
\begin{equation}
  \displaystyle\frac{d \varphi_{m}^{(1)} (x,\lambda_{m})}{dx} =
  \sqrt{2a \lambda_{m} (\lambda_{m}+1)} k_{m}^{2}
    \displaystyle\frac{\cos{k_{m}x}
        \bigl(k_{m}x - \sin{k_{m}x}\cos{k_{m}x} + k_{m}a\lambda_{m}\bigr) -
      2\sin^{3}{k_{m}x}}
    {\Bigl(k_{m}x - \sin{k_{m}x}\cos{k_{m}x} + k_{m}a\lambda_{m}\Bigr)^{2}}.
\label{eq.7.4.5}
\end{equation}

At $x$-coordinates $x=0$ and $x=a$ of the walls we obtain
($\cos{k_{m}a} = (-1)^{m}$):
\begin{equation}
\begin{array}{cc}
  \displaystyle\frac{d \varphi_{m}^{(1)} (0,\lambda_{m})}{dx} =
    k_{m} \sqrt{\displaystyle\frac
      {2 (1 + \lambda_{m})} {a\lambda_{m}}}, &

  \displaystyle\frac{d \varphi_{m}^{(1)} (a,\lambda_{m})}{dx} =
    (-1)^{m} k_{m} \sqrt{\displaystyle\frac
      {2 \lambda_{m}} {a(1 + \lambda_{m})}}.
\end{array}
\label{eq.7.4.6}
\end{equation}

\vspace{3mm}
\noindent
\underline{\bf Property 2:}
\emph{According to (\ref{eq.7.4.6}), angles of slope of WF relatively
axis $x$ at $x$-coordinates of the walls are changed at the
deformation of this WF by use of the parameter $\lambda_{m}$.}

\subsection{The deformation of wave functions of other bound states
\label{sec.7.5}}

According to sec.~\ref{sec.5.2}, we write the normalizable WF of the
arbitrary bound state for the deformed potential $V_{1}(x,\lambda_{m})$
at the level with the number $n$, not coincident with the energy of
factorization ($E_{n} \ne {\cal E} = E_{m}$):
\begin{equation}
\begin{array}{lcl}
  \varphi_{n}^{(1)}(\lambda_{m}) & = &
    \varphi_{n}^{(1)} +
    \Delta\varphi_{n}^{(1)}(\lambda_{m}), \\

  \Delta\varphi_{n}^{(1)}(\lambda_{m}) & = &
    \displaystyle\frac{\Delta W(\lambda_{m})}
    {|E_{n}^{(1)} - E_{m}^{(1)}|} A_{m} 
    \varphi_{n}^{(1)}, \\

  \Delta W(\lambda_{m}) & = &
    \displaystyle\frac{d}{dx} \log{\bigl|I_{m} + \lambda_{m} \bigr|}.
\end{array}
\label{eq.7.5.1}
\end{equation}
We see, that the deformation of this WF in result of change of the
parameter $\lambda_{m}$ is defined by the function
$\Delta\varphi_{k_{n}}^{(1)}(\lambda_{m})$ purely.
Taking into account (\ref{eq.7.2.1}), (\ref{eq.7.3.1}) and
(\ref{eq.7.1.7}), we find:
\begin{equation}
\begin{array}{c}
  \Delta W(\lambda_{m}) A_{m} \varphi_{n}^{(1)} =

    \displaystyle\frac{d}{dx} \log{\bigl|I_{m} + \lambda_{m} \bigr|} 
    \biggl(\displaystyle\frac{d}{dx} + W_{m} \biggr)
    \varphi_{n}^{(1)} = \\

%
%
%
%
%

  = \sqrt{\displaystyle\frac{2}{a}}
    \displaystyle\frac{4k_{m} \sin{k_{m}x}}
    {2k_{m}x - \sin{2k_{m}x} + 2k_{m}a\lambda_{m}}

    \Bigl(k_{n} \cos{k_{n}x}\sin{k_{m}x} -
      k_{m} \cos{k_{m}x} \sin{k_{n}x} \Bigr).
\end{array}
\label{eq.7.5.2}
\end{equation}
From here it follows, that the function
$\Delta\varphi_{n}^{(1)}(\lambda_{m})$ can be equal to zero only at
fulfillment of one from two following conditions:
\begin{equation}
\begin{array}{cc}
  \sin{k_{m}x} = 0, &
  k_{n} \cos{k_{n}x}\sin{k_{m}x} = k_{m} \cos{k_{m}x} \sin{k_{n}x}.
\end{array}
\label{eq.7.5.3}
\end{equation}
The first condition is fulfilled at such points:
\begin{equation}
\begin{array}{lcl}
  x_{l} = \displaystyle\frac{\pi l}{k_{m}} = \displaystyle\frac{al}{m},
    & \mbox{at } l = 0, 1, 2 \ldots m.
\end{array}
\label{eq.7.5.4}
\end{equation}

Thus, we have obtained the following properties of the deformation
of WF of the bound state at the level with arbitrary number $n$,
not coincident with the energy of factorization ($n \ne m$, for the
rectangular well these properties have obtained at the first time).

\vspace{3mm}
\noindent
\underline{\bf Properties 1:}
\begin{itemize}
\item
\emph{There are such points, where all deformed curves of WF of the
state with the selected number $n$ intersect between themselves and
with undeformed WF}. One can separate these points into two types:

\begin{itemize}
\item
The points of the first type, $x$-coordinates of which equal to
(\ref{eq.7.5.4}) and coincide with the nodes (\ref{eq.7.1.8}) or
(\ref{eq.7.4.5}) for WF of the bound state at the level with the
number $m$, coincident with the energy of factorization, and with
zero-points of the potential of the first type (\ref{eq.7.2.4}) also.

\item
The points of the second type, $x$-coordinates of which do not
coincide with the nodes of WF of the bound state at the level with
the number $m$, coincident with the energy of factorization. The
coordinates of these points can be determined from the second
equation of the system (\ref{eq.7.5.3}).
\end{itemize}

\item
The boundary points $x=0$ and $x=a$ are such points of both the
first type, and the second one.

\item
Such point of the first and second types do not displaced at the
deformation of WF using the parameter $\lambda_{m}$.

\item
WF of arbitrary bound state (in its deformation using the parameter
$\lambda_{m}$) have $m+1$ points of the first type and number of the
points of the second type, which can be determined from the second
equation from (\ref{eq.7.5.3}).

\end{itemize}

Now let's find a derivative of this WF. From (\ref{eq.7.5.1}) we
obtain:
\begin{equation}
\begin{array}{c}
  \displaystyle\frac{d \varphi_{n}^{(1)}(\lambda_{m})}{dx} =

  \displaystyle\frac{d \varphi_{n}^{(1)}}{dx} +
    \displaystyle\frac{d}{dx}
      \Delta\varphi_{n}^{(1)}(\lambda_{m}) = \\

%
%
%
%
%
%
  = \displaystyle\frac{d \varphi_{n}^{(1)}}{dx} +

    \sqrt{\displaystyle\frac{2}{a}}
      \displaystyle\frac
          {4k_{m}^{2} \cos{k_{m}x}
            \Bigl(k_{n} \cos{k_{n}x}\sin{k_{m}x} -
            k_{m} \cos{k_{m}x} \sin{k_{n}x} \Bigr)
            }
          {|E_{n}^{(1)} - E_{m}^{(1)}|
            \Bigl(2k_{m}x - \sin{2k_{m}x} + 2k_{m}a\lambda_{m}\Bigr)} + \\

   + \sqrt{\displaystyle\frac{2}{a}}
       \displaystyle\frac
          {4k_{m} \Bigl(k_{m}^{2}-k_{n}^{2}\Bigr)
            \sin^{2}{k_{m}x} \sin{k_{n}x}}
          {|E_{n}^{(1)} - E_{m}^{(1)}|
            \Bigl(2k_{m}x - \sin{2k_{m}x} + 2k_{m}a\lambda_{m}\Bigr)}

   - \sqrt{\displaystyle\frac{2}{a}}
      \displaystyle\frac
          {16k_{m}^{2} \sin^{3}{k_{m}x}
            \Bigl(k_{n} \cos{k_{n}x}\sin{k_{m}x} -
               k_{m} \cos{k_{m}x} \sin{k_{n}x} \Bigr)}
          {|E_{n}^{(1)} - E_{m}^{(1)}|
          \Bigl(2k_{m}x - \sin{2k_{m}x} + 2k_{m}a\lambda_{m}\Bigr)^{2}}.
\end{array}
\label{eq.7.5.5}
\end{equation}
From here we have in the coordinates $x=0$ and $x=a$ of the walls:
\begin{equation}
\begin{array}{cc}
  \displaystyle\frac{d \varphi_{n}^{(1)}(\lambda_{m})}{dx} =
    \displaystyle\frac{d \varphi_{n}^{(1)}}{dx}, &
  \displaystyle\frac{d}{dx} \Delta\varphi_{n}^{(1)}(\lambda_{m}) = 0.
\end{array}
\label{eq.7.5.6}
\end{equation}

Thus, we have proved the properties, pointed out in 
\cite{Zakhariev.1990.PEPAN} (p.~918):

\vspace{3mm}
\noindent
\underline{\bf Properties 2:}

\begin{itemize}
\item
At the coordinates $x=0$ and $x=a$ of the walls WF of the arbitrary
bound state at the level with the number $n$, which does not coincide
with the energy of factorization ($E_{n} \ne {\cal E} = E_{m}$),
leaves at the same angle of the slope to the axis $x$ under its
deformation using the parameter $\lambda_{m}$ (or under change of a
derivative $\gamma_{m}$ in the approach of the inverse problem).

\item
The angles of leaving of this WF from $x$-coordinates of the walls
are determined by the derivative of the undeformed WF in the form
(\ref{eq.7.5.6}).
\end{itemize}

\subsection{Divergences in solutions
\label{sec.7.6}}

The found solution (\ref{eq.7.4.5}) for WF at the level with the
number $m$, which coincides with the energy of factorization, or the
solution (\ref{eq.7.5.1}) for WF of the state with the number $n$ at
another level for the deformed potential describes the
\underline{bound state} only in such a case, when this WF has not
divergences and is continuous on the whole region of its definition
at $x$.
According to these solutions, the divergences for WF with the selected
numbers $m$ and $n$ appear only in the case, when:
\begin{equation}
  2k_{m}x - \sin{2k_{m}x} + 2k_{m}a\lambda_{m} = 0.
\label{eq.7.6.1}
\end{equation}
One can exclude these divergences (at any $x$) from the found
solutions for WF, in particular, by imposing on a choice of the
parameter $\lambda_{m}$ the following requirements:
\[
\begin{array}{cc}
  2k_{m}a\lambda_{m} < -2k_{m}a -1, &
  2k_{m}a\lambda_{m} > 1
\end{array}
\]
or
\begin{equation}
\begin{array}{cc}
  \lambda_{m} < -\displaystyle\frac{1+2k_{m}a}{2k_{m}a}, &
  \lambda_{m} > \displaystyle\frac{1}{2k_{m}a}.
\end{array}
\label{eq.7.6.2}
\end{equation}

From here we conclude the following.

\vspace{3mm}
\noindent
\underline{\bf Properties:}

\begin{itemize}
\item
If the parameter $\lambda_{m}$ is in one from regions, determined in
(\ref{eq.7.6.2}), then WF of the form (\ref{eq.7.4.5}) for the state
at the level with arbitrary number $m$, which coincides with the
energy of factorization, does not have the divergences, it is the
continuous function at $0 \le x \le a$ and equals to zero at
coordinates $x=0$ and $x=a$ of the walls.
Therefore, it describes the bound state with the number $m$ (in spite
of the fact that the superpotential, constructed on the basis of WF
of the excited bound state with the number $m$, has the divergences
at the nodes $x_{l} = a l/m$ of this WF, $l = 0, 1, 2 \ldots m$).

\item
If the parameter $\lambda_{m}$ is in one from the regions, determined
in (\ref{eq.7.6.2}), then WF of the form (\ref{eq.7.5.1})--(\ref{eq.7.5.2})
for the state at arbitrary level with the number $n$, which does not
coincide with the energy of factorization, does not have the
divergences, it is the continuous function at $0 \le x \le a$ and
equals to zero at the coordinates $x=0$ and $x=a$ of the walls.
Therefore, it describes the bound state with the number $n$ (at
arbitrary $m$, in spite of the fact that the superpotential,
constructed on the basis of WF of the excited bound state with the
number $m$, has the divergences at the nodes
$x_{l} = a l/m$ of this WF, $l = 0, 1, 2 \ldots m$).

\item
If the parameter $\lambda_{m}$ is in one from the regions, determined
in (\ref{eq.7.6.2}), then the deformed potential of the form
(\ref{eq.7.2.3}) does not have the divergences and is the continuous
function at $0 \le x \le a$ (in spite of the fact that the
superpotential, constructed on the basis of WF of the excited bound
states with the number $m$, has the divergences at the nodes
$x_{l} = a l/m$ of this WF, $l = 0, 1, 2 \ldots m$).
\end{itemize}

\vspace{3mm}
\noindent
\underline{\bf Conclusion:}

\noindent
\emph{%
We have proved an appropriateness of the method (from
sec.~\ref{sec.5.1}) of construction of new isospectral potentials
(with own bound states and without the divergences) on the basis of
one known potential $V_{1}$ in the form of the rectangular well with
infinitely high external walls, if in the definition of the
superpotential as the energy of factorization the arbitrary selected
level of the discrete energy spectrum of the given $V_{1}$ is used,
and as the function of factorization the WF of the bound state at
such level is used}.

\subsection{Analysis
\label{sec.7.7}}

Let's consider, how in the frameworks of the supersymmetric approach
one can deform the rectangular well with infinitely high external
walls and finite width $a$ (further, for the convenience we choose
$a=1$).
As a method of deformation we shall use the method from sec.~\ref{sec.5.1},
where in construction of the superpotential
as the energy of factorization we choose an arbitrary selected 
level with number $m$ of the energy spectrum of this well,
and as the function of factorization we use WF of the bound state 
at this level.
We shall consider each new potential, obtained by help of this method, 
as a deformation of the old rectangular well (i.~e. the starting potential 
$V_{1}(x, \lambda_{m})$ in its formalism) using parameters of the
deformation --- the number $m$ of the level, equals to the energy of
factorization, and the parameter $\lambda_{m}$.
We shall obtain the deformed potential from (\ref{eq.7.2.2}),
the superpotential --- from (\ref{eq.7.3.1}),
WFs --- from (\ref{eq.7.4.5}) and (\ref{eq.7.5.1}).

\begin{figure}[htbp]
\centerline{
\includegraphics[width=75mm]{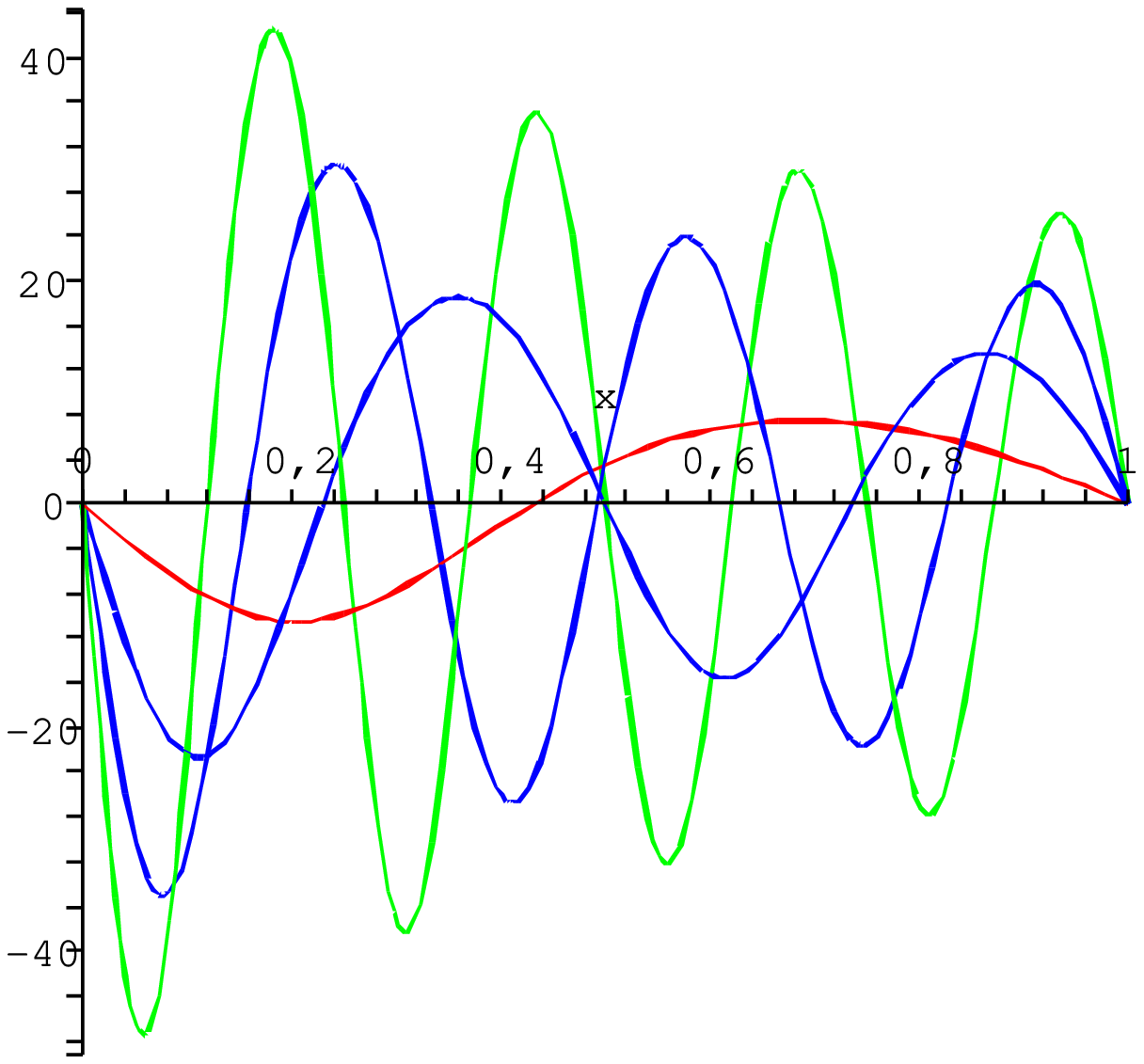}
\includegraphics[width=75mm]{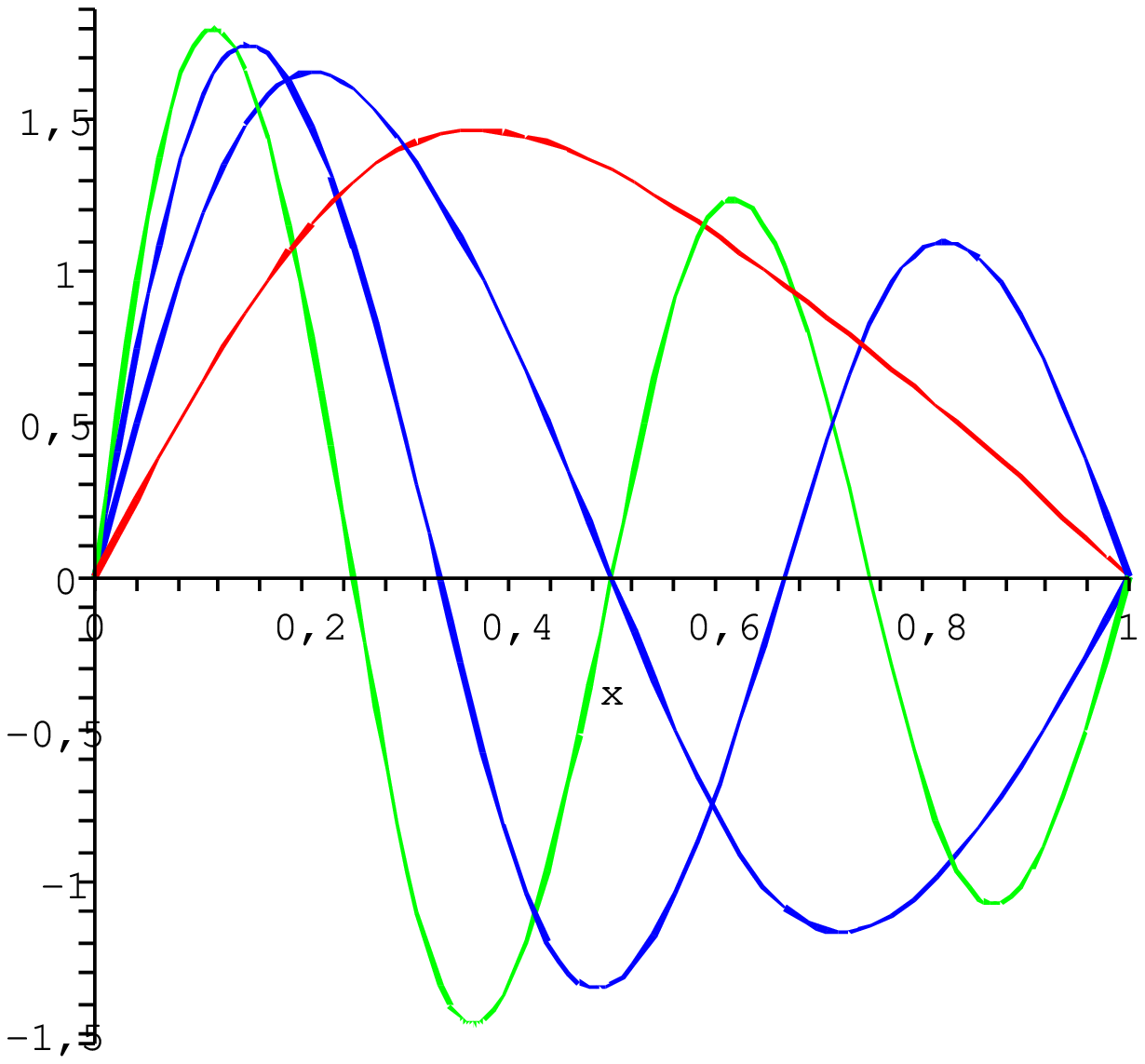}}
\caption{
\small
The deformation of the shapes of the potential and WF of the bound
state at the level with the number $m$, coincident with the energy
of factorization, in dependence on a change of the number $m$
(values $m = 1, 2, 3, 4$ at $\lambda_{m}=1$ are used):
(a) --- the deformation of the potential;
(b) --- the deformation of WF.
\label{fig.723.1}}
\end{figure}

The found deformed potential (a) and its FW of the bound state at the
level with the number $m$ (b), coincident with the energy of
factorization, are shown in Fig.~\ref{fig.723.1} at choice of the
different values $m$.
In the first figure (a) one can see, that the selected level $m$
determines number of intersections of the potential with axis $x$
(or \emph{number of ``zero-points'' of this potential}); increasing
it, one can increase the number of such points of intersections.
We also see, that by increasing $m$ both an angle of a slope of the
potential in its leaving from the left boundary at point $x=0$, and
an angle of a slope of WF at its leaving from this boundary, increase
(that follows from the choice of the definition of the function
$I_{m}(x)$ for calculation of the deformed WFs and the potential).
From these figures one can see, that at the selected values of the
parameter $\lambda_{m}$ there are no any divergence on the whole
region $x \in [0, a]$ both in the deformed potential, and in WF at
the different values $m$, and also that WF equals to zero at the
boundary points $x=0$ and $x=a$ (that points out the bounding of
this state).
\emph{It demonstrates an appropriateness of the method from
sec.~\ref{sec.5.1} for construction of new isospectral potentials
using the superpotential, defined on the basis of WF of not only
the ground state at $m=1$, but also \underline{the arbitrary excited
state at $m \ne 1$}.}

\begin{figure}[htbp]
\centerline{
\includegraphics[width=57mm]{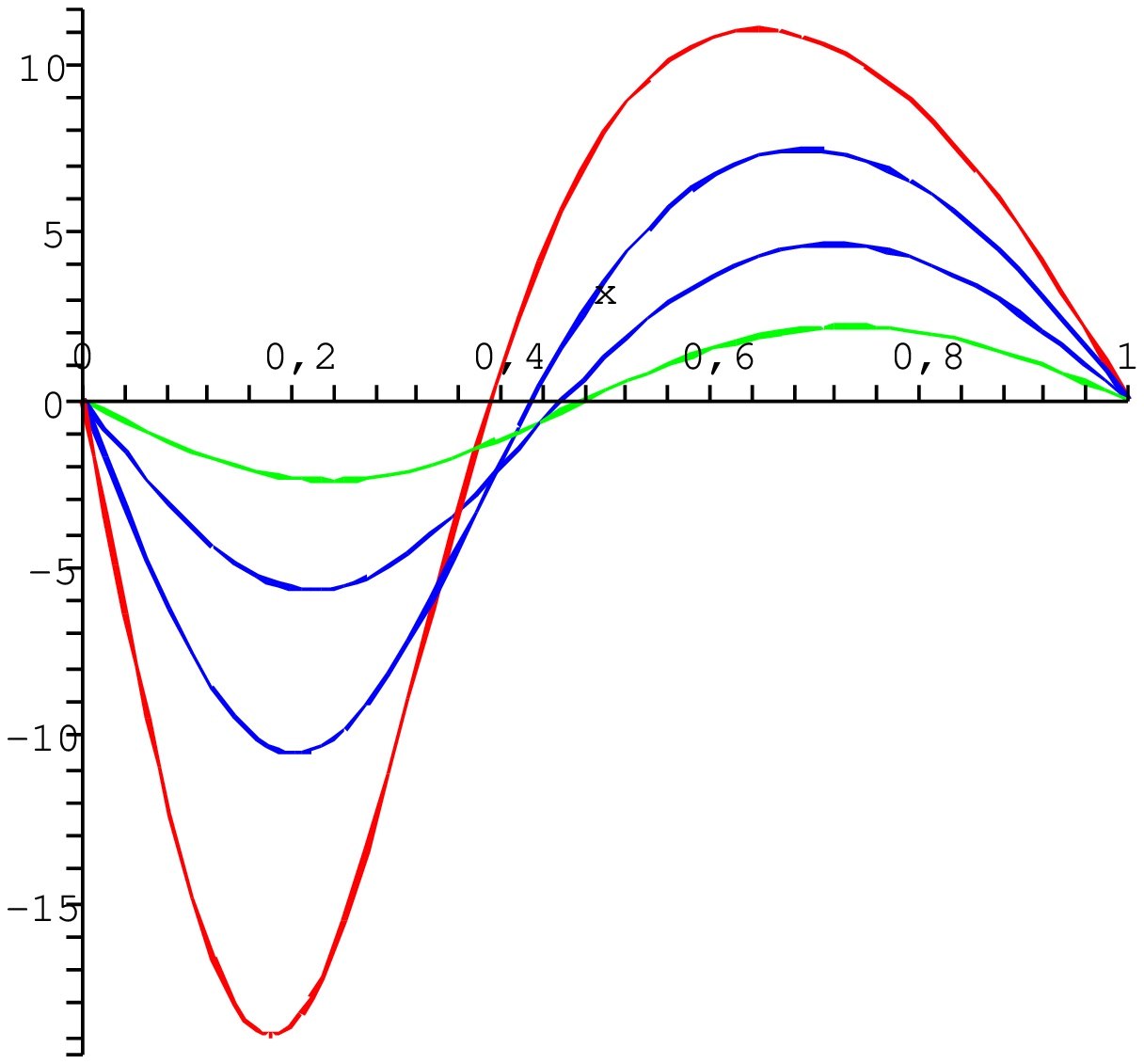}
\includegraphics[width=57mm]{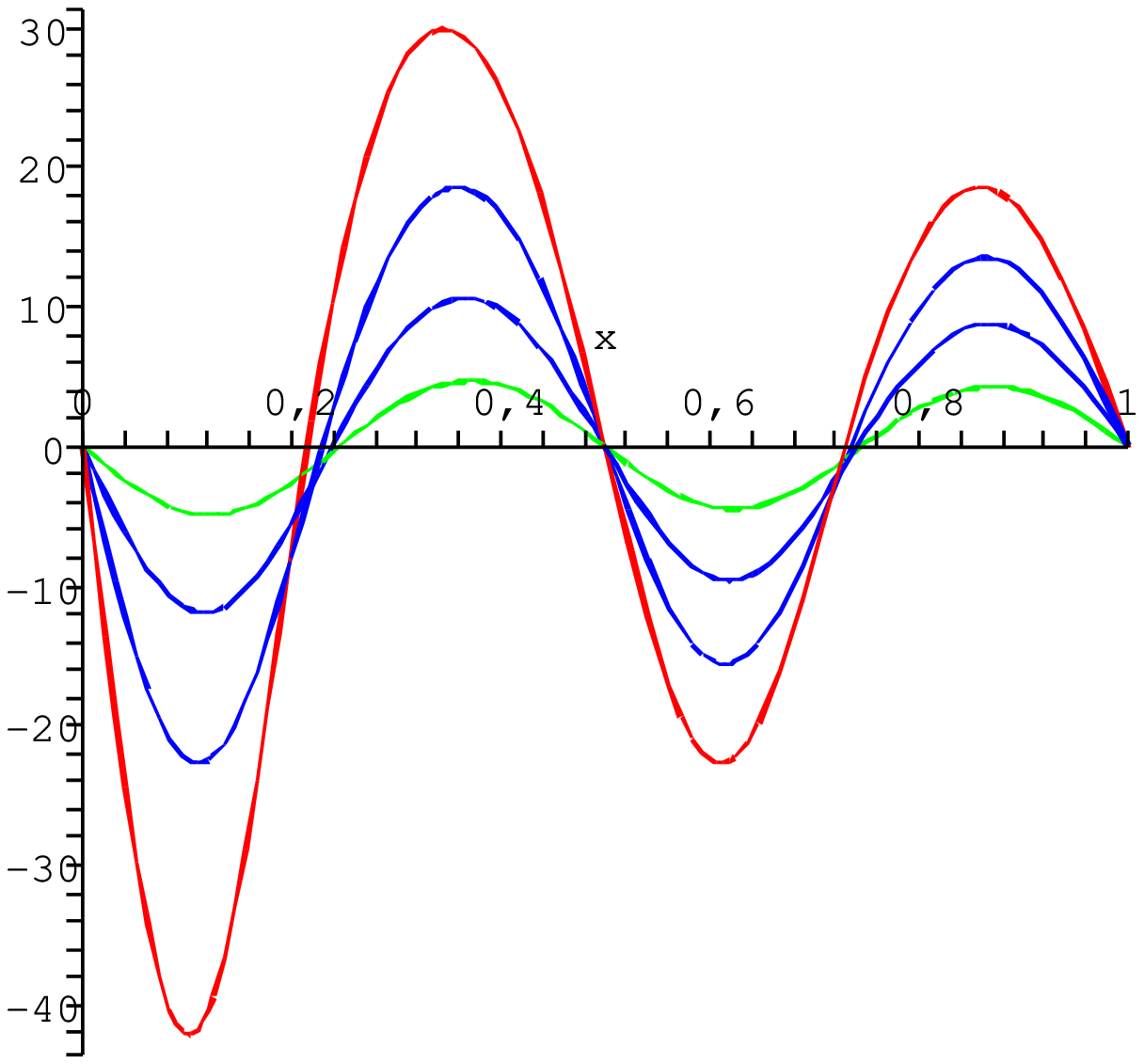}
\includegraphics[width=57mm]{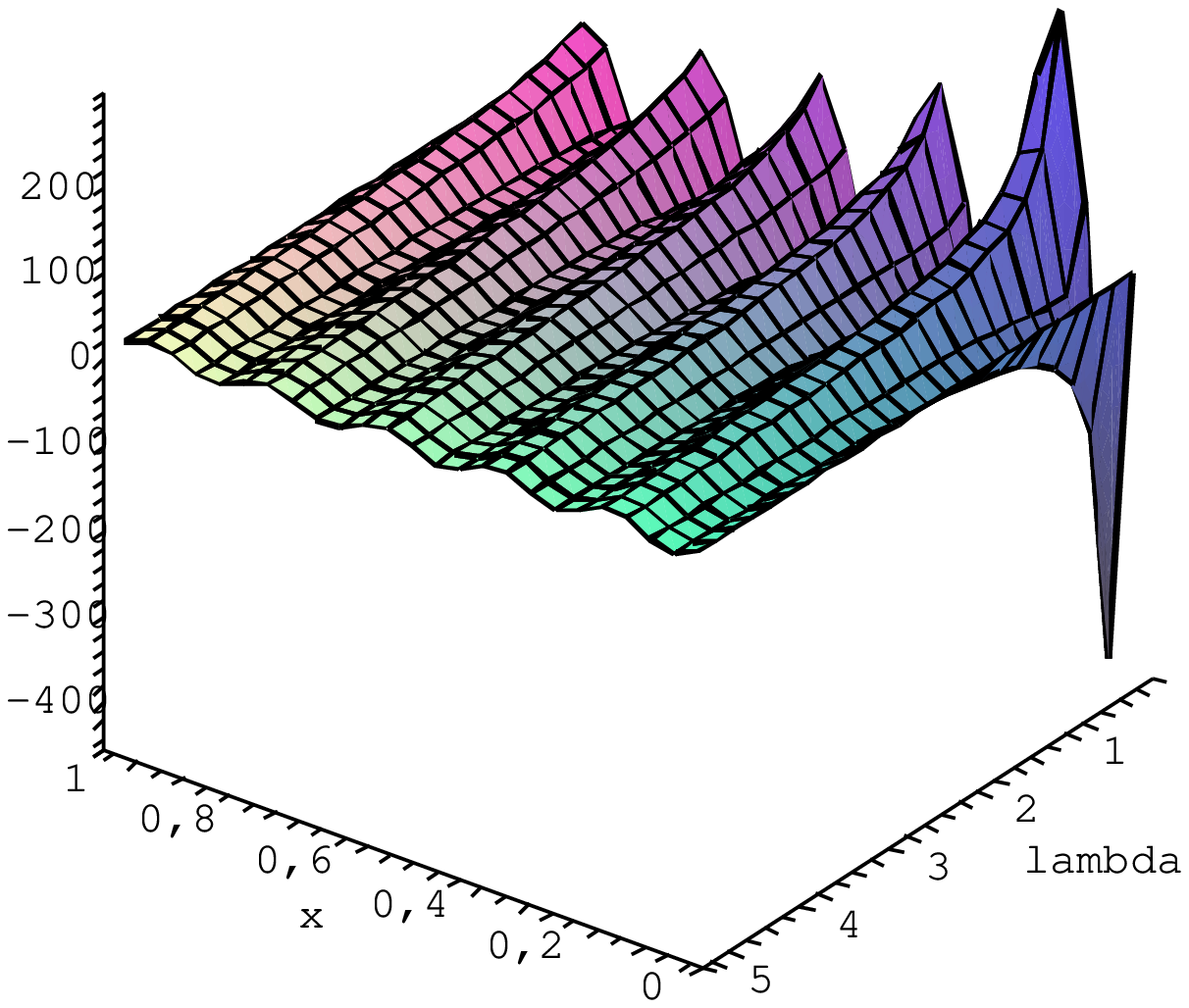}}
\caption{
\small
The deformation of the potential at the change of the parameter
$\lambda_{m}$
(the values $\lambda_{m} = 0.5, 1, 2, 5$ are used):
(a) --- at $m=1$,
(b) --- at $m=2$,
(c) --- the shape of the potential is deformed continuously at the
change of the parameter $\lambda_{m}$ (without the divergences)
\label{fig.723.2}}
\end{figure}

The deformation of the potential at the change of the parameter
$\lambda_{m}$ is shown in Fig.~\ref{fig.723.2}.
In the first figure (a) we see, how the potential is deformed, if as
the energy of factorization the lowest level at $m=1$ is used.
Here, we note the following:

\vspace{0mm}
{\small
\begin{itemize}
\item
We see, that this figure qualitatively looks like Fig.~1 (a) from
\cite{Zakhariev.1990.PEPAN} (see~p.~917), in which the deformation of
such potential by help of the methods of the inverse problem is shown 
(here, varying the parameter $\lambda_{1}$ and changing the
definition of the function $I_{m}(x)$, one can obtain a full
similarity between these figures).

\vspace{0mm}
\item
We see, that the increase of the slope of leaving of the potential in
the left wall (and the increase of minimums and maximums of the
potential) at increasing of the parameter $\lambda_{1}$ coincides
with the increase of the slope of leaving of the potential in the
right wall (and with the increase of its minimums and maximums) in
Fig.~1 (a) in~\cite{Zakhariev.1990.PEPAN} (see~p.~917), which was
caused by a change of the derivative $\gamma_{1}$ of WF of the ground
state in the right wall at $x=a$.
\emph{It points out a similarity between the parameter $\lambda_{1}$
in the considered method in the frameworks of SUSY QM and the
coefficient $\gamma_{1}$ in the approach of the inverse problem.}

\vspace{0mm}
\item
The following property is fulfilled: the deformed curves of the
potential $V_{1}(\lambda_{m})$ at the different values $\lambda_{m}$
intersect the axis $x$ at different points, i.~e. \emph{we obviously
observe a displacement of the central zero-point of the potential at
its deformation by $\lambda_{m}$}. It is explained clearly (and at
first time) by the found property from sec.~\ref{sec.7.2}, and this
property is observed also in Fig.~1 (a) in~\cite{Zakhariev.1990.PEPAN}
(see~p.~917) (the considered node is the node of the second type, its
coordinate is determined by equation (\ref{eq.7.2.5}) at the given
$\lambda_{m}$).
\end{itemize}}

\vspace{0mm}
\noindent
In the second figure (b) we show the deformation of the well, if in
the definition of the energy of factorization the level of the first
excited state at $m=2$ is used. Here, we note the following:

\vspace{0mm}
{\small
\begin{itemize}
\item
The deformed well after its mirror reflection coincides practically
with Fig.~1 (5) from \cite{Zakhariev.1990.PEPAN} (see~p.~917), where
the deformation of such well, obtained by the methods of the inverse
problem at the change of the derivative $\gamma_{2}$ of WF of the
first excited state at the right wall at $x=a$, is shown.

\vspace{0mm}
\item
One can see, that the increase of the leaving slope of the potential
in the left wall, caused by increasing the parameter $\lambda_{2}$,
coincides with the increase of the leaving slope of the potential in
the right wall in Fig.~1 (5) in~\cite{Zakhariev.1990.PEPAN}
(see~p.~917), caused by a change of the derivative $\gamma_{2}$.
It proves the conclusion about that \emph{the parameters $\lambda_{m}$
with the different $m$ in the method from~\ref{sec.5.1} play the same
role in the deformation of the potential, as the coefficients
$\gamma_{m}$ (defined at the same levels with the numbers $m$) in the
approach from~\cite{Zakhariev.1990.PEPAN}}.

\vspace{0mm}
\item
From the figure one can see, that the well after its deformation
has ``zero-points'' of both types, described in sec.~\ref{sec.7.2}
(this property has observed and explained at the first time):

\vspace{0mm}
\begin{itemize}
\item
three zero-points, located at points of the walls of the well and at
its center, which are not displaced under the change of $\lambda_{m}$
(they coincide with all nodes of WF of the first excited state at
$m=2$; coordinates of these zero-points can be found from
(\ref{eq.7.2.4}));

\vspace{0mm}
\item
two zero-points, located between the previous three zero-points,
which are displaced under the change of $\lambda_{m}$ (their
coordinates can be found from (\ref{eq.7.2.5}) at the given
$\lambda_{m}$).
\end{itemize}

\vspace{0mm}
\noindent
As we see, this property is displayed also in Fig.~1 (5)
in~\cite{Zakhariev.1990.PEPAN} (see~p.~917).
\end{itemize}}

\vspace{0mm}
\noindent
According to the figures (a,b), the $m$ is larger, the number of
zero-points on the region $x \in [0, a]$ is larger and the
displacement of zero-points of the second type in the deformation
of the well is smaller.

\begin{figure}[htbp]
\centerline{
\includegraphics[width=80mm]{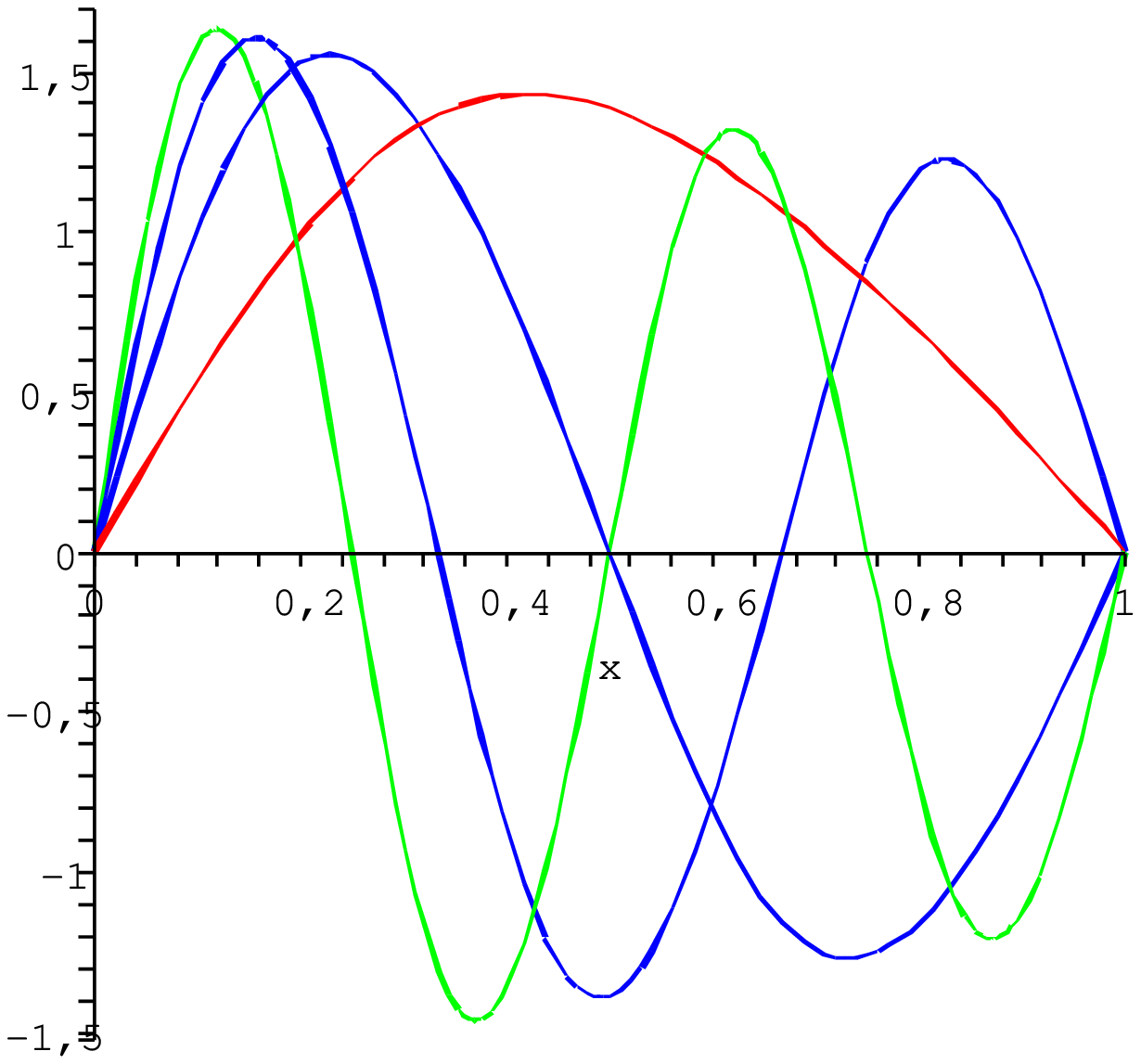}
\includegraphics[width=80mm]{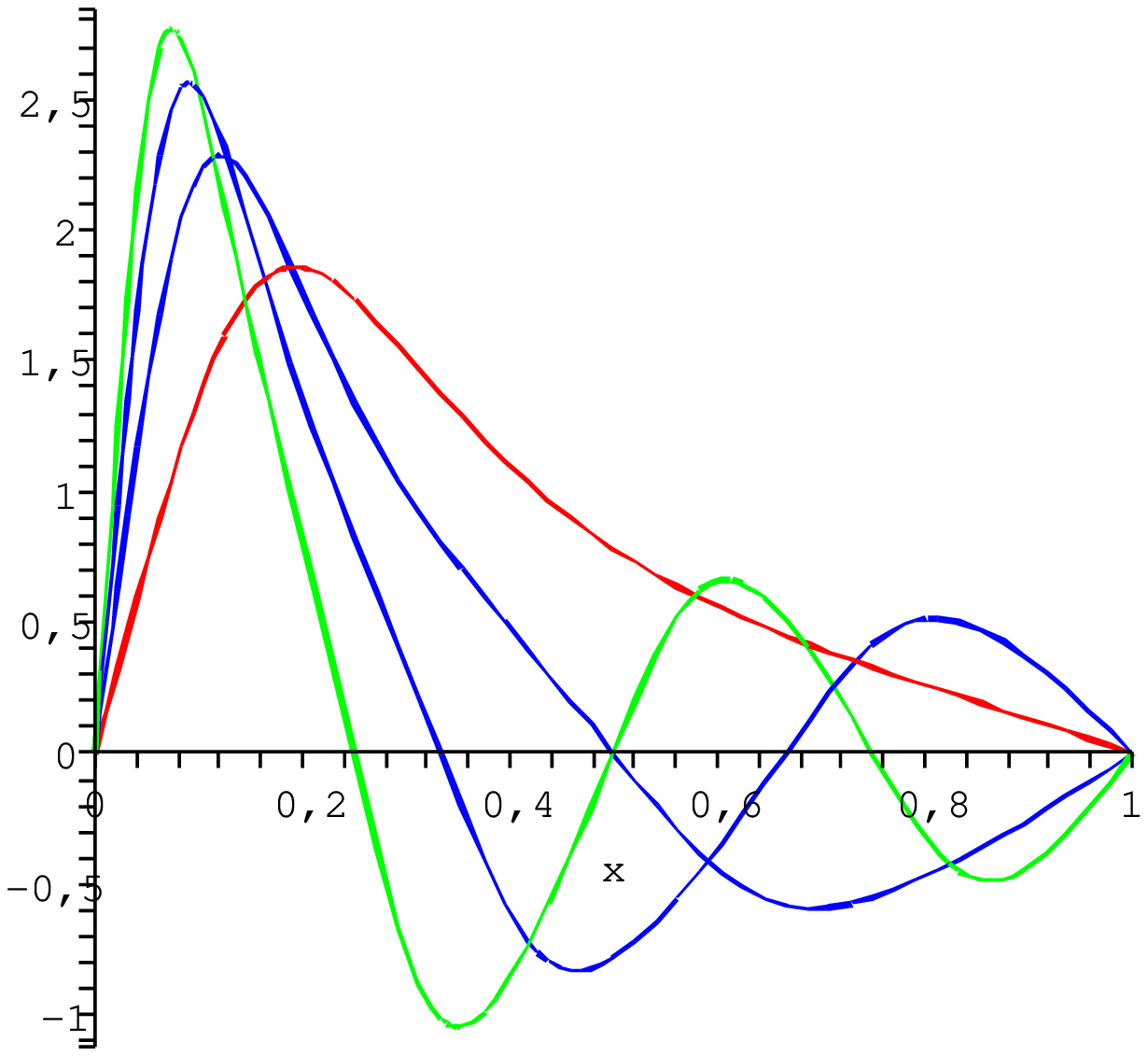}}
\caption{
\small
The deformation of the shape of WF at the level with the number $m$,
coincident with the energy of factorization, caused by the change of
the number $m$ (the values $m = 1, 2, 3, 4$ are used):
(a) --- at $\lambda_{m}=2$;
(b) --- at $\lambda_{m}=0.1$
\label{fig.723.3}}
\end{figure}

Now let's analyze, how the shape of WF is deformed at the level with
the number $m$, coincident with the energy of factorization.
In Fig.~\ref{fig.723.3} such deformation of WF at the change of the
number $m$ of such level is shown.
From the figures one can see the following:

\vspace{0mm}
{\small
\begin{itemize}
\item
Distances between arbitrary two neighboring nodes of WF with the
arbitrary number $m$ are equal, they decrease at increasing of the
number $m$ and their coordinates can be found from (\ref{eq.7.1.8})
(with substitution of index $n$ into $m$).

\vspace{0mm}
\item
For any $m$ at decreasing of the module of $\lambda_{m}$ a relative
weight (amplitude) of WF near one wall is increased and near another
wall is decreased, at $\lambda_{m} \to \pm\infty$ a smoothing of
the deformation of WF takes place and WF tends to its undeformed form.

\end{itemize}}

\begin{figure}[hbtp]
\centerline{
\includegraphics[width=57mm]{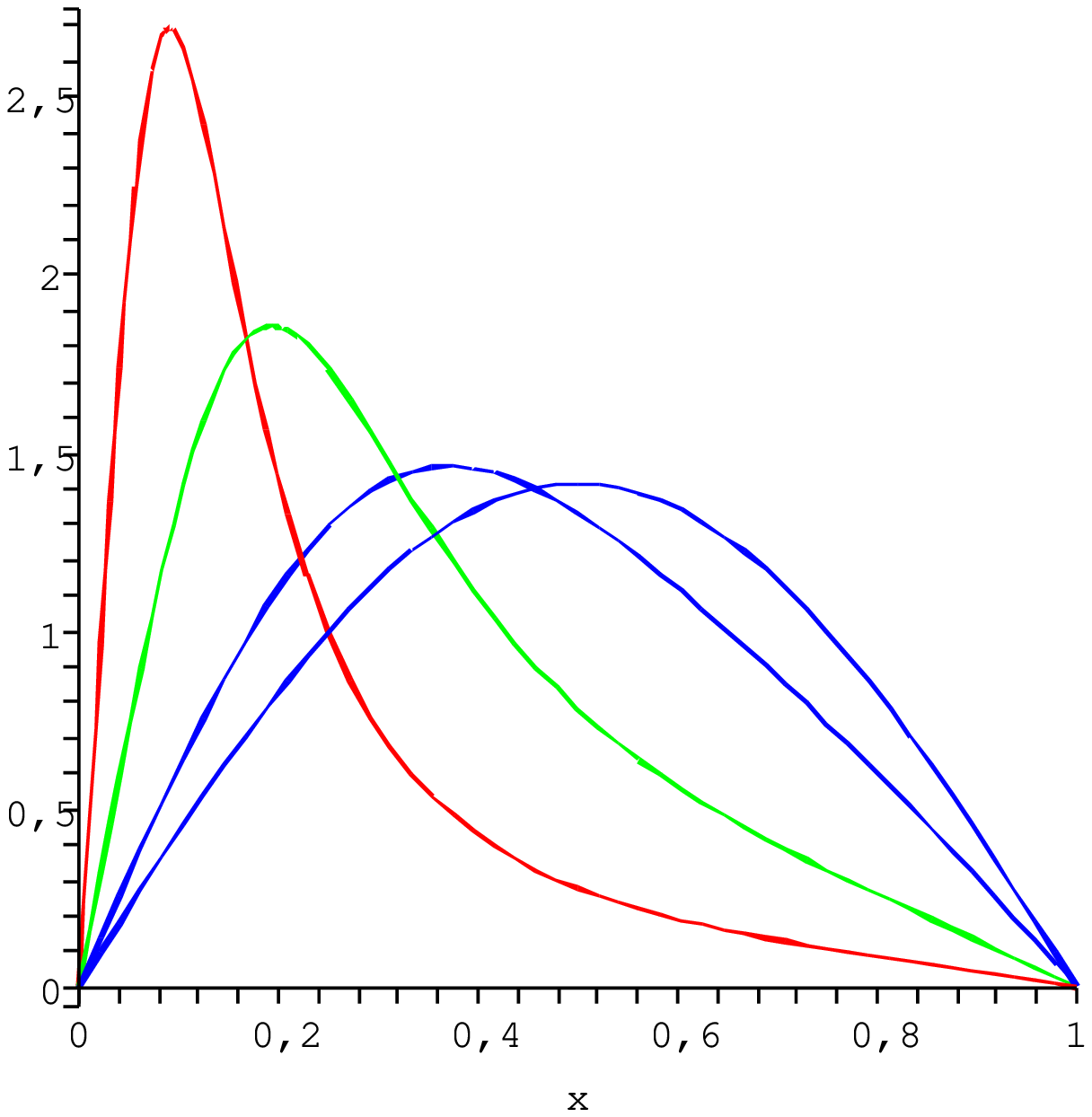}
\includegraphics[width=57mm]{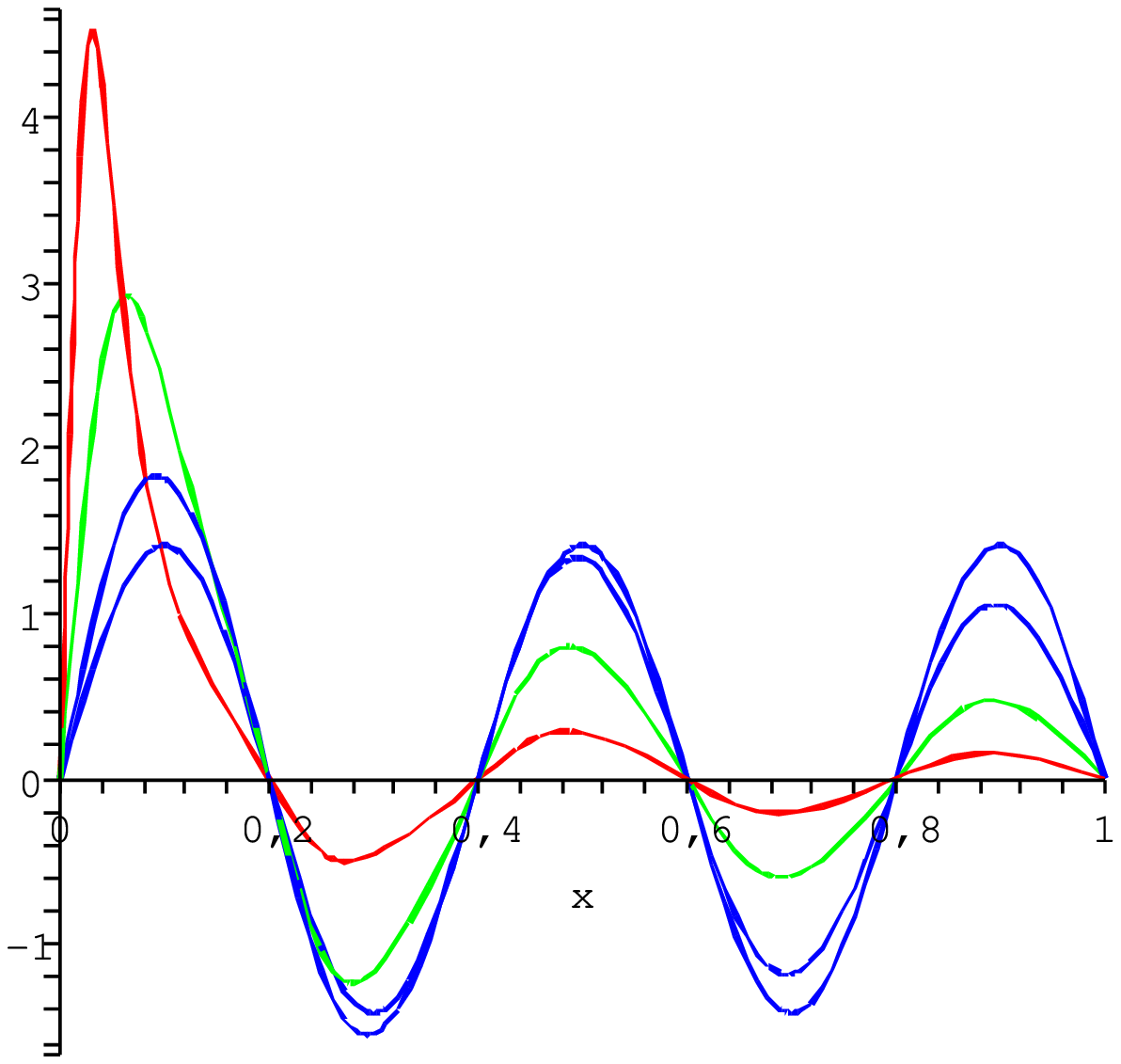}
\includegraphics[width=57mm]{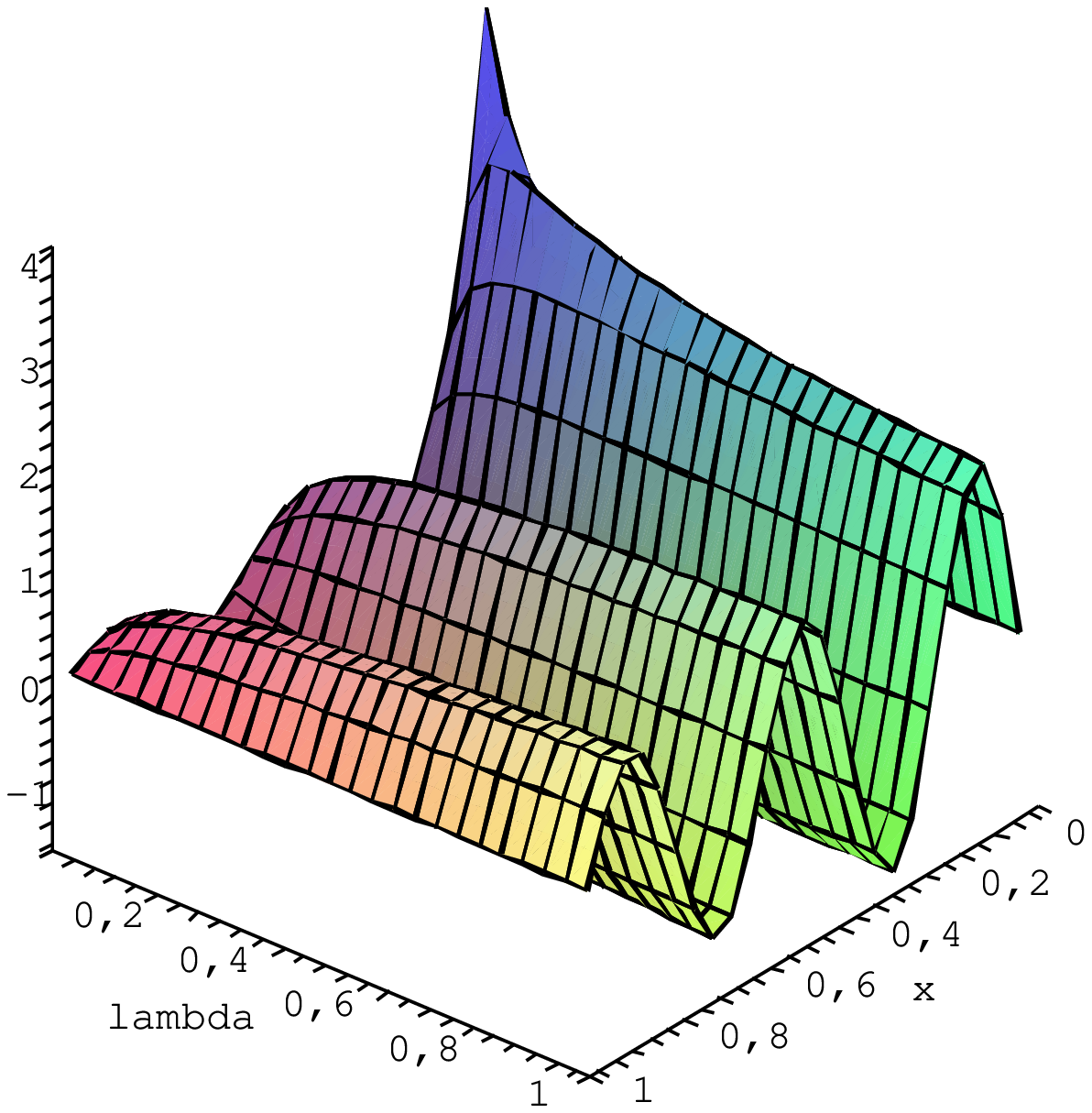}}
\caption{
\small
The deformation of the shape of WF of the bound state at the level
with the number $m$ at changing of the parameter $\lambda_{m}$
(at $\lambda_{m} = 1, 2, 3, 4$):
(a) --- for the ground state at $m=1$;
(b) --- for the state at $m=5$;
(c) --- the shape of WF is deformed continuously (without divergences)
at change of the parameter $\lambda_{m}$
\label{fig.723.4}}
\end{figure}

In Fig.~\ref{fig.723.4} a dependence of the shape of WF at the level
with the number $m$, coincident with the energy of factorization, on
the parameter $\lambda_{m}$ is shown.
In the first figure (a) the deformation of WF of the ground state at
$m=1$ is shown. We see, that this figure after mirror reflection
looks like Fig.~1 (2) in~\cite{Zakhariev.1990.PEPAN} (see~p.~917),
where the deformation of WF of the ground state, caused by the
methods of the inverse problem at change of the derivative
$\gamma_{1}$ of this WF in the right wall, is shown!
Here, we reproduce explicitly a demonstration of a deforming
property of this WF, which was described
in~\cite{Zakhariev.1990.PEPAN} (see~p.~917--918, and obtained by
the approach of the inverse problem):
\emph{``... acceleration of increase of $\Phi_{1}$ with moving off
from $x=a$ must be compensated by quicker passing to falling, because
the integral from the square of wave function of the ground state
remains normalizable to one. Just the hollow of the potential near
$x=a$ provides such a behavior $\Phi_{1}(x)$ ...
But, further, it needs to compel the function, with not changing a
sign, --- nodes for the ground state are unallowed, --- to tend to
zero at point $x=0$.
This is achieved by reflected part of addition to the initial potential
in the left part of the well: under the barrier the function is
smaller essentially then unperturbed one.''}
(in original text:
\emph{``... ускорение роста $\Phi_{1}$ при удалении от $x=a$ должно
быть скомпенсировано более быстрым переходом к спаданию, чтобы
интеграл от квадрата волновой функции основного состояния оставался
нормирован на единицу. Как раз углублуние потенциала вблизи $x=a$ и
обеспечивает такое поведение $\Phi_{1}(x)$ ...
Но дальше нужно заставить функцию, не меняя знака, --- узлы у основного
состояния запрещены, --- спуститься к нулю в точке $x=0$. Это
достигается отталкивающей частью добавки к исходному потенциалу в
левой части ямы: под барьером функция существенно меньше
невозмущенной.''}).
Note, that increase of the slope of WF in the left wall (and,
correspondingly, decrease in the right wall) at decreasing of the
parameter $\lambda_{1}$ in Fig.~\ref{fig.723.4} (a) coincides with
increase of the slope of WF in the right wall (and, correspondingly,
with decrease in the left wall) at changing of $\gamma_{1}$ in Fig.~1 (2)
in~\cite{Zakhariev.1990.PEPAN}
(a possibility to change the angle of the slope of WF at
$x$-coordinates of the walls is proved analytically in
sec.~\ref{sec.7.4}).
In the second figure (b) the deformation of the shape of WF at the
level with the number $m=5$ is shown. Here, we note the following:

\vspace{0mm}
{\small
\begin{itemize}
\item
We see, that the deformation of this WF looks qualitatively like the
deformation of WF of the first excited state in Fig .~1 (7)
from~\cite{Zakhariev.1990.PEPAN}, caused by change of the derivative
$\gamma_{2}$ in the right wall (with taking into account of decrease
of $m$).
From here one can see, that \emph{the parameters $\lambda_{m}$ in the
method from~\ref{sec.5.1} play the same role \underline{in the
deformation of WF} as coefficients $\gamma_{m}$ (defined at the same
levels) in the approach from~\cite{Zakhariev.1990.PEPAN}}.

\vspace{0mm}
\item
All deformed curves of WF with the selected number $m$ at the
different values of the parameter $\lambda_{m}$ intersect through the
same points --- the nodes of the undeformed WF of the form
(\ref{eq.7.1.8}). These nodes of WF are not displaced at change of
the parameter $\lambda_{m}$, two external nodes are located at
$x$-coordinates of the walls and total number of the nodes equals
to $m+1$.
This property is confirmed analytically by our formula
(\ref{eq.7.4.5}) and it is observed in Fig.~1 (2, 7)
in~\cite{Zakhariev.1990.PEPAN}.
\end{itemize}}

\begin{figure}[htbp]
\centerline{
\includegraphics[width=57mm]{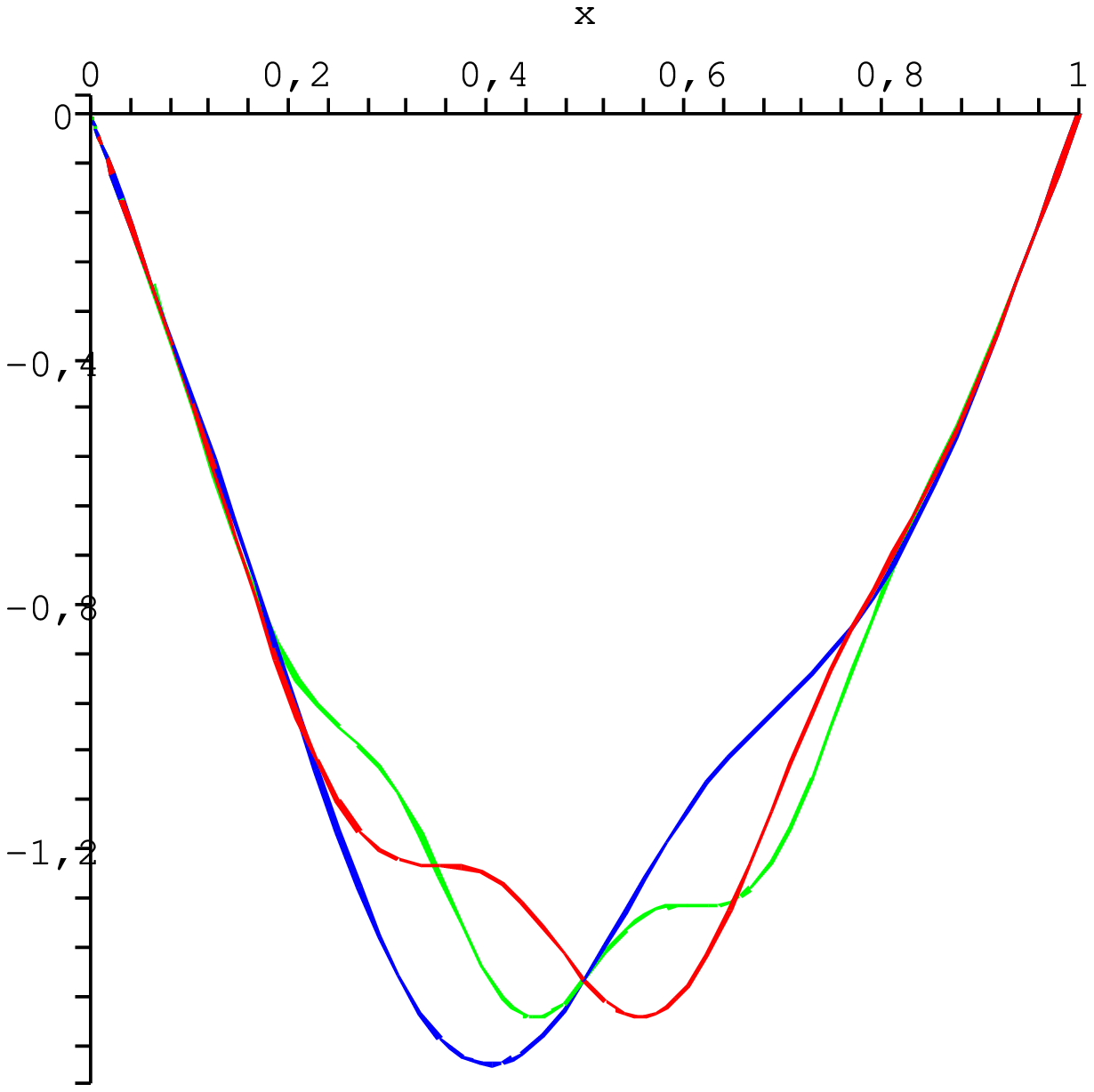}
\includegraphics[width=57mm]{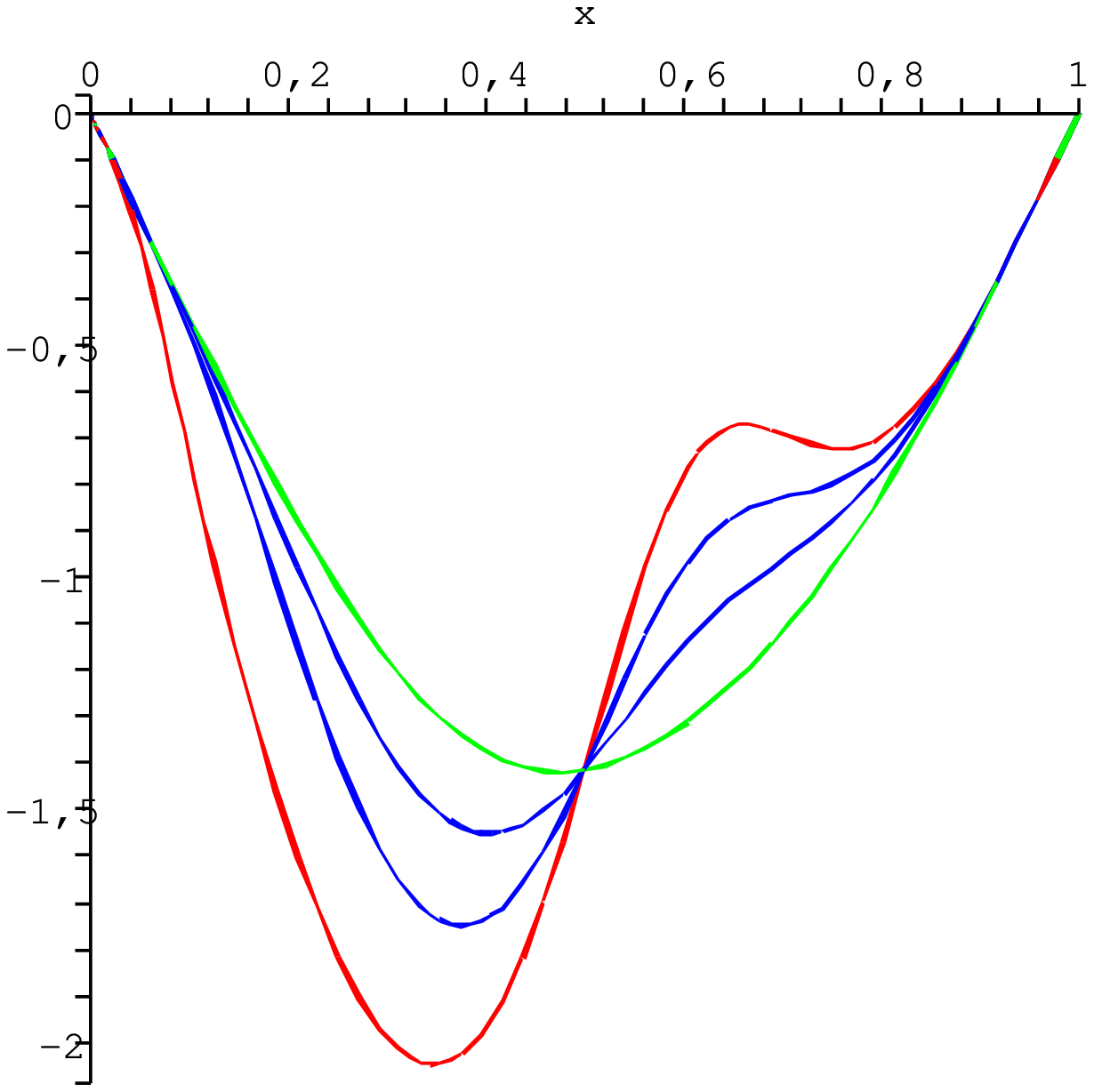}
\includegraphics[width=57mm]{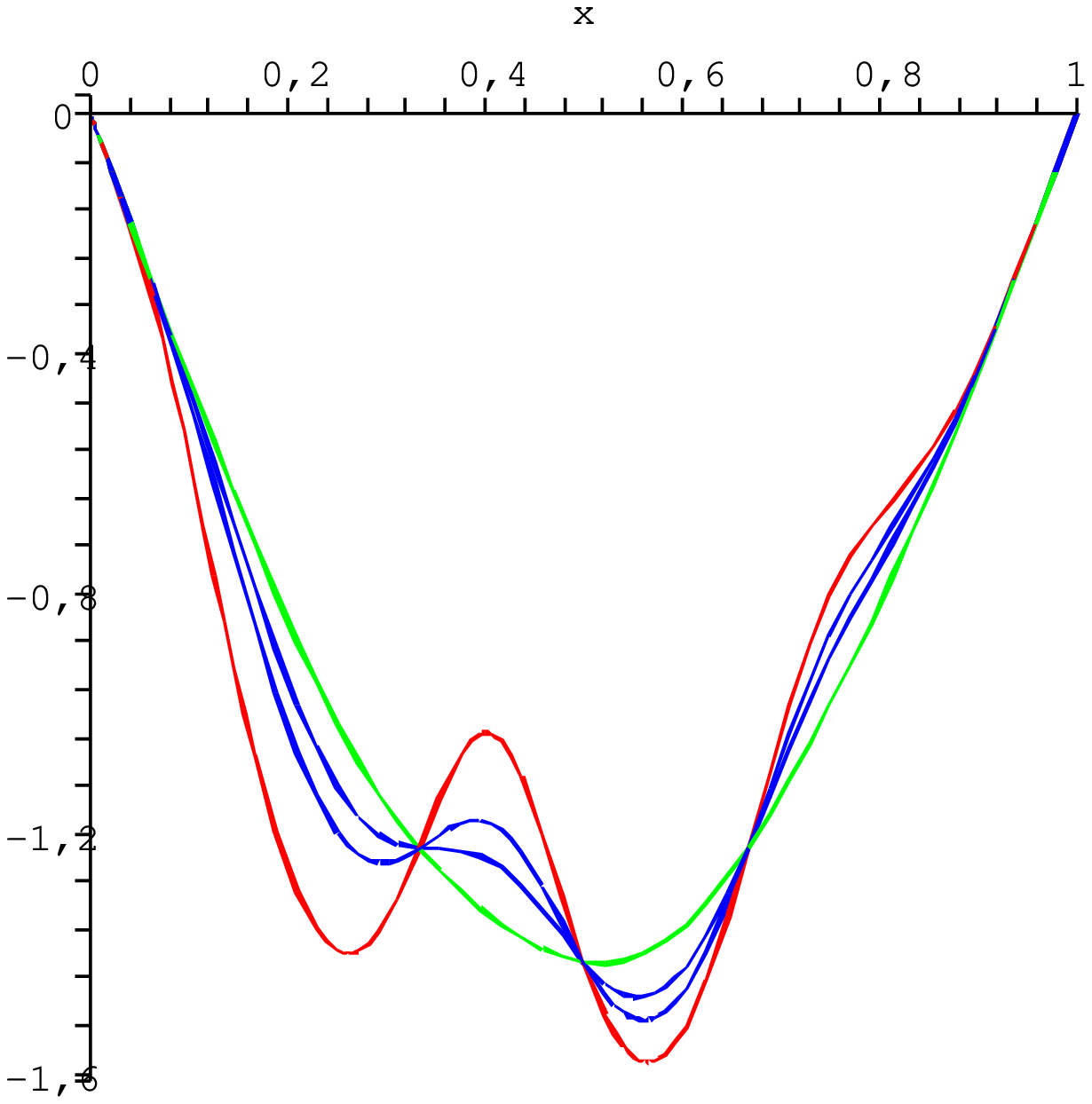}}
\caption{
\small
The deformation of the shape of WF of the ground bound state:
(a) --- at change of the number $m$ of the level, coincident with
the energy of factorization ($m = 2, 3, 4$ at $\lambda_{m}=1$);
(b) --- at change of the parameter $\lambda_{2}$
($\lambda_{2} = 0.01, 0.3, 1, 10$,
at definition of the energy of factorization the level with the
number $m=2$ is used);
(c) --- at change of the parameter $\lambda_{3}$
($\lambda_{3} = 0.1, 0.5, 1, 10$,
at definition of the energy of factorization the level with the number
$m=3$ is used)
\label{fig.723.5}}
\end{figure}

Now let's consider, how the shape of WF of the ground state can be
deformed, if in definition of the energy of factorization the level
of the excited state at $m \ne 1$ is used.
In Fig.~\ref{fig.723.5} the deformation of the shape of WF of the
ground state at change of the number $m$ of the level, coincident
with the energy of factorization (a), and at change of the parameter
$\lambda_{m}$ at $m=2$ (b) and at $m=3$ (c), is shown.
In the first figure (a) the deformation of WF of the ground state at
change of the number $m$ is shown (such a case of deformation is
absent in~\cite{Zakhariev.1990.PEPAN} see~p.~917).
Here, we note the following:

\vspace{0mm}
{\small
\begin{itemize}
\item
increase of the number $m$ by 1 introduces one new perturbation (or
\emph{``oscillation}) into the shape of WF;

\vspace{0mm}
\item
angles of leaving of WF of the ground state at points of the right
and left walls keep safe at change of the number $m$, that
demonstrates clearly properties 2 found in sec.~\ref{sec.7.5}.
\end{itemize}}

\vspace{0mm}
\noindent
In the second figure (b) the deformation of WF of the ground state
at change of the parameter $\lambda_{2}$ at $m=2$ is shown.
We see, that this figure after mirror reflection coincides with
Fig.~1 (6) from~\cite{Zakhariev.1990.PEPAN} (see~p.~917), where
the deformation of WF of the ground state, caused by the methods of
the inverse problem on the basis of change of the derivative
$\gamma_{2}$ of WF of the first excited state in the right wall, is
shown. Comparing these figures, we conclude:

\vspace{0mm}
{\small
\begin{itemize}
\item
angles of leaving of WF of the ground state at points of the right
and left walls at change of the parameter $\lambda_{2}$ in the
method from~\ref{sec.5.1} and at change of the coefficient
$\gamma_{2}$ in the method of the inverse problem
from~\cite{Zakhariev.1990.PEPAN} \underline{are not changed} ---
this demonstrates the properties 2 found in sec.~\ref{sec.7.5};

\vspace{0mm}
\item
\emph{there is only one point, in which all deformed curves of WF
of the ground state intersect between themselves and with the
undeformed curve of this WF} (without coordinates of the walls);
the coordinate of this point is located exactly in the center of the
well and coincides with the node of WF of the first excited state
with the number $m=2$, and also with zero-point of the first type of
the deformed potential, --- this visibly demonstrates properties 1
found in sec.~\ref{sec.7.5} (this is the point of the first type);

\vspace{0mm}
\item
there is the same increase of the angle between curves of the
deformed WF and undeformed WF at point of their intersection (in
the center of the well) at decreasing of the parameter $\lambda_{2}$
in the method from sec.~\ref{sec.5.1} and at change of the
coefficient $\gamma_{2}$ in the approach of the inverse problem
from~\cite{Zakhariev.1990.PEPAN}.
\end{itemize}}

\vspace{0mm}
\noindent
In the third figure (c) the deformation of WF of the ground state at
change of the parameter $\lambda_{3}$ at $m=3$ is shown, from here
one can see, that all considered above properties of the deformation
of WF at $m=2$ are carried out and in this case (however, here WF has
three points, in which it is not deformed: besides two points of the
first type, the coordinates of which coincide with the nodes of WF of
the excited state at the level with the number $m=3$, which coincides
with the energy of factorization, else one point of the second type
appears in the center of the well, that confirms the properties 1
from sec.~\ref{sec.7.5}).
From analysis of all figures one can conclude the following:

\vspace{0mm}
{\small
\begin{itemize}
\item
For arbitrary selected level with the number $m$, coincident with the
energy of factorization, there are different numbers of the points of
intersection between the deformed curves for the same WFs (using the
parameter $\lambda_{m}$) for other levels with the other numbers then
$m$ (it confirms the properties found in sec.~\ref{sec.7.5} and
formula (\ref{eq.7.5.3}) for coordinates of these points).

\vspace{0mm}
\item
If the numbers $m$ and $n$ determine the number of oscillations of
the curve of the deformed WF relatively its undeformed form, then
using $\lambda_{m}$ one can reinforce or enfeeble amplitude of these
oscillations.
\end{itemize}}

\begin{figure}[htbp]
\centerline{
\includegraphics[width=57mm]{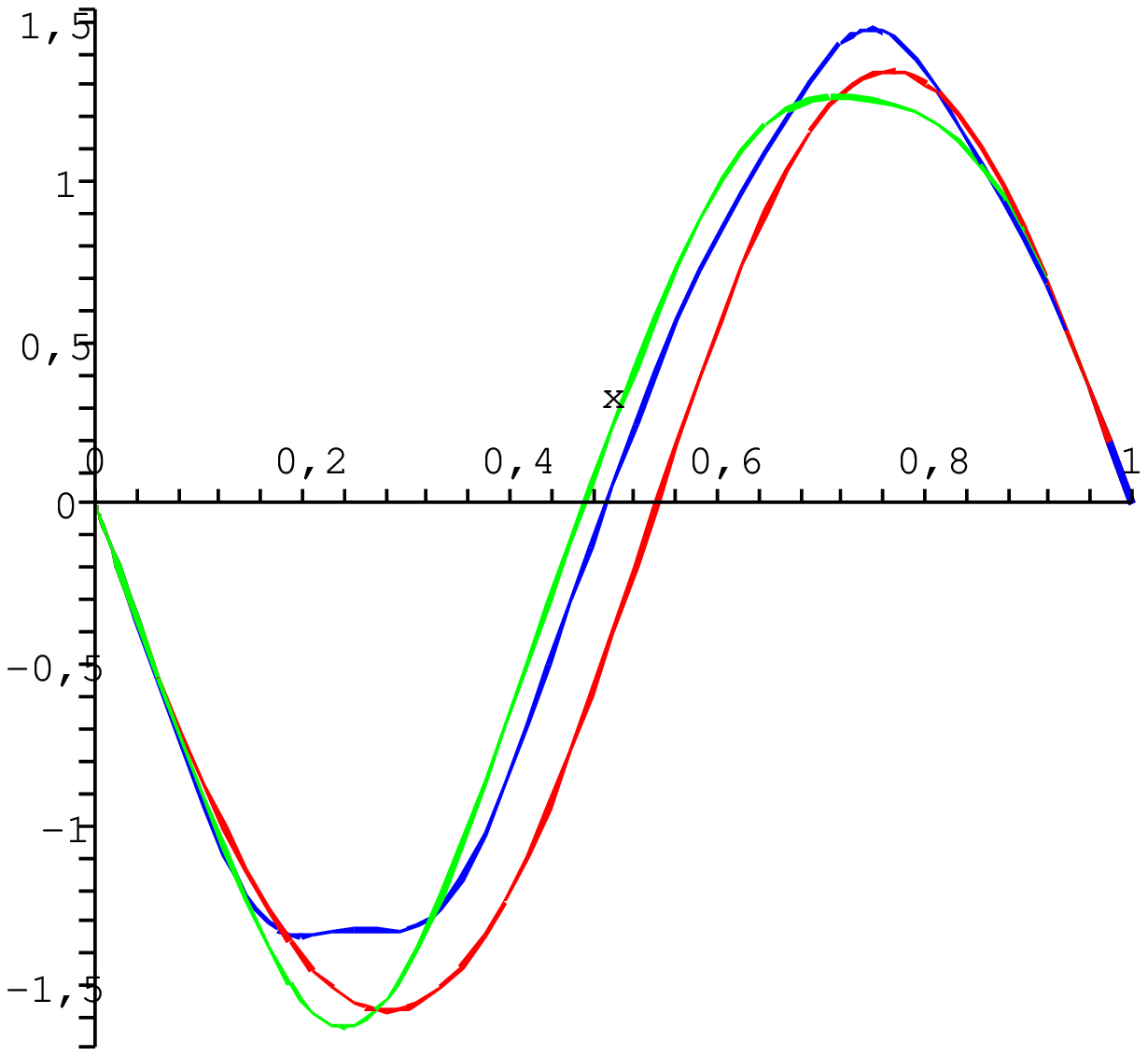}
\includegraphics[width=57mm]{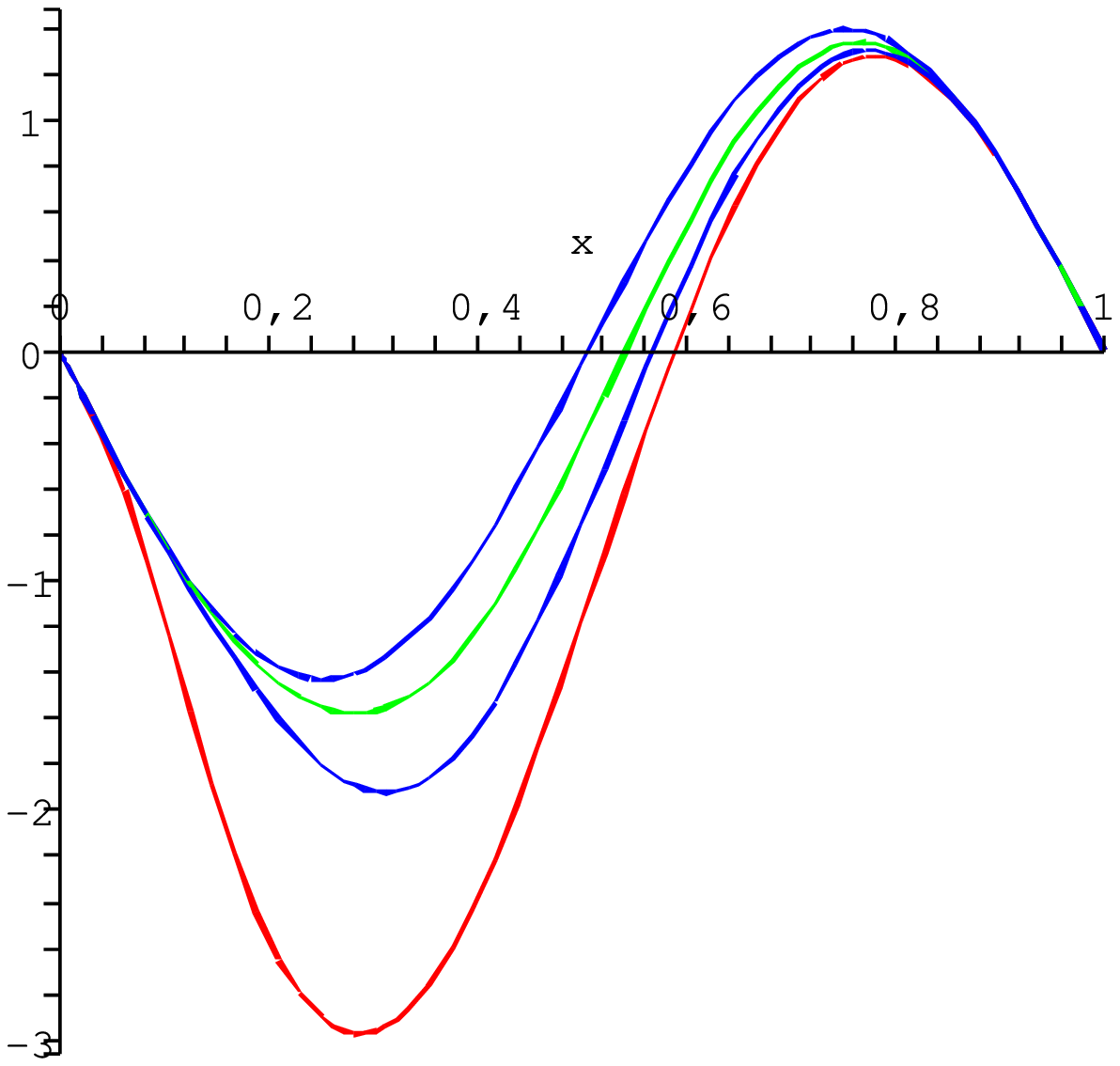}
\includegraphics[width=57mm]{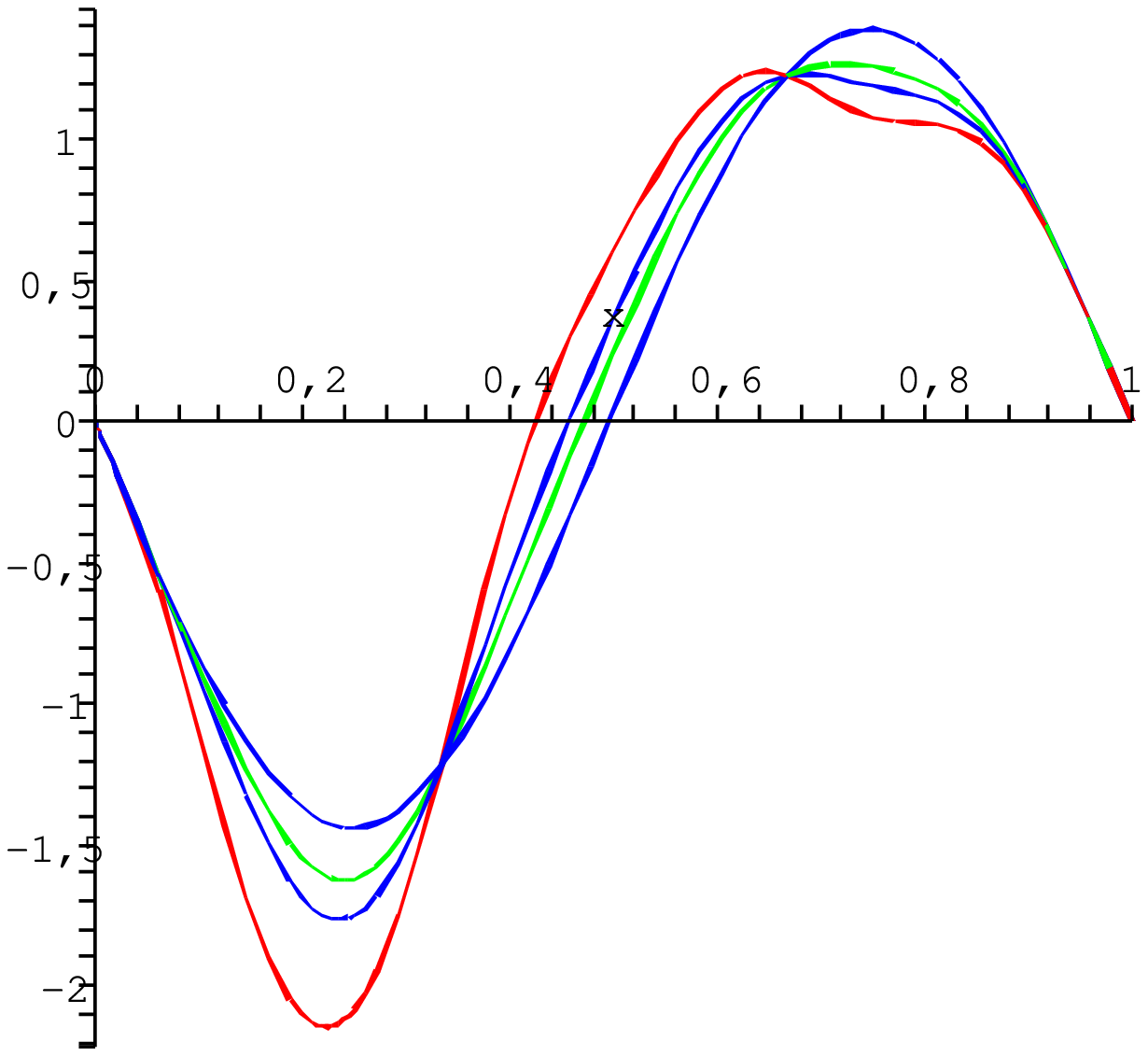}}
\caption{
\small
The deformation of the shape of WF of the first excited state:
(a) --- at change of the number $m$ of the level of the deformation
($m = 1, 3, 5$,
at $\lambda_{m}=1$);
(b) --- at change of the parameter $\lambda_{m}$
($\lambda_{m} = 0.01, 0.3, 1, 10$, in definition of the energy of 
factorization the level with the number $m=1$ is used);
(c) --- at change of the parameter $\lambda_{m}$
($\lambda_{m} = 0.1, 0.5, 1, 10$, in definition of the energy of
factorization the level with the number $m=3$ is used)
\label{fig.723.6}}
\end{figure}

The deformation of WF of the first excited state is shown in
Fig.~\ref{fig.723.6}.
In the first figure (a) the deformation of this WF at change of the
number $m$ of the level is shown (there is no such a case in Fig.~1
in~\cite{Zakhariev.1990.PEPAN}, see p.~917).
We see, that a behavior of this WF in such deformation looks like WF
of the ground state (see Fig.~\ref{fig.723.5} (a)):

\vspace{0mm}
{\small
\begin{itemize}
\item
increase of the number $m$ by 1 introduce one new perturbation (or
\emph{``oscillation}) into the shape of WF;

\vspace{0mm}
\item
angles of leaving of WF of the ground state at points of the right
and left walls are not changed under change of the number $m$.
\end{itemize}}

\vspace{0mm}
\noindent
In the second figure (b) the deformation of WF of the first excited 
state at change of the parameter $\lambda_{1}$ at $m=1$ is shown. We
see, that this figure at mirror reflection coincides with Fig.~1 (3)
in~\cite{Zakhariev.1990.PEPAN} (see p.~917), where the deformation of
WF of the first excited state, caused by change of the derivative
$\gamma_{1}$ of WF of the ground state in the right wall using the
methods of the inverse problem, is shown. Comparing these figures,
we conclude:

\vspace{0mm}
{\small
\begin{itemize}
\item
angles of leaving of WF of the first excited state at points of the
right and left walls at change of the parameter $\lambda_{1}$ in the
method from~\ref{sec.5.1} and at change of the coefficient
$\gamma_{1}$ in the method of the inverse problem
from~\cite{Zakhariev.1990.PEPAN} are not changed;

\vspace{0mm}
\item
curves of the deformed WFs at change of $\lambda_{1}$ in the method
from~\ref{sec.5.1} and at change of $\gamma_{1}$ in the method of the
inverse problem from~\cite{Zakhariev.1990.PEPAN} are not intersected
(it agrees with the formula (\ref{eq.7.5.3})).
\end{itemize}}

\vspace{0mm}
\noindent
From the third figure (c) one can see, that the considered above
properties of the deformation of WF of the ground state at change of
$\lambda_{m}$ at the selected number $m$, are fulfilled also for WF
of the excited state (with increasing of $m$, the number of points of
intersections of the deformed curves of WF between themselves and
with the undeformed shape of this WF for each selected excited state
is increased, that follows also from (\ref{eq.7.5.3})).

\begin{figure}[htbp]
\centerline{
\includegraphics[width=75mm]{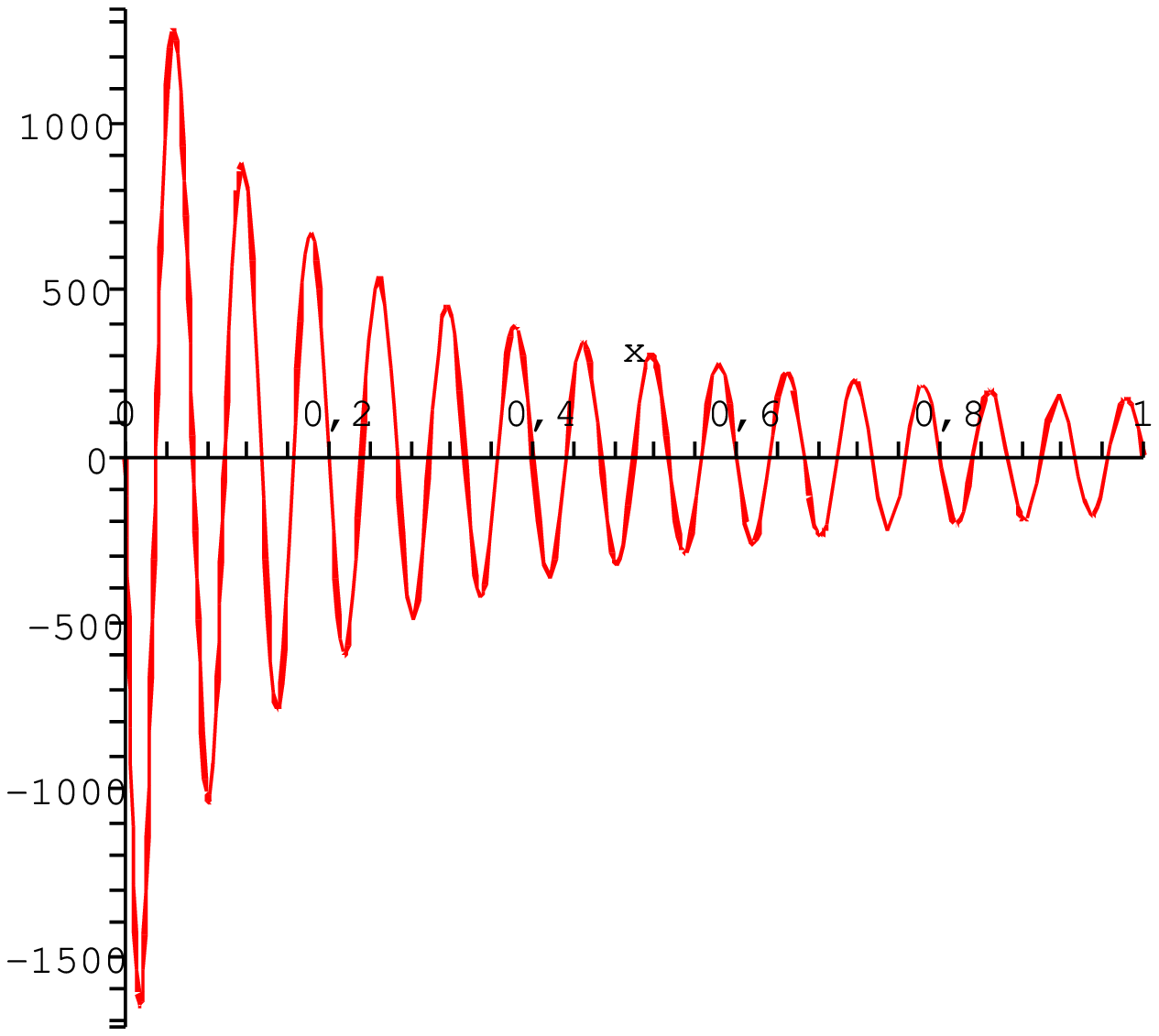}
\includegraphics[width=75mm]{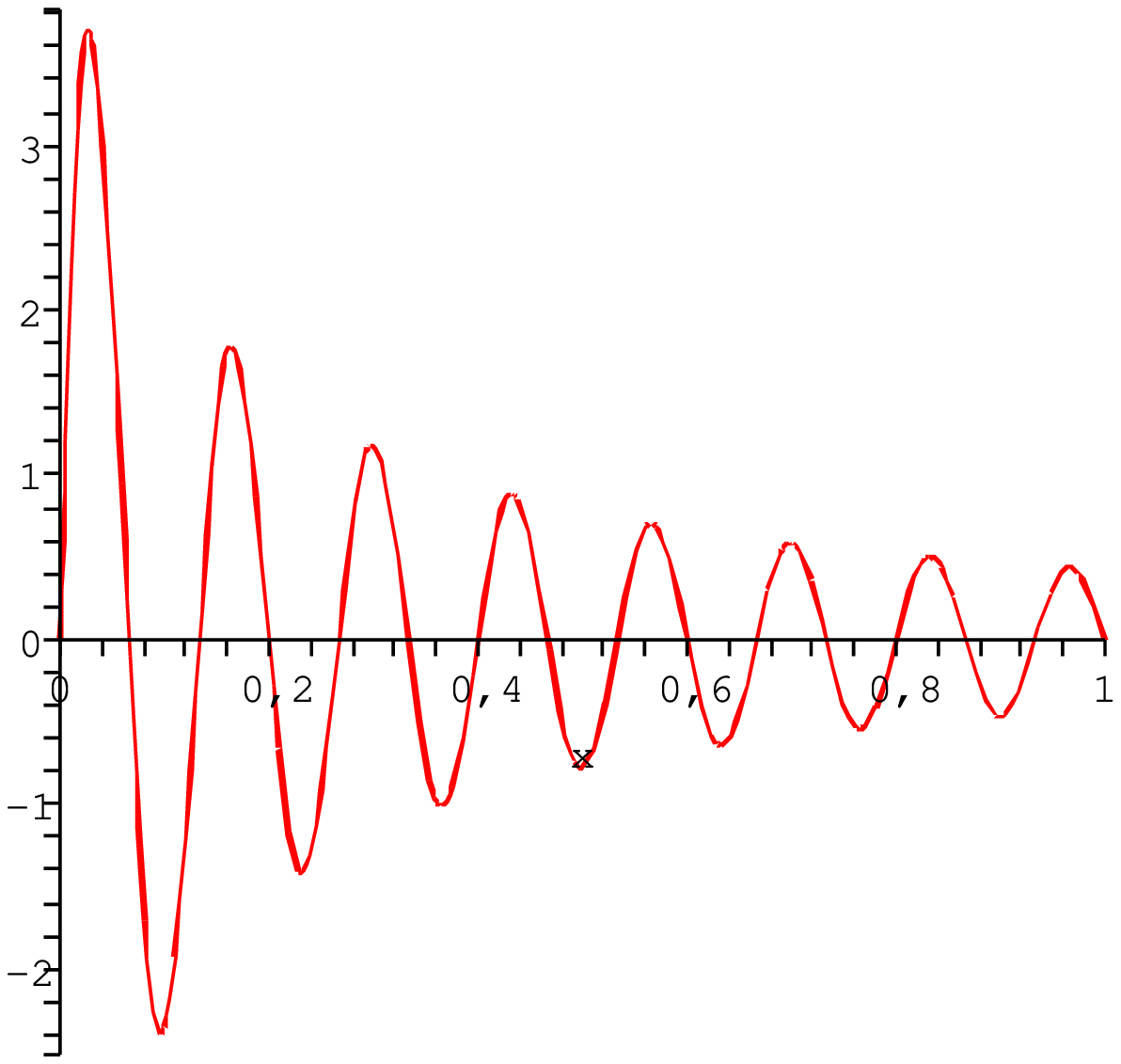}}
\caption{
\small
Shapes of the potential (a) and its WF of the bound state (b) at the
level, coincident with the energy of factorization, 
at $m=15$ and $\lambda_{m} = 0.1$
\label{fig.723.7}}
\end{figure}

In Fig.~\ref{fig.723.7} one can see, which forms the deformed
potential (a) and its WF of the bound state (b) at the level with a
larger number $m$, coincident with the energy of factorization, have
(the value $m=15$ is chosen).

\vspace{3mm}
\noindent
\underline{\bf Conclusions:}

Using the supersymmetric method from sec.~\ref{sec.5.1}, we have
obtained all pictures of the deformation (without displacement of
the levels in the spectrum) for the rectangular well with infinitely
high walls and its WFs of the lowest bound states, which were
obtained early in the review \cite{Zakhariev.1990.PEPAN}
(see p.~916--919) on the basis of the methods of the inverse problem.
Note the following:

\begin{itemize}
\item
Appropriateness of the method from sec.~\ref{sec.5.1} for construction
of new isospectral potentials on the basis of one given potential
$V_{1}$ in the form of the rectangular well with infinitely high
external walls has proved (one can consider such a method as the
method of deformation of the given $V_{1}$), in which for definition
of the function of factorization in the construction of the
superpotential the WF of arbitrary \underline{excited} bound state
(but not the ground state only) is used. 
All possible divergences in the found solutions can be excluded by
change of the parameter $\lambda_{m}$ inside one from regions in
(\ref{eq.7.6.2}).

\item
The parameter $\lambda_{m}$ with arbitrary number $m$ of the level
for the bound state in the supersymmetric method from~\ref{sec.5.1}
plays the same role in the deformation of the potential and its WFs,
as the derivative $\gamma_{m}$ of WF of the bound state at the same
level at point of one from the boundaries of the potential, which is
used for the deformation of the potential and its WFs by the methods
of inverse problem (see \cite{Zakhariev.1990.PEPAN} p.~917--927)
(it has obtained at the first time).

\item
Let's summarize, how WFs of the different bound states are changed
under the deformation by the parameter $\lambda_{m}$:

For WF at the arbitrary level with the number $m$, coincident with
the energy of factorization:

\vspace{-2mm}
{\small
\begin{itemize}
\item
WF has $m+1$ nodes exactly, which are not displaced under change of
$\lambda_{m}$ and coincide with the nodes of the undeformed WF
(\ref{eq.7.1.7});

\item
$x$-coordinates $x=0$ and $x=a$ of the walls are such nodes also;

\item
angles of leaving of this WF at points of the right and left walls
are changed at the deformation.

\end{itemize}}

For WF of the other bound state at the level with arbitrary number
$n$, not coincident with the energy of factorization ($n \ne m$):

\vspace{-2mm}
{\small
\begin{itemize}
\item
There are such points, in which \underline{all} deformed curves of
such WF intersect between themselves and with the undeformed WF
(i.~e. these points are not displaced at change of the parameter
$\lambda_{m}$).
\emph{This property (at the first time has been confirmed analytically
by formulas (\ref{eq.7.4.5}) and (\ref{eq.7.5.2})--(\ref{eq.7.5.3}),
with taking into account the next point in these Conclusions) gives
an answer on a question, stated in~\cite{Zakhariev.1990.PEPAN}
(see p.~918):
``it is not clear, why the functions $\Phi_{1}(x)$ and $\Phi_{2}(x)$
(i.~e. WFs of the ground and the first excited states) at all values
(i.~e. in any deformation using the coefficient $\gamma_{2}$ --- as
analogue of $\lambda_{2}$ in the methods of the inverse problem) pass
through only one point in the center of the well''}
(in the original text: ``не ясно также, почему функции $\Phi_{1}(x)$ и
$\Phi_{2}(x)$ при всех значениях проходят через одну и ту же точку в
центре ямы'').

\item
One can devide all these points into two types:

\begin{itemize}
\item
The points of the first type, $x$-coordinates of which equal to
(\ref{eq.7.5.4}) and coincide with the nodes (\ref{eq.7.1.8}) or
(\ref{eq.7.4.5}) of WF of the bound state at the level with the number
$m$, which coincides with the energy of factorization, and also with
zero-points of the first type of the form (\ref{eq.7.2.4}) of the
potential.

\item
The points of the second type, coordinates of which do not coincide
with the nodes of WF of the bound state at the level with the number
$m$, which coincides with the energy of factorization. The
coordinates of these points are determined from the second equation
of the system (\ref{eq.7.5.3}).
\end{itemize}

\item
The boundary points $x=0$ and $x=a$ are such points both the first
type, and the second one.

\item
WF of arbitrary bound state (at its deformation using the parameter
$\lambda_{m}$) has exactly $m+1$ such points of the first type, and
number of the points of the second type which can be determined from
the second equation of (\ref{eq.7.5.4}).

\item
Angles of leaving of such WF at points $x=0$ and $x=a$ of the walls
\underline{are not changed} under the deformation of this WF, caused
by change of the parameter $\lambda_{m}$ (or by change of the
coefficient $\gamma_{m}$ in the methods of the inverse problem) ---
in contrast to WF of the state with the number $m$!
\emph{This property (at the first time has been confirmed
analytically in sec.~\ref{sec.7.5}) gives an answer on a question,
stated in~\cite{Zakhariev.1990.PEPAN} (see p.~918):
how to explain an invariable slope of WF at $x$-coordinates of the
walls under its deformation by the coefficient $\gamma_{2}$ using the
method of the inverse problem.}
\end{itemize}}

\item
Now let's summarize, how the shape of the well is deformed under
change of the parameter $\lambda_{m}$ with the fixed number $m$.

According to (\ref{eq.7.2.3}), one can divide all zero-points of the
potential (i.~e. the points of the intersection of the potential with
axis $x$) into two types (for the deformed rectangular well this
property has been revealed and has been explained at the first time):

\vspace{-2mm}
{\small
\begin{itemize}
\item
Zero-points of the first type, which coincide with the nodes of WF of
the bound state at the level with the number $m$, which coincides
with the energy of factorization. These zero-points are not displaced
under the deformation of the potential by the parameter $\lambda_{m}$
and their coordinates are determined from (\ref{eq.7.2.4}).

\item
Zero-points of the second type, which do not coincide with the nodes
of WF of the bound state at the level with the number $m$, which
coincides with the energy of factorization. These points are
displaced under the deformation of the potential by the parameter
$\lambda_{m}$ and their coordinates are determined from (\ref{eq.7.2.5}).
\end{itemize}}

\end{itemize}

\section{Construction of new potentials, where unbound state is used
in definition of the function of factorization
\label{sec.8}}

On the example of the deformation of the rectangular well we have seen,
that the supersymmetric method (with taking into account not only
ground bound states, but \emph{excited} ones also in construction
of the superpotential) allows to obtain a whole variety of the
deformations (without displacement of levels in spectrum), which were
obtained early by the methods of the inverse problem.

\emph{It turns out, that if to do not restrict oneself by the bound
states only in the definition of the superpotential, but to construct
it using a function (we shall name it as WF of the unbound state),
which is a general solution (without implying of boundary conditions)
of the Schr\"{o}dinger equation with the starting potential $V_{1}$ at
arbitrary selected energy level (which can be not coincident with the
energy levels of the potential $V_{1}$), then one can obtain new
exactly solvable potentials (which have own energy spectra with the
\underline{bound states}).}
Such a way for the construction of the superpotential enlarges further
a class of exactly solvable potentials with only one given starting
$V_{1}$ (and gives more rich possibilities for the deformation of
$V_{1}$), which was obtained early by the method from
Sec.~\ref{sec.5.1}--\ref{sec.5.2}.
In particular, one can construct a new potential, energy spectrum of
which is shifted in accordence with a chosen rule relatively the
energy spectrum of the starting potential $V_{1}$ (that has not been
obtained else by the other known SUSY QM methods).
This points out to a possibility and advisability of development of
SUSY QM methods, where in the construction of the superpotential as
the energy of factorization the arbitrary energy (which can be not
coincident with the energy spectrum of $V_{1}$) is used, and the
function of factorization is defined concerning the unbound state at
this energy.
As a demonstration, further we shall present two new approaches of
construction of new potentials, when as the starting potential
$V_{1}$ the rectangular well with infinitely high external walls
in the form (\ref{eq.7.1.1}) is used.

Let's consider an \emph{arbitrary unbound state at an arbitrary
selected level} of the energy spectrum of the rectangular well $V_{1}$.
We denote these WF and the level by index $n$. One can write such
WF in a form (at $E_{n}^{(1)} \ge 0$):
\begin{equation}
  \varphi_{n}^{(1)}(x) = C_{n} \sin(k_{n} x + \delta_{n}),
\label{eq.8.1.1}
\end{equation}
where
$k_{n}$ is a wave vector, corresponding to the level $E_{n}$
($k_{n}^{2}=E_{n}$),
$C_{n}$ and $\delta_{n}$ are constants, which have complex values.

Choosing the arbitrary energy for the definition of the energy of
factorization, and WF of the unbound state at this energy for the
definition of the function of factorization, we construct the
superpotential (we add index $w$ to designations of such energy, its
wave vector and WF):
\[
  W(x) = 
  -\displaystyle\frac{d}{dx} \log {|\varphi_{w}^{(1)}|} =
  -\displaystyle\frac{d \sin{(k_{w}x+\delta_{w})}}{dx} \Big/
    \sin{(k_{w}x+\delta_{w})} =
  -k_{w} \cot{(k_{w}x+\delta_{w})}
\]
or
\begin{equation}
  W(x) = -k_{w} \cot{(k_{w}x+\delta_{w})}.
\label{eq.8.1.2}
\end{equation}

\subsection{A potential-partner to the rectangular well with a number
``up''
\label{sec.8.2}}

In the first case, we shall find a new potential $V_{2}$, which is
a partner to the rectangular well with the number ``up'', obtained
using direct SUSY-transformation with the superpotential in the
form (\ref{eq.8.1.2}).

Let's find a \underline{\emph{form of the new potential $V_{2}$}}:
\[
  V_{2} (x) =
    2 \displaystyle\frac{d W}{dx} =
    -2k_{w} \displaystyle\frac{d}{dx} \cot{(k_{w}x+\delta_{w})} =
    \displaystyle\frac{2k_{w}^{2}}{ \sin^{2}{(k_{w}x+\delta_{w})}}
\]
or
\begin{equation}
  V_{2} (x) =
    \displaystyle\frac{2k_{w}^{2}}{ \sin^{2}{(k_{w}x+\delta_{w})}}.
\label{eq.8.2.1.1}
\end{equation}
One can see, that:
\begin{itemize}
\item

\emph{at $E_{w}>E_{1}^{(1)}$ the new potential $V_{2}$ has infinitely
high barriers with maximums at points $x_{barrier}$:}
\begin{equation}
\begin{array}{cc}
  x_{barrier} = \displaystyle\frac{\pi l -\delta_{w}}{k_{w}}, &
  \mbox{at } l = 0, 1, 2 \ldots
\end{array}
\label{eq.8.2.1.2}
\end{equation}

\item
at $E_{w}<E_{1}^{(1)}$ (and wiht selected $\delta_{w}$) the potential
$V_{2}$ becomes continuous function and has no any divergence in the
region $x \in [0,a]$;

\item
the coordinates (\ref{eq.8.2.1.2}) of the divergences for $V_{2}$
coincide with the coordinates of the divergences of the superpotential
(\ref{eq.8.1.2}).

\end{itemize}
However, we can take the solution (\ref{eq.8.2.1.1}) as the solution
for the new exactly solvable potential $V_{2}$, only if it has the
bound states (without the divergences).
Therefore, it is important to analyze WFs of $V_{2}$.

\vspace{5mm}
At first, let's analyze, \emph{which states for the new potential
$V_{2}$ are possible at the level, coinciding with the energy of
factorization $E_{w}$}.
Direct action of operator $\hat{A}_{w} = \frac{d}{dx} + W$ on wave
function $\varphi_{w}^{(1)}$ at the energy $E_{w}$ gives zero
solution.
According to Sec.~\ref{sec.4.2}, there is the following
non-normalizable solution for $V_{2}$, concerned with the function
$\varphi_{w}^{(1)}$:
\begin{equation}
\begin{array}{cc}
  \varphi_{w}^{(2)} =
    \displaystyle\frac{I_{w} + \lambda_{w}} {\varphi_{w}^{(1)}}, &
  I_{w} (x) = \displaystyle\int\limits_{-\infty}^{x}
    (\varphi_{w}^{(1)}(x'))^{2} dx',
\end{array}
\label{eq.8.2.1.3}
\end{equation}
where $\lambda_{w}$ is constant. We find a function $I_{w}$:
\[
\begin{array}{c}
  I_{w} (x) =
  \displaystyle\int\limits_{-\infty}^{x}
    (\varphi_{w}^{(1)}(x'))^{2} dx' =
  C_{w}^{2} \displaystyle\int\limits_{0}^{x}
    \sin^{2}{(k_{w}x'+\delta_{w})} dx' =
  C_{w}^{2} \displaystyle\int\limits_{0}^{x}
    \displaystyle\frac{1}{2} \Bigl(1 - \cos(2k_{w}x'+2\delta_{w}) \Bigr) dx' = \\

  = \displaystyle\frac{C_{w}^{2}}{2}
    \biggl(x' -
           \displaystyle\frac{\sin{(2k_{w}x'+2\delta_{w})}}{2k_{w}} \biggr)
    \Biggr|_{x'=0}^{x'=x} =
    C_{w}^{2} \displaystyle\frac
      {2k_{w}x + \sin{2\delta_{w}} - \sin{(2k_{w}x+2\delta_{w})} }
      {4k_{w}}
\end{array}
\]
or
\begin{equation}
  I_{w} (x) =
    C_{w}^{2} \displaystyle\frac
      {2k_{w}x + \sin{2\delta_{w}} - \sin{(2k_{w}x+2\delta_{w})} }
      {4k_{w}}.
\label{eq.8.2.1.4}
\end{equation}

Taking into account this, we obtain:
\begin{equation}
  \varphi_{w}^{(2)} (x) =
    \displaystyle\frac{C_{w}}{4k_{w}}
    \displaystyle\frac
      {2k_{w}x + \sin{2\delta_{w}} - \sin{(2k_{w}x+2\delta_{w})}
        + \displaystyle\frac{4k_{w}\lambda_{w}} {C_{w}^{2}} }
      {\sin{(k_{w}x+\delta_{w})}}.
\label{eq.8.2.1.5}
\end{equation}
The solution (\ref{eq.8.2.1.5}) defines bound state, if the function
$\varphi_{w}^{(2)}$ is continuous in the region $x \in [0,a]$ and
equals to zero at $x=0$ and $x=a$.

Let's analyze a behavior of $\varphi_{w}^{(2)}$ at point $x=0$.
At $\delta_{w} \ne \pi l$ we obtain:
\[
  \varphi_{w}^{(2)} (0) =
  \displaystyle\frac{C_{w}}{4k_{w}}
    \displaystyle\frac{4k_{w}\lambda_{w}} {C_{w}^{2} \sin{\delta_{w}}} =
  \displaystyle\frac{\lambda_{w}}{C_{w} \sin{\delta_{w}}}.
\]
One can see, that at $\delta_{w} \ne \pi l$ the function
$\varphi_{w}^{(2)}$ equals to zero at $\lambda_{w}=0$ only.
At $\delta_{w} = \pi l$, the function $\varphi_{w}^{(0)}(0)$ equals
to zero also at $\lambda_{w}=0$. In deed, we find:
\[
\begin{array}{lcl}
  \varphi_{w}^{(2)} (0) & = &
  \displaystyle\frac{C_{w}}{4k_{w}}
    \displaystyle\frac
      {2k_{w} \Bigl(1 - \cos{(2k_{w}x+2\delta_{w})} \Bigr)}
      {k_{w} \cos{(k_{w}x+\delta_{w})}} \Biggr|_{x \to 0} =
  \displaystyle\frac{C_{w}}{2k_{w}}
    \displaystyle\frac
      {1 - \cos{2\delta_{w}}} {\cos{\delta_{w}}} =
  \displaystyle\frac{C_{w}}{2k_{w}}
    \displaystyle\frac{1 - 1} {(-1)^{l}} = 0.
\end{array}
\]
We write down values $\delta_{w}$ and $\lambda_{w}$, when the
function $\varphi_{w}^{(2)}$ equals to zero at point $x=0$:
\begin{equation}
\begin{array}{ccc}
  \lambda_{w} = 0, &
  \delta_{w} \mbox{ is arbitrary}, &
  k_{w} \mbox{ is arbitrary } (k_{w} \ne 0).
\end{array}
\label{eq.8.2.1.6}
\end{equation}

Now let's analyze a behavior of the function $\varphi_{w}^{(2)}$ at
point $x=a$. If $k_{w}a+\delta_{w} \ne \pi l$, then from
(\ref{eq.8.2.1.5}) at $\lambda_{w}=0$ we obtain a condition, when
the function $\varphi_{w}^{(2)}$ equals to zero:
\begin{equation}
  2k_{w}a + \sin{2\delta_{w}} - \sin{(2k_{w}a+2\delta_{w})} = 0.
\label{eq.8.2.1.7}
\end{equation}
Such dependence defines a needed value of the phase $\delta_{w}$ at
selected $k_{w}$, which can be arbitrary.
At $k_{w}a+\delta_{w} = \pi l$ the function $\varphi_{w}^{(2)}(a)$
equals to zero only at 
\begin{equation}
  \lambda_{w} =
    -\displaystyle\frac{C_{w}}{4k_{w}}
    \Bigl(2k_{w}a + \sin{2\delta_{w}}\Bigr) = 0.
\label{eq.8.2.1.8}
\end{equation}
In deed, we find:
\[
\begin{array}{lcl}
  \varphi_{w}^{(2)} (a) & = &
  \displaystyle\frac{C_{w}}{4k_{w}}
    \displaystyle\frac
      {2k_{w} \Bigl(1 - \cos{(2k_{w}x+2\delta_{w})} \Bigr)}
      {k_{w} \cos{(k_{w}x+\delta_{w})}} \Biggr|_{x \to a} =
  \displaystyle\frac{C_{w}}{2 k_{w}}
    \displaystyle\frac{1 - 1} {(-1)^{l}} = 0.
\end{array}
\]
From (\ref{eq.8.2.1.8}) we obtain the necessary condition for $k_{w}$:
\begin{equation}
  \sin{2\delta_{w}} = -2k_{w}a.
\label{eq.8.2.1.9}
\end{equation}
In this case the function $\varphi_{w}^{(2)}$ equals to zero at
point $x=a$ at some values of $k_{w}$ and $\delta_{w}$.

We write down such values of $\delta_{w}$ and $\lambda_{w}$,
when the function $\varphi_{w}^{(2)}$ equals to zero at points $x=0$
and $x=a$ with selected $k_{w}$, which can be arbitrary:
\begin{equation}
\begin{array}{lllll}
  1) &
    k_{w}a+\delta_{w} \ne \pi l, &
    2k_{w}a + \sin{2\delta_{w}} - \sin{(2k_{w}a+2\delta_{w})} = 0, &
    \lambda_{w} = 0; \\

  2) &
    k_{w}a+\delta_{w} = \pi l, &
    \sin{2\delta_{w}} = -2k_{w}a, &
    \lambda_{w} = 0,
\end{array}
\label{eq.8.2.1.10}
\end{equation}
where $l=0,1,2\ldots$.

\vspace{5mm}
At second, let's find \emph{WF of the arbitrary (unbound) state for
$V_{2}$ at the level $E_{n}$ with the number $n$, not coinciding
with the energy of factorization}:
\[
\begin{array}{c}
  \varphi_{n}^{(2)} (x) =
    \bar{C}_{n} \hat{A}_{w} \varphi_{n}^{(1)} =
    \bar{C}_{n} \biggl(\displaystyle\frac{d}{dx} + W \biggr)
      \varphi_{n}^{(1)} = \\
   = \bar{C}_{n} C_{n} \Bigl(k_{n} \cos{(k_{n}x+\delta_{n})} -
      k_{w} \cot{(k_{w}x+\delta_{w})} \sin{(k_{n}x+\delta_{w})} \Bigr)
\end{array}
\]
or
\begin{equation}
  \varphi_{n}^{(2)} (x) =
  \bar{C}_{n} C_{n} \Bigl(k_{n} \cos{(k_{n}x+\delta_{n})} -
      k_{w} \cot{(k_{w}x+\delta_{w})} \sin{(k_{n}x+\delta_{n})} \Bigr),
\label{eq.8.2.1.11}
\end{equation}
where
$\hat{A}_{w} = \frac{d}{dx} + W$,
$\bar{C}_{n}$ is arbitrary constant, which we shall use instead of
the factor $|E_{n}^{(1)} - E_{w}^{(1)}|^{-1}$, because we consider
the non-normalizable state.

In order to the function (\ref{eq.8.2.1.11}) describes the bound state,
it needs, that it must be continuous in the region $x \in [0,a]$ and
equals to zero at points $x=0$ and $x=a$.
Let's analyze, when this function can be continuous.
According to (\ref{eq.8.2.1.11}), the divergences appear at point 
$x_{break}$ at
\begin{equation}
  \sin{(k_{w}x_{break} + \delta_{w})} = 0.
\label{eq.8.2.1.12}
\end{equation}
However, if at all points $x_{break}$ the following condition is
fulfilled:
\begin{equation}
  \sin{(k_{n}x_{break} + \delta_{n})} = 0,
\label{eq.8.2.1.13}
\end{equation}
then the function (\ref{eq.8.2.1.11}) is continuous and has no
divergences.
In deed, assuming, that the conditions (\ref{eq.8.2.1.12}) and
(\ref{eq.8.2.1.13}) are fulfilled at point $x=x_{break}$, we analyze
the second item of the function (\ref{eq.8.2.1.11}) at this point.
We obtain:
\[
\begin{array}{c}
  -k_{w} \cot{(k_{w}x_{break}+\delta_{w})}
         \sin{(k_{n}x_{break}+\delta_{n})} =
  -k_{w} \displaystyle\frac
     {\cos{(k_{w}x+\delta_{w})} \sin{(k_{n}x+\delta_{n})}}
     {\sin{(k_{w}x+\delta_{w})}} \Biggr|_{x \to x_{break}} = \\
  = k_{w} \displaystyle\frac
     {k_{w}\sin{(k_{w}x+\delta_{w})} \sin{(k_{n}x+\delta_{n})} -
      k_{n}\cos{(k_{w}x+\delta_{w})} \cos{(k_{n}x+\delta_{n})}}
     {k_{w}\cos{(k_{w}x+\delta_{w})}} \Biggr|_{x \to x_{break}} = \\
  = -k_{w} \displaystyle\frac
     {k_{n}\cos{(k_{w}x_{break}+\delta_{w})}
           \cos{(k_{n}x_{break}+\delta_{n})}}
     {k_{w}\cos{(k_{w}x_{break}+\delta_{w})}} =
  - k_{n}\cos{(k_{n}x_{break}+\delta_{n})} = (-1)^{l+1} k_{n}.
\end{array}
\]

\vspace{0mm}
\noindent
\underline{\bf Continuity condition of WF:}
\emph{the function $\varphi_{n}^{(2)} (x)$, describing a state with 
number $n$ and having the form (\ref{eq.8.2.1.11}), is continuous in
the region $x \in [0,a]$, if each point $x_{break}$, defined by
(\ref{eq.8.2.1.12}) at selected $k_{w}$ and $\delta_{w}$, satisfies
to the condition (\ref{eq.8.2.1.13}) for the corresponding $k_{n}$
and $\delta_{n}$}.
According to (\ref{eq.8.2.1.11}), such points $x_{break}$ are nodes 
of this WF (at even $l$).

Now let's find out, when the function (\ref{eq.8.2.1.11}) equals to
zero at points $x=0$ and $x=a$.
From a condition of equality to zero we obtain:
\begin{equation}
\begin{array}{lcl}
 (x=0) & \to &
   k_{n} \cos{\delta_{n}} - k_{w} \cot{\delta_{w}} \sin{\delta_{n}} = 0; \\

 (x=a) & \to &
   k_{n} \cos{(k_{n}a + \delta_{n})} -
     k_{w} \cot{(k_{w}a + \delta_{w})} \sin{(k_{n}a + \delta_{n})} = 0.
\end{array}
\label{eq.8.2.1.14}
\end{equation}
Let's $\cos{\delta_{w}} \ne 0$. From the first equation we obtain:
\begin{equation}
\begin{array}{ccc}
  \tan{\delta_{n}} = \displaystyle\frac{k_{n}}{k_{w}} \tan{\delta_{w}}, &
  \cos{\delta_{n}} \ne 0, &
  \cos{\delta_{w}} \ne 0.
\end{array}
\label{eq.8.2.1.15}
\end{equation}
Rewrite the second equation of the system (\ref{eq.8.2.1.14}) by such
a way:
\begin{equation}
\begin{array}{lcl}
  k_{n} \cos{\delta_{n}} \Bigl(\cos{k_{n}a} -
        \displaystyle\frac{k_{n}}{k_{w}}
        \sin{k_{n}a} \tan{\delta_{w}} \Bigr) =

  k_{w} \cot{(k_{w}a + \delta_{w})} \cos{\delta_{n}}
        \Bigl(\sin{k_{n}a} +
          \displaystyle\frac{k_{n}}{k_{w}}
          \cos{k_{n}a} \tan{\delta_{w}} \Bigr).
\end{array}
\label{eq.8.2.1.16}
\end{equation}
From here we obtain:
\begin{equation}
  \tan{k_{n}a} =
    \displaystyle\frac{k_{n}}{k_{w}}
    \displaystyle\frac{1 - \cot{(k_{w}a+\delta_{w})} \tan{\delta_{w}}}
      {\cot{(k_{w}a+\delta_{w})} +
       \displaystyle\frac{k_{n}^{2}}{k_{w}^{2}} \tan{\delta_{w}}}.
\label{eq.8.2.1.17}
\end{equation}

Transform this equation so:
\begin{equation}
  \tan{k_{n}a} =
    \displaystyle\frac{k_{n}}{k_{w}}
    \displaystyle\frac{\sin{k_{w}a}}
      {\cos{(k_{w}a+\delta_{w})}\cos{\delta_{w}} +
       \displaystyle\frac{k_{n}^{2}}{k_{w}^{2}}
       \sin{(k_{w}a+\delta_{w})}\sin{\delta_{w}}}.
\label{eq.8.2.1.18}
\end{equation}
In contrast to (\ref{eq.8.2.1.17}), numerator and denominator of the
right part of the equation (\ref{eq.8.2.1.18}) have no any infinitely
large value at arbitrary $k_{n}$, $k_{w}$ ($k_{w} \ne 0$) and
$\delta_{w}$. Such expression is more convenient for analysis of
zeros.

Now let's consider a case $\cos{\delta_{w}} = 0$. From
(\ref{eq.8.2.1.14}) we obtain instead of the equations
(\ref{eq.8.2.1.15}) an (\ref{eq.8.2.1.18}) the following:
\begin{equation}
\begin{array}{ccc}
  \tan{k_{n}a} = \displaystyle\frac{k_{w}}{k_{n}} \tan{k_{w}a}, &
  \cos{\delta_{n}} = 0, &
  \cos{\delta_{w}} = 0.
\end{array}
\label{eq.8.2.1.19}
\end{equation}

We see, that the equations (\ref{eq.8.2.1.18}) are (\ref{eq.8.2.1.19})
transcendental and in a general case it needs to resolve each from
them numerically, that makes a problem of obtaining of new types of
exactly solvable potentials as very difficult.
In result, we shall obtain a spectrum of all possible values of
$k_{n}$ (energy spectrum $E_{n}^{(2)}$) in dependence on selected 
values of $k_{w}$ and $\delta_{w}$.
However, some partial solutions of the new potentials (with WFs and
energy spectra, in the explicit form) can be found more simply. Let's
consider two cases.

\subsubsection{New isospectral potentials with shift of their shape
along axis $x$
\label{sec.8.2.2}}

In the first case, let's assume, that all energy levels of the new
potential $V_{2}$ coincide with the levels of the spectrum of the
rectangular well $V_{1}$. Then we have:
\begin{equation}
  \sin{k_{n}a} = 0.
\label{eq.8.2.2.1}
\end{equation}
From (\ref{eq.8.2.1.18}) at $\cos{\delta_{w}} \ne 0$ or from
(\ref{eq.8.2.1.19}) at $\cos{\delta_{w}}=0$ we obtain:
\begin{equation}
  \sin{k_{w}a} = 0.
\label{eq.8.2.2.2} 
\end{equation}
We see, that the energy of factorization $E_{w}$ coincides with one
level from the spectrum $E_{n}^{(1)}$ for the rectangular well (we
have an arbitrariness in a choice of the parameter $\delta_{w}$).

Now we assume, that the new potential $V_{2}$, in addition to the
found energy spectrum, has some other levels.
From (\ref{eq.8.2.1.19}) we see, that at
$\delta_{w}=\frac{\pi}{2} + \pi l$ it is not possible ($l = 0,1,2\ldots$).
At $\delta_{w} \ne \frac{\pi}{2} + \pi l$ equation (\ref{eq.8.2.1.18})
can give such solution. As we found early, the level $E_{w}$ belongs
to the spectrum $E_{n}^{(1)}$, therefore: $\sin{k_{w}a}=0$.
From (\ref{eq.8.2.1.18}) we obtain:
\begin{equation}
  k_{n}^{2} = -k_{w}^{2} \cot^{2}{\delta_{w}}.
\label{eq.8.2.2.3}
\end{equation}
As for WF of the starting potential we have chosen the solutions
(\ref{eq.8.1.1}) at the levels $E_{n}^{(1)}>0$, then $k_{n}^{2}>0$
and $k_{w}^{2}>0$.
Therefore, the solution (\ref{eq.8.2.2.3}) is possible only at
complex values of $\delta_{w}$ (at $Im \delta_{w} \ne 0$).
\emph{The complex phase $\delta_{w}$ gives complex values of the
superpotential $W$ and can lead to new types of SUSY-transformations
(which one can find in literature)}.

Taking into account the found energy spectrum for the potential
$V_{2}$, the condition of continuity of WF of the bound state with
the number $n$ leads to the following additional interdependence
between the phases $\delta_{w}$ and $\delta_{n}$ ($w$ is the number
of the level $E_{w}$ in the spectrum $E_{n}^{(2)}$):
\begin{equation}
\begin{array}{cc}
  \delta_{n} = \displaystyle\frac{n}{w}\delta_{w} + 2\pi l, &
  l=0,1,2\ldots
\end{array}
\label{eq.8.2.2.4}
\end{equation}

\vspace{5mm}
\noindent
\underline{\bf Conclusion:}
At the real phase $\delta_{w}$, the new potential $V_{2}$ has the
energy spectrum $E_{n}^{(2)}$, coinciding completely with the energy
spectrum $E_{n}^{(1)}$ of the rectangular well (with possible
exclusion of one level $E_{w}$).
At $E_{w}=E_{1}^{(1)}$ (at $k_{n} = n k_{w}$) we have the following:

{\small
\begin{itemize}
\item
At $\delta_{w} = \pi l$ ($l=0,1,2\ldots$) the potential $V_{2}$
coincides with the potential (24) from \cite{Cooper.1995.PRPLC}
without shift down by value
$\Delta E = E_{1} = \frac{\hbar^{2}\pi^{2}}{2mL^{2}}$
(see p.~278--279, Fig.~2.2;
for exact comparison it needs to use $\hbar=1$, $2m=1$, $L=a$,
$n_{\mbox{\scriptsize my}} =
  n_{\mbox{\tiny Cooper, Khare, Sukhatme}} + 1$,
and we obtain the potential and spectrum without displacement:
$V_{2} = \frac{\hbar^{2}\pi^{2}}{mL^{2}} \mbox{cosec}^{2} {(\pi x/L)}$,
$E_{n} = \frac{\hbar^{2}\pi^{2}}{2mL^{2}} n^{2}$).
\begin{itemize}
\item
In contrast to \cite{Cooper.1995.PRPLC} (see p.~278--279), we connect
the potentials $V_{1}$ and $V_{2}$ by SUSY-transformation with
non-zero lowest level $E_{1} \ne 0$
($E_{1} \ne 0$ в ~\cite{Cooper.1995.PRPLC}).

\item
At $\delta_{n} = \pi l$ the condition (\ref{eq.8.2.2.4}) is
fulfilled jointly with (\ref{eq.8.2.1.15}). Therefore, WF of the
form (\ref{eq.8.2.1.11}) is continuous in the region $x \in [0,a]$
and equals to zero at points $x=0$ and $x=a$. Therefore, a state
described by this WF at arbitrary level with the number $n$ (not
coincident with $E_{w}$) of the energy spectrum $E_{n}^{(2)}$ for
the new potential $V_{2}$ must be bound.

\item
At $\delta_{n} = \pi l$ the conditions (1) and (2) from
(\ref{eq.8.2.1.10}) are not fulfilled (besides $k_{w}=0$).
Therefore, the state for $V_{2}$ at the lowest level $E_{1}^{(1)}$
cannot be bound and we must to exclude this level from the energy
spectrum of the new potential $V_{2}$ (this agrees with
\cite{Cooper.1995.PRPLC}).
\end{itemize}

\item
At $\delta_{w} \ne \pi l$ the potential $V_{2}$ displaces along axis
$x$ (using $\delta_{w}$) with appearance of one infinitely high
barrier, which has a maximum with coordinate determined from
(\ref{eq.8.2.1.2}). Here, we have two cases:

\begin{itemize}
\item
At $\cos{\delta_{w}}=0$ the equation (\ref{eq.8.2.2.4}) is fulfilled
jointly with the condition $\cos{\delta_{n}}=0$ from
(\ref{eq.8.2.1.19}) at $\delta_{n}=\frac{\pi}{2} + \pi l$ and at
arbitrary odd number $n$. This defines a bound excited state at such
$n$ (continuity of WF of this state in the region $x \in [0,a]$, its
zero values at points $x=0$ and $x=a$).
Therefore, \emph{we obtain a new potential $V_{2}$ with infinitely 
high barrier, maximum of which is located exactly in the center of
the well at $x_{barrier} = a/2$ and all levels of the spectrum (with
the bound states) coincide with the levels with odd numbers of the
spectrum $E_{n}^{(1)}$ of the starting potential $V_{1}$}.

\item
At $\cos{\delta_{w}} \ne 0$, for analysis of existence of excited
bound states we must to analyze a compatibility of the conditions
(\ref{eq.8.2.2.4}) and (\ref{eq.8.2.1.15}), that leads to the
transcendental equation (\ref{eq.8.2.1.15}).
A barrier of $V_{2}$ can be displaced along axis $x$, its maximum has
a coordinate:
\begin{equation}
  x_{barrier} = \displaystyle\frac{\pi-\delta_{w}}{k_{w}} =
    a \Bigl(1 - \displaystyle\frac{\delta_{w}}{\pi} \Bigr).
\label{eq.8.2.2.5}
\end{equation}

\item
There is no bound state at the level $E_{w}$ in the potential $V_{2}$.
Necessary conditions (1)--(2) from (\ref{eq.8.2.1.10}) are not
fulfilled.
\end{itemize}

\end{itemize}}

\subsubsection{New potentials with shift of the energy spectrum 
\label{sec.8.2.3}}

Now let's consider another case.
We assume, that the levels of the new potential $V_{2}$ do not
coincide with the levels of the energy spectrum of the starting
potential $V_{1}$ and instead of the equation (\ref{eq.8.2.2.1})
satisfy to the following:
\begin{equation}
  \cos{k_{n}a} = 0.
\label{eq.8.2.3.1}
\end{equation}
Let's analyze, whether the bound states in the potential $V_{2}$ with
such energy spectrum are possible.

At $\cos{\delta_{w}} \ne 0$ from (\ref{eq.8.2.1.18}) with taking
into account (\ref{eq.8.2.3.1}) we obtain:
\begin{equation}
  \cos{(k_{w}a+\delta_{w})}\cos{\delta_{w}} =
    - \displaystyle\frac{k_{n}^{2}}{k_{w}^{2}}
    \sin{(k_{w}a+\delta_{w})}\sin{\delta_{w}}.
\label{eq.8.2.3.2}
\end{equation}
This equation has a solution for a spectrum of values of $k_{n}$
with different numbers $n$ at one pair of $\delta_{w}$ and $k_{w}$,
if it equals to zero. From here we find:
\begin{equation}
\begin{array}{c}
  \Bigl(\cos{(k_{w}a+\delta_{w})} = 0 \Bigr) \to
    \Bigl(\sin{(k_{w}a+\delta_{w})} \ne 0 \Bigr) \to
    \Bigl(\sin{\delta_{w}} = 0 \Bigr) \to \\
    \to
    \Bigl(\cos{\delta_{w}} \ne 0 \Bigr) \to
    \Bigl(\cos{k_{w}a} = 0, \; \delta_{w} = \pi l \Bigr),
\end{array}
\label{eq.8.2.3.3}
\end{equation}
where $l=0,1,2 \ldots$.

At $\cos{\delta_{w}} = 0$ from (\ref{eq.8.2.1.19}) with taking into
account (\ref{eq.8.2.3.1}) we obtain:
\begin{equation}
\begin{array}{cc}
  \cos{k_{w}a} = 0, &
  \delta_{w} = \displaystyle\frac{\pi}{2} + \pi l.
\end{array}
\label{eq.8.2.3.4}
\end{equation}

Let's write down a full solution:
\begin{equation}
\begin{array}{cc}
  \cos{k_{w}a} = 0, & 
  \delta_{w} = \displaystyle\frac{\pi}{2} l,
\end{array}
\label{eq.8.2.3.5}
\end{equation}
and the energy spectrum for the new potential $V_{2}$:
\begin{equation}
\begin{array}{cc}
  k_{n} = \displaystyle\frac{\pi}{2a} (-1 + 2n), &
  E_{n}^{(2)} = k_{n}^{2} =
          \displaystyle\frac{\pi^{2}}{4a^{2}} (-1 + 2n)^{2},
\end{array}
\label{eq.8.2.3.6}
\end{equation}
where $n = 1,2,\ldots$.

\begin{figure}[htbp]
\centerline{
\includegraphics[width=57mm]{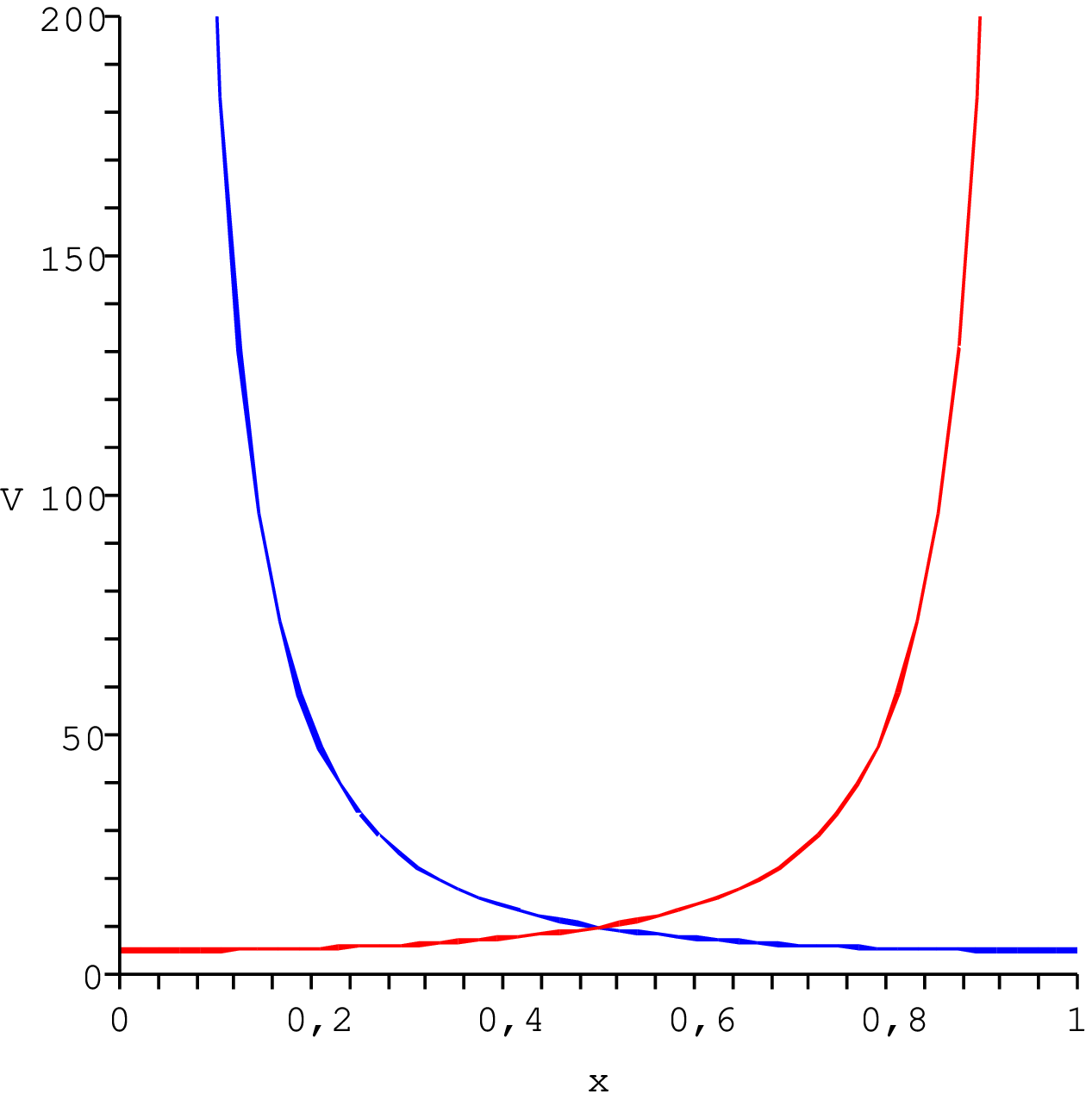}
\includegraphics[width=57mm]{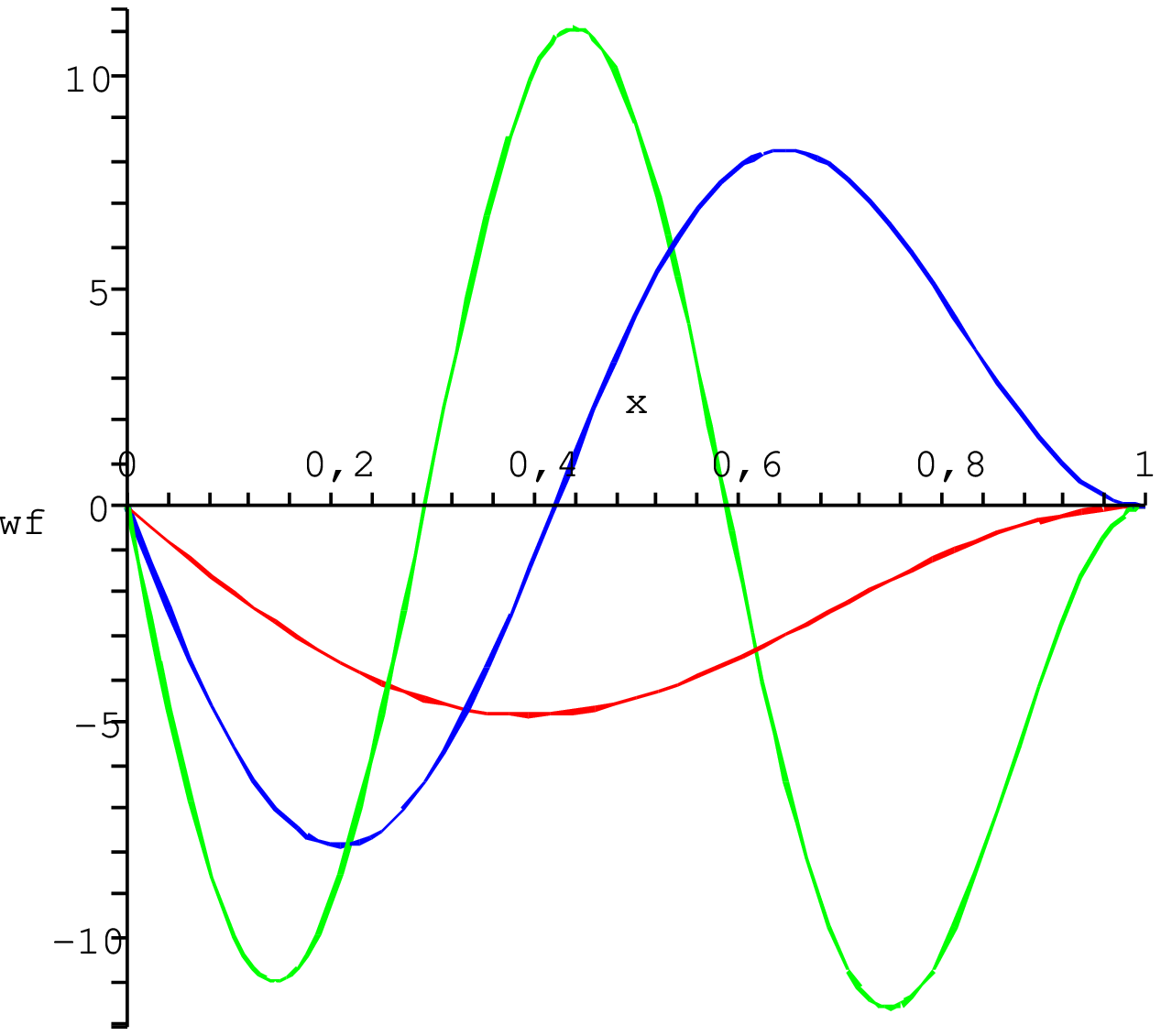}
\includegraphics[width=57mm]{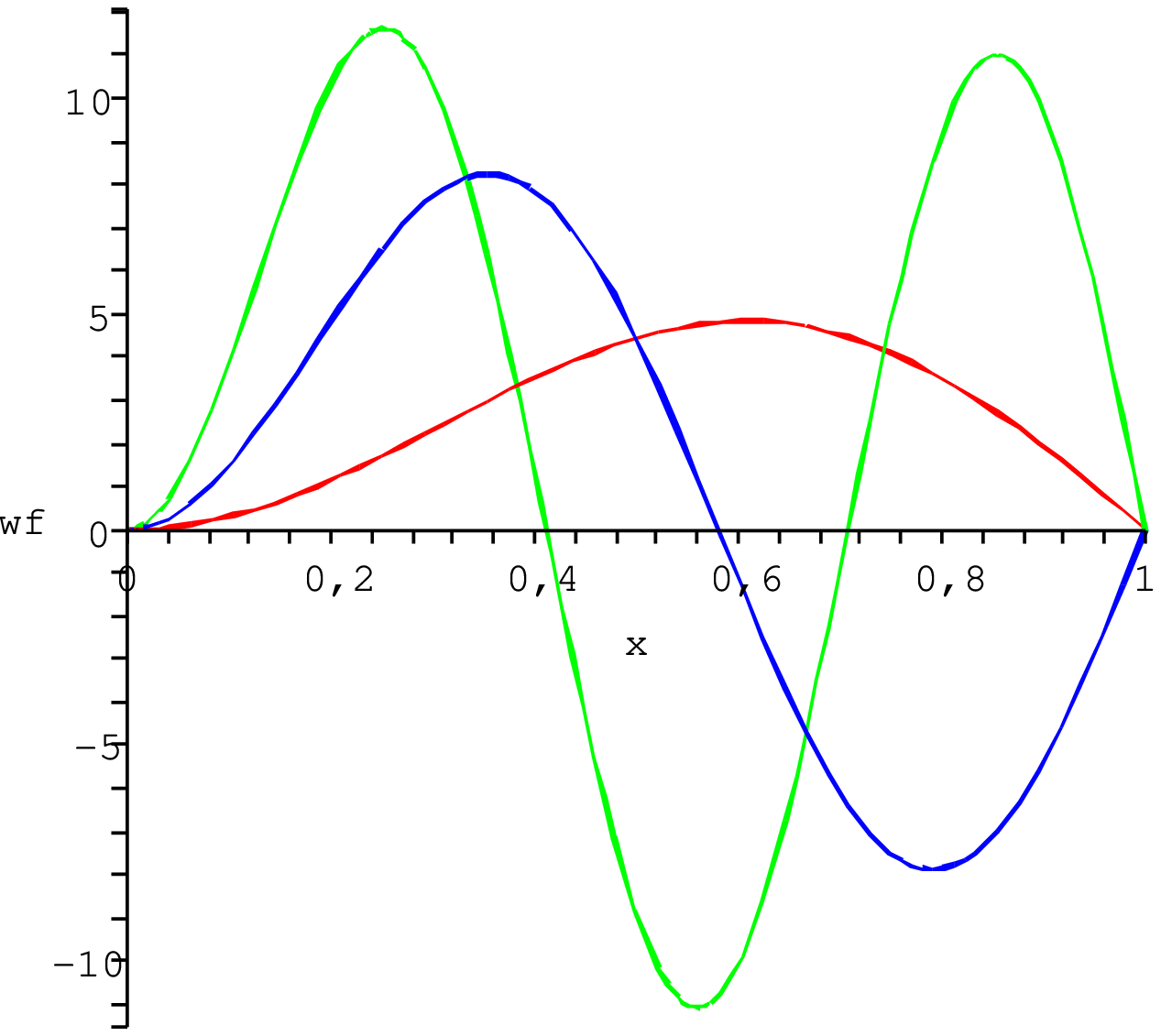}}
\caption{
\small
The potential and its WF of the first three bound states (l=0,1,2\ldots):
(a) is a shape of the potential at $\delta_{w}=\pi/2 + \pi l$
(red curve, increasing to the right) and at $\delta_{w}=\pi l$
(blue curve, increasing to the left),
(b) are WFs of the first three bound states (for the first three
lowest levels) at $\delta_{w}=\pi/2 + \pi l$,
(c) are WFs of the first three bound states at these levels at
$\delta_{w}=\pi l$
\label{fig.823}}
\end{figure}

If at $E_{w}$ to choose the lowest level from the found spectrum 
$E_{n}^{(2)}$, then one can make sure, that each level of the
spectrum $E_{n}^{(2)}$ has own bound state.
A form of the potential $V_{2}$ with such spectrum is shown in
Fig.~\ref{fig.823} (a), a shape of WF of the ground and first two 
excited states (which correspond to three lowest levels of the
spectrum (\ref{eq.8.2.3.6})) --- in Fig.~\ref{fig.823} (b,c).
From here one can see, that these WF are continuous in the region
$x \in [0,a]$ and equal to zero at points $x=0$ and $x=a$, that
allows on their basis to define the bound states at three lowest
levels for the potential $V_{2}$.
Thus, \emph{the found potential $V_{2}$ in the form 
(\ref{eq.8.2.1.1}) with the parameters (\ref{eq.8.2.3.5}) and
$k_{w} = \frac{\pi}{2a}$ is the new exactly solvable potential,
which has own energy spectrum with the bound states}.

Note, that the spectrum of $V_{2}$ is shifted relatively the
spectrum of the rectangular well. Taking into account (\ref{eq.7.1.4}),
one can write:
\begin{equation}
  k_{n} =
    \displaystyle\frac{\pi n}{a} - \displaystyle\frac{\pi}{2a} =
    k_{n, \; well} - \displaystyle\frac{\pi}{2a},
\label{eq.8.2.3.7}
\end{equation}
where $k_{n, \; well}$ is wave vector of the rectangular well with
the number $n$.

\subsection{New potentials with insertion of one additional level in
the spectrum
\label{sec.8.3}}

Now let's consider another example of the deformation of the
rectangular well with infinitely high walls (\ref{eq.7.1.1}).
We shall use the approach for the deformation of the rectangular
well in Sec.~\ref{sec.7}, where at first we fulfill using direct
SUSY-transformation a transition from the rectangular well (as the
starting potential $V_{1}$) into a new potential $V_{2}$ with
transformation of one bound state into unbound one, and then we
fulfill the inverse transition with the transformation of this
unbound state into bound one, returning to the deformed ``rectangular
well'' $V_{1}(\lambda)$.
However, in contrast to the approach in Sec.~\ref{sec.7}, now in the
construction of the superpotential we shall use an arbitrary energy
$E_{w}$ (which can be not coincident with the levels of the spectrum
for $V_{1}$) for the definition of the energy of factorization,
and the general solution of WF of the unbound state at this energy 
for the definition of the function of factorization.
So, we have the function of factorization in the form (\ref{eq.8.1.1})
(with substitution of indexes $n \to w$) and the superpotential ---
in the form (\ref{eq.8.1.2}).

Further, we shall restrict ourselves by finding such new potential,
the energy spectrum of which coincides with the energy spectrum of
the rectangular well with a possible insertion of one new level $E_{w}$.
\emph{We shall consider existence of the bound states in the found
potential as a condition of obtaining of the new exactly solvable
potential $V_{1}(\lambda)$}.
Therefore, first of all for the new potential we must to analyze a
form of WF of the arbitrary state at the arbitrary level $E_{n}^{(1)}$,
which is not coincident with the energy of factorization $E_{w}$.
One can write such WF in the form (\ref{eq.7.5.1}).

In contrast to the approach from Sec.~\ref{sec.7}, finding new
solutions we shall analyze for each level of the starting potential
$V_{1}$ not WF of the bound state, but a general solution of WF of
the unbound state (as we do it in Sec.~\ref{sec.8}),
i.~e. we write such WF at the level $E_{n}^{(1)}$ with the number
$n$ in the form (\ref{eq.8.1.1}).

In result, we obtain a condition of equality of WF to zero at the
level $E_{n}$ for the new potential $V_{1}(\lambda)$ at points $x=0$
and $x=a$ (we write it without detailed computations):
\begin{equation}
  \tan{\delta_{n}} =
    \displaystyle\frac{2 k_{n} k_{w}}{|E_{n}^{(1)} - E_{w}|}
    \displaystyle\frac
      {\cos{2(k_{w}a+\delta_{w})} - \cos{2\delta_{w}}}
      {2k_{w}a + \Bigl(\sin{2\delta_{w}} - \sin{2(k_{w}a+\delta_{w})} \Bigr)
      \biggl(1 + \displaystyle\frac{2k_{w}^{2}}
             {|E_{n}^{(1)} - E_{w}|} \biggr)}.
\label{eq.8.3.1}
\end{equation}
Choosing values for the parameters $k_{w}$ and $\delta_{w}$, from
(\ref{eq.8.3.1}) for each level $E_{n}$ with the number $n$ one can
calculate a value of $\delta_{n}$, providing the bound state at this
level for the new potential.
Therefore, we obtain the new potential, which has the energy spectrum,
coinciding completely with the energy spectrum of the old potential
with a possible insertion of one additional level $E_{w}$, if the
state at such level is bound at selected $k_{w}$ and $\delta_{w}$.

For obtaining a shape of the new potential $V_{1}(\lambda_{w})$, the
expression (\ref{eq.7.2.2}) must be modified with inclusion of the
phase $\delta_{w}$:
\begin{equation}
  V_{1}(x, \lambda_{w}, \delta_{w}) = 
  -2 \displaystyle\frac{d^{2}}{dx^{2}} \log
    \Bigl|2k_{w}x + \sin{2\delta_{w}} -
          \sin{2(k_{w}x +\delta_{w})} +
          2k_{w}a\lambda_{w} \Bigr|.
\label{eq.8.3.2}
\end{equation}

\vspace{3mm}
\noindent
\underline{\bf Conclusion:}
\emph{Inclusion of the phase $\delta_{w}$ (caused by use of WF of the
unbound state at the arbitrary energy level $E_{w}$ in the
construction of the superpotential) introduces an additional
arbitrariness into the deformation of the rectangular well.}

\section{Conclusions
\label{sec.conclusions}}

In paper a generalized definition of superpotential has proposed,
which connects two one-dimensional potentials $V_{1}$ and $V_{2}$
with discrete energy spectra completely and where:
1) for definition of the energy of factorization an arbitrary level of
the energy spectrum of the given $V_{1}$ is used and the function of
factorization is defined concerning a bound (ground or excited) state
at this energy level,
2) for the definition of the energy of factorization an arbitrary energy
(which can be not coincident with levels of the spectrum of $V_{1}$)
is used and the function of factorization is defined concerning an
unbound (or non-normalizable) state at this energy.
In finishing, let's summarize main conclusions and perspectives.
\begin{itemize}

\item
It has shown, that for the unknown superpotential such its definition
follows directly from a solution of \emph{Riccati equation} at the
given potential $V_{1}$, own independent variant of which we obtain
in the paper. The found solution for the superpotential admits the
arbitrariness on the choice of boundary conditions, imposing on wave
function, which defines the function of factorization, and on the
choice of the energy of factorization. This points out to
appropriateness of the proposed generalization of the definition of
the superpotential.

\item
It has found, that the proposed generalization of the superpotential
allows to construct new exactly solvable potentials with discrete
energy spectra (and with own bound states) on the basis of one given
$V_{1}$. This points out to perspectives of such generalization.

\item
The generalized formalism of construction of the superpotential uses
wave functions of \emph{unbound} and \emph{non-normalizable states}.
This points out perspective of further investigations of such states
for potentials with discrete energy spectra.

\item
We improve the formalism of construction of \emph{$n$-parametric
family of isospectral potentials}, developed early
in~\cite{Cooper.1995.PRPLC} (see p.~326--328), on the basis of the
proposed generalization of the construction of the superpotential.

\item
Using the rectangular well of finite width with infinitely high
external walls as the starting potential $V_{1}$ and constructing the
superpotential, where
the energy of factorization coincides with arbitrary (not only the
lowest) level of the energy spectrum of the given $V_{1}$ and
the function of factorization is defined as WF of bound state at such
level,
by SUSY QM approach (the method from Sec.~\ref{sec.5.1}) we
reconstruct exactly all pictures of deformation (without displacement
of the levels) of this potential and its WFs of the lowest bound
states, which were obtained early in review
\cite{Zakhariev.1990.PEPAN} (see p.~916--919) by the methods of the
inverse problem. Note the following:

\begin{itemize}
\item
An appropriateness of the SUSY QM method from Sec.~\ref{sec.5.1} for
construction of new isospectral potentials (having own bound states
without divergences) has been shown.

\item
The parameter $\lambda_{m}$ at arbitrary $m$ of the level of the
bound state in the SUSY QM method from Sec.~\ref{sec.5.1} plays the
same role in the deformation of the potential and its WFs,
as derivative $\gamma_{m}$ of WF of the bound state at such level in
one boundary of the potential, used for the deformation of the
potential and its WFs in the approach of the inverse problem
(see \cite{Zakhariev.1990.PEPAN} p.~917--927).

\item
For WF of the bound state at the arbitrary level $E_{m}$ with the
number $m$ of the given $V_{1}$, if the energy of factorization
$\cal E$ coincides with this level (${\cal E} = E_{m}$), under its
deformation using the parameter $\lambda_{m}$ an analysis of behavior
of nodes of this WF is fulfilled, their coordinates are found, an
analysis of angles of its leaving at points of right and left walls
is fulfilled.

\item
For WF of the bound state at the arbitrary level $E_{n}$ with the
number $n$, if the energy of factorization $\cal E$ does not coincide
with this level ($E_{n} \ne {\cal E} = E_{m}$), under the deformation
by the parameter $\lambda_{m}$ such points are found, where all
deformed curves of this WF intersect between themselves and with
the non-deformed WF, analysis for these points is fulfilled and their
classification is proposed, it has shown, that angles of leaving of
this WF at points $x=0$ and $x=a$ are not changed (in contrast to WF
of the state with the number $m$, see point above).

\item
An analysis of the deformation of the shape of the well under change
of the parameter $\lambda_{m}$ has fulfilled, an analysis of
zero-points of the well (i.~e. points of intersection of the potential
with axis $x$) has fulfilled, their classification has proposed.
\end{itemize}
This all makes SUSY QM methods at their efficiency practically
achieved the level of the methods of the inverse problem.

\item
Constructing the superpotential on the basis of WF of the unbound
state at arbitrary energy of factorization (which can be located
higher then the lowest level), we obtain new types of the deformation
of the rectangular well with infinitely high external walls and its
FWs of the lowest bound states.
Using such deformation one can keep all levels of the energy spectrum
for the given $V_{1}$ at their locations (obtaining new isospectral
potentials), or one can displace all levels of this spectrum by given
rule.
In particular, using only one superpotential, one can join two
potentials, energy spectra of which (with own bound states) have no
one coincident level (it has found at the first time in the
supersymmetric approach)!
Perhaps, such a way will open a possibility to connect together
boson and fermion components with absolutely different real energy
spectra in the supersymmetric quantum theory of fields, using only
one superpotential.

\end{itemize}

{\small
\bibliographystyle{h-physrev4}
\bibliography{Wgen}}

\end{sloppypar}
\end{document}